\documentclass[12pt]{article}

\usepackage{geometry}
\usepackage{mathrsfs} 
\usepackage{indentfirst}

\usepackage{amsmath}
\usepackage{amssymb}
\usepackage{amsthm}
\newtheorem{lemma}{Lemma}
\theoremstyle{remark}
\newtheorem{remark}{Remark}
\theoremstyle{plain}
\newtheorem{proposition}{Proposition}
\newtheorem{assumption}{Assumption}

\newtheorem{theorem}{Theorem}
\newtheorem{corollary}{Corollary}

\usepackage[round]{natbib}

\usepackage[table]{xcolor}
\usepackage{tikz}

\usepackage{setspace}
\usepackage{hyperref}
\hypersetup{colorlinks=true, linkcolor=blue, citecolor=blue, urlcolor=blue}
\usepackage{url}
\usepackage{booktabs}
\usepackage{amsfonts}
\usepackage{nicefrac}
\usepackage{float}
\usepackage{caption}
\usepackage{enumitem}

\usepackage{graphicx}
\graphicspath{{graphs_tables/}}

\usepackage[english]{babel}

\begin{document}
	
	\doublespacing
	
	\begin{titlepage}
		\title{\vspace{-1cm} Sharp Identification for Regressions \\ with Interval-Observed and Missing Covariates\thanks{We thank Francesca Molinari, Arie Beresteanu, Federico Bugni, Toru Kitagawa, Ivan Canay, Charles Manski, Filip Obradovic, Pamela Giustinelli, Massimiliano Marcellino, V\'ictor Sancibri\'an, Joachim Freyberger, David Pacini, Chris Muris, Ismael Mourifié, Marc Henry, Daniel Wilhelm, Joachim Winter, Raffaella Giacomini and participants at the IAAE 2026 in Lisbon, as well as the 31st Panel Data Conference in Exeter for helpful discussions.} \\
			{\small }}
		\vspace{-0.5cm}
		\author{Martin Fankhauser\thanks{Bocconi University},
			Gil Peled\thanks{Cornell University}}
		\date{November 2026}
		\maketitle
		\vspace{-1cm} 
		\begin{abstract}
			Researchers frequently estimate regressions when key variables are missing or observed only through intervals. Imputation or assumptions about the observation process can yield seemingly precise but misleading conclusions. We develop methods for identification and inference that remain robust to such incomplete data. For a broad class of conditional moment models, we characterize the sharp identified set, the set of all parameter values consistent with the observed data and maintained regression model, without restricting how incomplete observations arise. We establish Hausdorff-consistent estimation and uniformly valid confidence regions, and develop computationally tractable implementations for common regression models. In an application relating interval-elicited discount rates to on-time high-school graduation, conventional inference excludes zero, whereas our interval-data-robust confidence regions include it. Exploiting sharp identification nevertheless yields substantially tighter confidence regions than non-sharp alternatives. \\
			\par\vspace{-0.2cm}
			
			\begingroup
			\setstretch{1}
			 \par
			\endgroup
		\end{abstract}
		\setcounter{page}{0}
		\thispagestyle{empty}
	\end{titlepage}
	\pagebreak \newpage

	\section{Introduction}
	\label{sec:introduction}
	
	\noindent Regression models are ubiquitous in applied economics, yet the variables entering them are often not observed exactly. Outcomes and covariates may be missing because of nonresponse or attrition, reported only in brackets in household surveys, top-coded in administrative data, or observed through intervals generated by experimental and survey designs. In each case, the observed data leave ambiguity about the latent variables entering the regression, so that the parameters of interest are typically only partially identified. \\
	\noindent Empirical work commonly addresses incomplete data through sample restrictions, imputation, or assumptions about the data generating process. These choices reflect the trade-off emphasized in \citet{manski2003} in his \emph{law of decreasing credibility}: stronger assumptions can deliver more informative conclusions at the cost of credibility. This paper therefore asks: what does the maintained regression model identify from the incomplete data alone, without additional assumptions about how missing or interval observations are generated? \\
	\noindent As formalized in Section \ref{sec:identification}, we study conditional moment models of the form
	\begin{align*}
		\mathbf{E}\left[m(Y,X,\theta_0)\mid X\right]=0,
	\end{align*}
	where some components of the non-conditioning variables $Y$, the conditioning variables $X$, or both are not observed exactly. This class includes, for instance, linear, probit, or logit regressions as well as more general econometric models characterized by conditional moment restrictions. Let $W\equiv(Y,X)$ denote the latent variables entering the model and let $W^O\equiv(Y_l,Y_u,X_l,X_u)$ denote the observed intervals. The observed data imply the containment restriction $Y\in[Y_l,Y_u]$ and $X\in[X_l,X_u]$ almost surely. Our setting nests the missing data case if the observed intervals span the full support of the respective variables.\\ 
	\noindent Our central identifying restriction is deliberately weak. Beyond this containment requirement, we leave the process generating incomplete observations unrestricted: whether a variable is missing, as well as the location and width of an observed interval, may depend on the latent variables themselves. In particular, we do not impose missingness-at-random or other restrictions on the relationship between the observation process and the latent variables. This approach gives a benchmark for assessing what additional assumptions contribute. Whether such weak restrictions yield informative conclusions is an empirical question, which our application addresses.\\
	\noindent Our object of interest is the sharp identified set $\Theta_I^*$: the set of parameter values for which a single completion of the latent data reproduces the distribution of the observed data, respects the reported intervals, and satisfies the conditional moment restriction.\footnote{$\Theta_I^*$ contains $\theta_0$ whenever the containment restriction and the maintained model hold. An empty identified set therefore rejects the model, whereas a nonempty set establishes only compatibility with at least one admissible latent distribution, not that the model holds under the true one.} Characterizing this set is difficult for two reasons. The conditional moment restriction is equivalent to infinitely many unconditional moment restrictions, and a single completion must satisfy all of them jointly. Existing procedures for missing covariates \citep{aucejo2017} check these restrictions separately and therefore deliver outer sets. In addition, the restriction conditions on covariates that are themselves incompletely observed. Random-set methods characterize sharp sets for moment restrictions that are unconditional or condition on observed variables, including best linear prediction with interval covariates \citep{beresteanu2011}, but not a conditional mean restriction given an incompletely observed covariate.\footnote{In the linear model, the difference is between $\mathbf E[(1,X)^\top(Y-\alpha-\beta X)]=0$ and $\mathbf E[Y-\alpha-\beta X\mid X]=0$ when $X$ is only interval-observed.} We make three contributions. \\
	\noindent First, we characterize the sharp identified set for the full conditional moment restriction in terms of the observed-data distribution alone. The key step is a single random set that encodes the entire conditional restriction: we interact the moment function with a countable class of instruments, evaluated at the latent covariates, and collect the resulting moments for all latent values consistent with the reported intervals. Weighting the instruments makes the infinitely many moments summable, so that this random set is bounded and well defined. A parameter belongs to the identified set if and only if the Aumann expectation of this random set contains zero, or equivalently, if a family of support-function inequalities holds under the observed-data distribution. This extends the characterization of \citet{beresteanu2011} to restrictions that condition on incompletely observed covariates and therefore involve infinitely many moments. The extension requires handling discontinuous instruments, such as indicators of cells partitioning the covariate space, which are standard in practice and enable exact linear-programming computation. With such instruments, the Aumann expectation equals only the closure of the set of moment vectors attainable by admissible completions; we show that this closure is harmless at zero. \\
	\noindent Second, estimation and inference require additional arguments as the instrument collection expands. For set estimation, we establish a triangular-array law of large numbers for random sets of growing dimension and use it to verify the consistency conditions of \citet{chernozhukov2007}. For uniformly valid confidence regions for $\theta_0$, we verify the manageability conditions required by \citet{andrews2017} for the support-function process indexed by an \emph{uncountable} class of directions and the instrument class. \\
	\noindent Third, we make the procedure operational. The theory involves infinitely many moments and an uncountable class of support-function directions; in practice we use finitely many of each, characterize the resulting outer sets, and show that inference remains valid even when the directions are selected from the data.\footnote{Finite approximations are standard in the many-moment inequality literature; see, for example, \citet{andrews2017}.} For linear, probit, and logit regressions, the nested minimax problem underlying the estimation criterion admits an exact linear-programming representation, so it can be solved with standard solvers. We propose two ways to choose directions for inference. An adaptive procedure selects informative directions on one half of the sample, tests on the other half, swaps the roles of the halves, and intersects the two confidence regions. A fixed procedure instead uses a prespecified set of directions and the full sample. The first concentrates on the most informative inequalities; the second uses all observations for testing. In Monte Carlo simulations, both maintain approximately nominal coverage and gain power as the sample size grows. Under random censoring, the power gap relative to complete-case OLS narrows with sample size; under outcome-dependent censoring, our procedures remain reliable while conventional methods lose coverage. \\
	\noindent Our empirical application examines how conclusions change
	as we progressively account for different forms of incomplete
	observation. We revisit \citet{castillo2018}, who use a multiple
	price list (MPL) to elicit discount rates and study their
	relationship with on-time high-school graduation. Because the
	offered rates lie on a discrete grid, even consistent responses
	reveal only an interval containing the latent discount rate.
	A conventional probit based on midpoint imputation yields a
	marginal effect of $-0.086$ with a $95\%$ confidence interval
	of $[-0.152,-0.020]$. Allowing discount rates to vary within their elicited MPL intervals
	yields a $95\%$ confidence set of $[-0.226,0.020]$, approximately
	$27\%$ wider than the corresponding midpoint-imputation benchmark
	under the same inference procedure, and renders the previously
	significant association inconclusive. Allowing for inconsistent switching
	patterns and missing graduation outcomes widens the confidence
	sets further. Nevertheless, our sharp approach remains
	substantially more informative than non-sharp outer
	approximations: for the basic MPL intervals, the procedure
	of \citet{aucejo2017} yields $[-0.396,0.403]$.
	Finally, sensitivity exercises show how the gains from more
	precise measurement depend on whether uncertainty is reduced
	for all respondents or eliminated for a subset. This
	illustrates how our framework can assess both the robustness
	of empirical conclusions and the potential value of
	additional identifying assumptions or improved data collection. \\
	\noindent Our work is most closely related to the partial-identification literature on regressions with interval or missing covariates. \citet{manski2002} study regression models with interval-observed variables and obtain sharp identification results under an auxiliary mean-independence restriction linking the latent variable to its observed interval. \citet{asher2018} build on their work and impose further assumptions. In contrast, we retain only the containment restriction and otherwise leave the observation process unrestricted. \citet{aucejo2017} study conditional moment models with missing covariates and develop estimation and inference procedures for outer sets, i.e., supersets of the sharp identified set. In comparison, our work imposes very similar assumptions but delivers results for the sharp identified set and can handle interval-valued data. \citet{schennach2014} develops a more general entropy-based approach to latent-variable models and extends it to conditional moment restrictions through a growing collection of unconditional moments.\footnote{Applying this framework to our setting would still require developing estimation, inference, and computational procedures for the resulting growing-moment problem. Some of the tools developed in this paper for handling an expanding collection of moments could potentially also be used for such an implementation.} Our more specialized support-function approach instead exploits the structure of interval and missing observations directly, yielding sharp identification together with estimation, inference, and computational procedures tailored to regression models as delineated in Section \ref{sec:implementation}. \\
	\noindent A related literature obtains sharp identification when the relevant restrictions are finite-dimensional. \citet{horowitz1998} derive sharp worst-case bounds for conditional expectations of the form $\mathbf{E}[g(Y)\mid X\in A]$ when outcomes, covariates, or both may be missing, while \citet{horowitz2000} specialize this approach to randomized experiments and bound conditional response probabilities and treatment contrasts. In treatment-effect settings, \citet{lee2009} obtains sharp trimming bounds under random assignment and monotonicity of treatment-induced selection, with subsequent work extending this approach under related restrictions on selection. A distinct methodological strand uses random sets and support functions to characterize sharp identification from finitely many moment restrictions. Our approach builds directly on \citet{beresteanu2008,beresteanu2011}, who characterize feasible moment predictions through Aumann expectations and support-function inequalities; \citet{beresteanu2011} apply this framework to best linear prediction with interval data, improving on the computationally demanding characterization in \citet{horowitz2006}. \citet{chesher2017,chesher2020} use random sets to characterize identified sets in generalized instrumental variable models, including models with an interval-censored endogenous regressor, under restrictions conditional on observed instruments. More recently, \citet{moon2026} extend the same logic to GMM models with unrestricted, potentially endogenous missingness. In contrast, our restriction conditions on covariates that are themselves incompletely observed, which requires infinitely many moments and, in practice, discontinuous instruments. More broadly, the paper follows the partial-identification approach pioneered by \citet{manski2005,manski2008}; see \citet{molinari2020}, \citet{canay2017}, and \citet{tamer2023} for reviews. \\
	\noindent A separate literature restores point identification by placing stronger restrictions on how incomplete observations are generated or by modeling the missing values themselves. Missing-data methods commonly invoke ignorability or missing-at-random assumptions and use likelihood, multiple imputation, or inverse-probability weighting \citep{rubin1976,rubin1987,robins1995,wooldridge2007}; in longitudinal settings, related weighting approaches are routinely used to address attrition \citep{fitzgerald1998}. Selection models instead specify how observation depends on latent outcomes or unobservables, ranging from the classical parametric correction of \citet{heckman1979} to semiparametric approaches such as \citet{newey2009}. For more general forms of partial observation, \citet{heitjan1991} formulate conditions under which the coarsening mechanism can be ignored. Other approaches exploit auxiliary information, validation samples, restrictions on reporting behavior, or explicit models for the latent values within observed intervals \citep{hsiao1983,rigobon2009,pacini2018,duan2025,freyberger2025,bretschneider2026}. These methods can be considerably more informative when their assumptions are credible. Our identified set instead provides an assumption-minimal benchmark: each additional restriction can be viewed as eliminating latent completions from the larger set $\Gamma$ of all completions feasible under the containment restriction; see Section~\ref{sec:identification}. \\
	The remainder of the paper is organized as follows. Section~\ref{sec:identification} derives the sharp identification result. Section~\ref{sec:estimation} develops estimation and inference, and Section \ref{sec:implementation} presents the computational implementation. Section~\ref{sec:simulation} studies finite-sample performance, Section~\ref{sec:application} presents the empirical application, and Section~\ref{sec:conclusion} concludes.
	
	\paragraph{Notation.}
	Throughout the paper, $\ell_1$ and $\ell_\infty$ denote the usual sequence spaces equipped with norms $|\cdot|_1$ and $|\cdot|_\infty$, respectively, and $\langle u,a\rangle$ denotes the dual pairing between $u\in\ell_\infty=(\ell_1)^*$ and $a\in\ell_1$. $h(A,u)\equiv\sup_{a\in A}\langle u,a\rangle$ denotes the support function of a set $A\subseteq\ell_1$. For a random set $Q$, $\mathbb E[Q]=\operatorname{cl} \left\{ \mathbf{E}[q]: \ q\text{ is an integrable selection of }Q \right\}$ denotes its Aumann expectation, whereas $\mathbf E_F[Z]$ denotes the ordinary expectation of a random element $Z$ under the probability law $F$, with the subscript suppressed when clear from context. We write $\rho_H(A,B)$ for the Hausdorff distance and $\vec\rho_H(A,B)$ for the directed Hausdorff distance. The symbols $A\oplus B$ and $\operatorname{co}(A)$ denote the Minkowski sum and convex hull, respectively. Finally, for a metric space $\mathcal S$, $\Delta(\mathcal S)$ denotes the set of Borel probability measures on $\mathcal S$, and $\mathcal L_F(Z)$ denotes the law of $Z$ under $F$.

	\section{Identification}
	\label{sec:identification}
	
	Formally, we study conditional moment models of the form
	\begin{align}
		\label{eq:model1}
		\mathbf{E}_{F_0}\left[m(Y,X,\theta_0)\mid X\right]=0
		\qquad F_X\text{-a.s.}
	\end{align}
	Let $(\Omega,\mathcal A,F_0)$ denote the underlying probability space. The random variables $Y$ and $X$ take values in $\mathcal Y\subseteq\mathbb R^{d_y}$ and $\mathcal X\subseteq\mathbb R^{d_x}$, respectively. The known moment function
	$m:\mathbb R^{d_y}\times\mathbb R^{d_x}\times \Theta
	\to\mathbb R^{d_m}$
	satisfies \eqref{eq:model1} at the true parameter $\theta_0\in\Theta$, should it exist. We assume $d_y,d_x,d_m<\infty$; we allow $\Theta$ to be arbitrary for the purpose of identification; however, in Section \ref{sec:estimation} we restrict attention to $\Theta \subset \mathbb R^{d_\theta}$.  The source of partial identification is not an incomplete model, but incomplete observation of the variables entering it.
	\begin{assumption}
		\label{A:assumption1}
		Let $W\equiv(Y,X)$ take values in
		$\mathcal W\equiv\mathcal Y\times\mathcal X$, where $\mathcal Y\subset\mathbb R^{d_y}$ and $\mathcal X\subset\mathbb R^{d_x}$ are compact. Let $W^O=(Y_l,Y_u,X_l,X_u)$ be observed on the same probability space as $W$ and take values in $\mathcal W^O\equiv\mathcal Y^2\times\mathcal X^2$, with nonempty componentwise intervals almost surely. Assume
		\begin{align*}
			P_{F_0}\left[
			Y\in[Y_l,Y_u],
			X\in[X_l,X_u]
			\right]=1.
		\end{align*}
		We find it convenient at times to write this last condition as 
		$P_{F_0}(W \in D(W^O))=1$, for $D(W^O)\equiv [Y_l,Y_u]\times [X_l,X_u]$.
	\end{assumption}
	Assumption~\ref{A:assumption1} contains the central identifying stance of the paper. Apart from requiring the latent realization to lie inside its observed interval, we place no further restrictions on how incompleteness is generated. A fully observed value is represented by a degenerate interval, while a missing value is represented by the variable's full support. The purpose of the analysis is precisely to determine what follows before adding any stronger assumptions. \\
	Let us first define the identified set for our setting.
	Assumption~\ref{A:assumption1} implicitly defines the set of admissible completions
	\begin{align*}
		\Gamma
		\equiv
		\left\{
		\eta\in\Delta(\mathcal W\times\mathcal W^O):
		\eta_{\mathcal W^O}=P_{W^O},
		\quad
		\eta\big(\{(w,w^o):w\in D(w^o)\}\big)=1
		\right\}
	\end{align*}
	An element of $\Gamma$ is a candidate joint law for the latent and observed data that has the observed marginal $P_{W^O}$ and assigns probability one to latent values compatible with the reported intervals. Such a law cannot be ruled out from the observed data alone. Assumption~\ref{A:assumption1} also guarantees nonemptiness, because $\eta_0\equiv\mathcal L_{F_0}(W,W^O)\in\Gamma$. \\
	The identified set being discussed is defined directly from the maintained model
	\eqref{eq:model1} and Assumption~\ref{A:assumption1} as
	\begin{align}
		\label{eq:sharp-set}
		\Theta_I^*
		\equiv
		\left\{
		\theta\in\Theta:
		\exists\eta\in\Gamma\text{ such that }
		\mathbf E_\eta[m(Y,X,\theta)\mid X]=0
		\quad \eta_X\text{-a.s.}
		\right\}.
	\end{align}
	The set $\Theta_I^*$ may be empty. This natural definition of the identified set may sometimes be redefined in convenient ways. In particular, \citet{schennach2014} considers a distance-based formulation for model \eqref{eq:model1} expressed with the condition $\Vert \mathbb{E}_{\eta}[m(Y,X,\theta)|X]\Vert =0$, $F_X$-almost surely. The precise relationship between the original identified set and that based on a distance-based formulation has not been previously examined to our knowledge. In Appendix Section~\ref{sec:schennach-definition} we establish that in our setting the two identified sets coincide, show that the relationship between them can be expressed well through random sets, and provide a sufficient condition for equivalence. We do not discuss this distance-based formulation any further in the main text of this paper.\\
	The characterization of $\Theta_I^*$ in terms of conditional expectations is inconvenient as the conditioning event is not observed. We begin by replacing the conditional equality by a (countably) infinite collection of unconditional equalities using instrument functions, as is standard in the conditional-moment literature; see, among others, \citet{andrews2013}, \citet{andrews2017} and \citet{aucejo2017}. In Lemma \ref{L:lemma1} we establish that this can be performed without changing the identified set. The following assumption gives general conditions on the instrument class used, whereas in the Appendix Section~\ref{sec:weigthedcubes} we verify them for the class of weighted cubes used in our applications.
	
	\begin{assumption}
		\label{A:assumption2}
		\leavevmode
		\begin{enumerate}[label=(\roman*)]
			\item Let $\mathcal H=\{h_k:k\geq1\}$ be a countable class of bounded Borel functions on $\mathcal X$ that span the class of continuous functions. Precisely, for every $f\in C(\mathcal X)$ and every $\varepsilon>0$, there exists $v_\varepsilon\in\operatorname{span}(\mathcal H)$   satisfying
			\[
			\|f-v_\varepsilon\|_\infty<\varepsilon.
			\]
			\item Let $\mathcal G=\{g_k:k\geq1\}$ be the weighted instrument class defined by $g_k=\omega_k h_k$,
			
			where the deterministic weights $\omega_k>0$ are chosen so that    \[
			\lim_{K\to\infty}
			\sup_{x\in\mathcal X}
			\sum_{k>K}|g_k(x)|=0.
			\]
			In particular, weights decaying at the rate $O(k^{-\beta})$ for some $\beta > 1$ achieve this.
		\end{enumerate}
	\end{assumption}
	
	Part~(i) ensures that the instruments reveal any nonzero conditional moment; note that $\text{span}(\mathcal H)=\text{span}(\mathcal G)$. This mimics Assumption MFA in \citet{andrews2017} and Assumption CI in \citet{andrews2013}, which both require the instrument class to be rich enough that the set of $x$ violating the conditional moments has positive measure if and only if some instrument reveals that fact.\footnote{These formulations are based on what \citet{stinchombe1998} refer to as \textit{comprehensively revealing}.} Part~(ii) ensures that the stacked moment vector lives in $\ell_1$ and, combined with a uniform integrability condition (supplied by Assumption \ref{A:bounded-moments}), is also compact. This setting handles the technical difficulties of the space and ensures finite truncations are uniformly accurate.  \\
	We reformulate the identified set using the unconditional-moment set
	\begin{align}
		\label{eq:unconditional-set}
		\Theta_I
		\equiv
		\left\{
		\theta\in\Theta:
		\begin{array}{l}
			\exists\eta\in\Gamma\text{ such that}\\[2pt]
			\displaystyle
			\int_{\mathcal W\times\mathcal W^O}
			m_j(y,x,\theta) g_k(x) 
			\eta(dy,dx,dw^o)=0,\\[-1pt]
			\hfill j=1,\ldots,d_m,\quad k\geq1,
		\end{array}
		\right\},
	\end{align}
	With this definition, we state the first auxiliary lemma.
	\begin{lemma}
		\label{L:lemma1}
		Let $\mathcal X$ be compact and let Assumption~\ref{A:assumption2}(i) hold. Suppose that
		\[
		\mathbf E_\eta\!\left[\|m(Y,X,\theta)\|_\infty\right]
		<\infty
		\qquad
		\text{for all }(\theta,\eta)\in\Theta\times\Gamma.
		\]
		Then, for every $(\theta,\eta)\in\Theta\times\Gamma$,
		\begin{align*}
			\mathbf E_\eta[m(Y,X,\theta)\mid X]=0
			\quad\eta_X\text{-a.s.}
			\qquad\Longleftrightarrow\qquad
			\int m_j(y,x,\theta) g_k(x)\,d\eta=0
			\ \text{ for every }(j,k).
		\end{align*}
		Consequently, $\Theta_I=\Theta_I^*$.
		
		Furthermore, given our weights $\omega_k>0$, one could equivalently define $\Theta_I$ and this lemma using the unweighted class $\mathcal H$ with elements $h_k(x)$ instead of $g_k(x)$. 
	\end{lemma}
	
	It is noted that the compactness and integrability conditions in Lemma~\ref{L:lemma1} are ensured by Assumptions~\ref{A:assumption1} and~\ref{A:bounded-moments}, respectively. Lemma~\ref{L:lemma1} is the well-known property that a conditional moment equality is equivalent to an infinite family of unconditional moment equalities, and it isolates the central feasibility requirement: one admissible completion must satisfy every unconditional restriction simultaneously. \citet{aucejo2017} derive a similar result in their Lemma~1 and note, as we do, that verifying the existence of such an $\eta$ is extremely difficult and hard to make progress on in this direct formulation. \citet{andrews2017} and \citet{chernozhukov2017} likewise work with infinite families of moment (in-)equalities, but typically assume that the variables entering those moments are observable, whereas here $\eta$ is not identified. The proof is in Appendix~\ref{sec:appendixA}. An immediate implication is instrument-class invariance: every class satisfying Assumption~\ref{A:assumption2} generates the same population set $\Theta_I^*$. \\
	Lemma~\ref{L:lemma1} removes searching over the unobserved conditioning event in the original model and thus makes a random-set formulation possible. For each candidate $\theta$, making use of our weighted instruments $\mathcal G$, we collect all moments in an $\ell_1$-valued random set and ask whether its expectation contains zero. This logic follows the seminal paper by \citet{beresteanu2011}. We extend their framework to accommodate an infinite-dimensional moment vector and a discontinuous instrument map. In general, random-set methods were introduced to econometrics by \citet{beresteanu2008}; see \citet{molchanov2017,molchanov2018} for
	textbook treatments. \\
	For $\theta\in\Theta$, define
	\begin{align*}
		M_\theta(y,x)
		\equiv
		\left(
		m_j(y,x,\theta)g_k(x)
		\right)_{j=1,\ldots,d_m,\ k\geq1}
		\in\ell_1.
	\end{align*}
	Let $\eta_0=\mathcal L_{F_0}(W,W^O)$ denote the true joint law of the latent and observed variables. By Lemma~\ref{L:lemma1}, the model in \eqref{eq:model1} can equivalently be written as
	\begin{align*}
		\int_{\mathcal W\times\mathcal W^O}
		M_{\theta_0}(y,x)\,\eta_0(dy,dx,dw^o)=0.
	\end{align*}
	Assumption~\ref{A:assumption1} implies $\eta_0\in\Gamma$, but the observed data generally do not determine which element of $\Gamma$ is the true joint law. For each candidate $\theta$, we therefore consider the set of attainable
	mean-moment vectors
	\begin{align}
		\label{eq:attainable-moments}
		I_\theta
		\equiv
		\left\{
		\int_{\mathcal W\times\mathcal W^O}
		M_\theta(y,x)\,\eta(dy,dx,dw^o):
		\eta\in\Gamma
		\right\}.
	\end{align}
	Thus,
	\begin{align*}
		\theta\in\Theta_I
		\quad\Longleftrightarrow\quad
		0\in I_\theta.
	\end{align*}
	
	Checking this condition directly requires finding an admissible joint law that satisfies all moment restrictions simultaneously. To obtain a more tractable characterization, we first collect the moment vectors compatible with each observed interval:
	\begin{align*}
		Q_\theta(W^O)
		\equiv
		\operatorname{cl}_{\ell_1}
		\left\{
		M_\theta(y,x):(y,x)\in D(W^O)
		\right\}.
	\end{align*}
	The set $Q_\theta(W^O)$ contains the moment vectors generated by feasible latent values, together with their limits in $\ell_1$. We next show how its Aumann expectation characterizes whether $I_\theta$ contains zero, leading to a support-function criterion evaluated under the observed-data distribution. We note that $Q_\theta(W^O)$ depends only on the observed data. 
	
	\begin{assumption}
		\label{A:bounded-moments}
		\leavevmode
		\begin{enumerate}[label=(\roman*)]
			\item For every $\theta\in\Theta$, the map
			$(y,x)\mapsto m(y,x,\theta)$ is continuous on $\mathcal W$.
			
			\item There exists a constant $C_1<\infty$ such that
			\[
			\sup_{(y,x,\theta)\in\mathcal W\times\Theta}
			\|m(y,x,\theta)\|_\infty\leq C_1.
			\]
		\end{enumerate}
	\end{assumption}
	
	Assumption~\ref{A:bounded-moments} is satisfied by common linear-mean, logit, and probit specifications on bounded domains. Because Assumption~\ref{A:assumption1} makes $\mathcal W$ compact, continuity already
	gives boundedness for each fixed $\theta$; Assumption~\ref{A:bounded-moments} supplies a common bound over $\Theta$ and therefore keeps the random sets integrably bounded under the conditions of Lemma~\ref{L:lemma1}. The Aumann expectations below are understood as the collection of expectations of ordinary random variables \citep{aumann1965}.
	
	\begin{theorem}
		\label{T:theorem1}
		Let Assumptions~\ref{A:assumption1}--\ref{A:bounded-moments} hold. Then $Q_\theta(W^O)$ is almost surely a nonempty compact random set in $\ell_1$, and $\mathbb E[Q_\theta(W^O)]$ is a nonempty closed convex subset of $\ell_1$.
		Moreover,
		\begin{align}
			\label{eq:aumann-identification}
			\theta\in\Theta_I^*
			\quad\Longleftrightarrow\quad
			0\in\mathbb E[Q_\theta(W^O)].
		\end{align}
		Therefore,
		\begin{align}
			\label{eq:support-identification}
			\Theta_I^*
			&=
			\left\{
			\theta\in\Theta:
			0\in\mathbb E[Q_\theta(W^O)]
			\right\}\notag\\
			&=
			\left\{
			\theta\in\Theta:
			\mathbf{E}_{P_{W^O}}
			[h(Q_\theta(W^O),u)]\geq0
			\quad\forall u\in B^\infty
			\right\} \\
			&=
			\left\{
			\theta\in\Theta:
			\inf_{u \in B^\infty}\mathbf{E}_{P_{W^O}}
			[h(Q_\theta(W^O),u)]\geq0 
			\right\} \\
			&=
			\left\{
			\theta\in\Theta:
			\inf_{u \in B^\infty}\mathbf{E}_{P_{W^O}}
			[\sup_{(x,y)\in D(W^O)} \langle M_\theta(y,x),u\rangle ]\geq0
			\right \},
		\end{align}
		where $B^\infty:=\{u\in\ell_\infty:\|u\|_\infty\leq1\}$.
	\end{theorem}
	
	The proof is in Appendix~\ref{sec:proof-theorem1}. Here we only sketch the arguments. First, we establish that the random-set construction is well defined and we relate it to the original feasibility problem. The appendix verifies the required random-set conditions, namely that $Q_\theta$ is measurable and integrably bounded. The Aumann expectation \(\mathbb E[Q_\theta(W^O)]\) is therefore well defined and nonempty; it is closed by definition and convex.\footnote{Aumann expectations on all appropriate non-atomic probability spaces are convex. Non-atomicity is often a trivial requirement.  Without loss of generality the Aumann expectation is taken on the uniform enrichment of the original probability space described in Appendix~\ref{sec:proof-theorem1}. A similar construction is used in the proof of the Artstein Selection Theorem in \citet{molchanov2017}.}  \\
	We next connect the random set to the original identification problem. By Lemma~\ref{L:lemma1} and the definition of $I_\theta$:
	\begin{align*}
		\theta\in\Theta_I^*
		\quad\Longleftrightarrow\quad
		\theta\in\Theta_I
		\quad\Longleftrightarrow\quad
		0\in I_\theta.
	\end{align*}
	The remaining step is to relate the attainable set \(I_\theta\) to the Aumann expectation of \(Q_\theta(W^O)\). This is also where the main infinite-dimensional subtlety enters. If \(M_\theta\) were continuous, compactness of \(D(W^O)\) would imply that \(M_\theta(D(W^O))\) is already closed, and the means of selections of \(Q_\theta\) could be matched directly with admissible completion laws. Here, however, the instruments \(g_k\) need only be (Borel) measurable. Thus \(M_\theta\) may be discontinuous and \(M_\theta(D(W^O))\) need not be closed. Consequently, one cannot in general assert that \(\mathbb E[Q_\theta(W^O)]=I_\theta\). Instead, the Appendix proves $\mathbb E[Q_\theta(W^O)] = \operatorname{cl}_{\ell_1}I_\theta$. \\
	In general, \(\operatorname{cl}_{\ell_1}I_\theta\) may contain vectors that are not themselves generated by any admissible completion. That is, $\operatorname{cl}_{\ell_1}I_\theta \neq I_\theta$; identification, however, only asks whether the origin is attainable. The key additional lemma establishes that $0 \in \operatorname{cl}_{\ell_1}I_\theta \iff 0 \in I_\theta$, i.e. closure is harmless precisely at this point. \\
	Combining the preceding steps gives
	\begin{align*}
		\theta\in\Theta_I^*
		\quad\Longleftrightarrow\quad
		0\in I_\theta
		\quad\Longleftrightarrow\quad
		0\in\operatorname{cl}_{\ell_1}I_\theta
		\quad\Longleftrightarrow\quad
		0\in\mathbb E[Q_\theta(W^O)].
	\end{align*}
	
	Having completed the difficult work, as in \citet{beresteanu2008} or \citet{beresteanu2011}, the Hahn--Banach separation theorem implies that the Aumann expectation contains zero if and only if its support function is nonnegative for every \(u\in B^\infty\). The support-function identity for Aumann expectations gives
	\begin{align*}
		h\left(\mathbb E[Q_\theta(W^O)],u\right)
		=
		\mathbf{E}_{P_{W^O}}
		\left[
		h(Q_\theta(W^O),u)
		\right],
	\end{align*}
	which yields the characterization in Theorem~\ref{T:theorem1}. \\

	The language of support functions transforms the original problem into a geometric one that involves a nested minimax optimization problem:
	\begin{align*}
		\Theta_{I}^{*} = \{\theta \in \Theta: \ \inf_{u \in B^\infty}\mathbf{E}_{P_{W^O}}
		[\sup_{(x,y)\in D(W^O)} \langle M_\theta(y,x),u\rangle ]\geq0 \}
	\end{align*}
	In general this minimax problem is nonconvex and non-trivial to solve. It is frequently the case that, for a fixed $u\in B^\infty$, the inner problem, that is, $h(Q_\theta,u)=\sup_{(x,y)\in D(W^O)} \langle M_\theta(y,x),u\rangle $, is easy to solve and, depending on the form of Equation~\eqref{eq:model1}, convex. However, the potential computability of the inner problem is not sufficient to claim superiority in formulating the problem through support functions, and we do not make such a claim. We note two properties of the characterization that have computational appeal. \\
	First, Theorem~\ref{T:theorem1} defines the identified set as an infinite collection of inequalities depending only on observables. Thus, existing inference procedures, e.g.  \citet{andrews2017} or \citet{chernozhukov2019}, can be used for consistency and inference arguments. We discuss the details in the following section. \\
	Second, although the general problem is computationally demanding, several observations simplify the problem substantially. First, Theorem~\ref{T:theorem1} implies that we can exclude parameter candidates by finding counterexample-directions. In practice, this means we can search a large parameter space effectively by first screening out candidate $\theta$ based on a collection of fixed directions. Second, as we will show in the respective implementation section, the nested optimization problem reduces to a linear program for the important subclass of monotone index models. \\
	While the dual space of $\ell_1$ is $\ell_\infty$, suggesting the natural use of the $\ell_\infty$-unit ball, the compactness of $\mathbb E[Q_\theta(W^O)]$ allows one to consider other unit balls $B^p$, which may offer computational advantages. The choice of unit ball is equivalent for identification, and we conjecture the same for consistency and inference, though the proofs would need to be extended accordingly. As the related literature discussed in Section~\ref{sec:introduction} makes clear, our formulation also has computational benefits and drawbacks. In particular, the scale of $\langle M_\theta(y,x),u\rangle$ has no intrinsic interpretation beyond determining whether zero belongs to $\mathbb E[Q_\theta(W^O)]$; we do not assign substantive or finite-sample meaning to a positive distance from zero. Nor do we claim that our choice of instrument class $\mathcal G$, norm $\|\cdot\|_\infty$, or support-function representation is optimal for estimation or the construction of critical values. We favor the formulation for its geometric clarity:
	\begin{align*}
		\theta\in\Theta_I^*
		\quad\Longleftrightarrow\quad
		0\in\mathbb E[Q_\theta(W^O)].
	\end{align*}
	Section~\ref{sec:estimation} develops its computational and inferential implementation.

	\section{Estimation and Computation}
	\label{sec:estimation}
	
	\subsection{Estimator and Confidence Sets}
	\label{sec:finite_approximation}
	
	Theorem~\ref{T:theorem1} is pointwise in $\theta$ and requires neither compactness of $\Theta$\footnote{In fact, no restriction is put on $\Theta$ whatsoever for the purposes of identification.} nor continuity of $m(\cdot)$ in $\theta$. For estimation and inference the following assumptions must be introduced. Compactness of $\Theta$ in Assumption~\ref{A:compact-theta} and the uniform Lipschitz bound in Assumption~\ref{A:thetalipschitz} control the sample criterion uniformly over $\theta$, as required for the Hausdorff-consistency result in Theorem~\ref{T:theorem2}. The coverage proof for Theorem~\ref{T:theorem3} uses a common moment envelope and the decay of the weighted instruments to obtain relative compactness of the covariance kernels over $(\theta,P)\in\mathcal F$ for which Assumptions~\ref{A:assumption1}--\ref{A:bounded-moments} are sufficient. Its sequential argument takes limits of covariance kernels and requires no continuity of $m$ in $\theta$.
	
	\begin{assumption}
		\label{A:compact-theta}
		The parameter space $\Theta$ is a nonempty compact subset of $\mathbb R^{d_\theta}$.
	\end{assumption}
	
	\begin{assumption}
		\label{A:thetalipschitz}
		$m(y,x,\theta)$ is uniformly Lipschitz continuous. That is, there exists a constant $L<\infty$ such that, for every
		$\theta,\theta'\in\Theta$,
		\begin{align*}
			\sup_{(y,x)\in\mathcal W}
			\left\|
			m(y,x,\theta)-m(y,x,\theta')
			\right\|_\infty
			\leq
			L\|\theta-\theta'\|.
		\end{align*}
		The norm over $\Theta$ is inconsequential.
	\end{assumption}
	
	Theorem~\ref{T:theorem1} characterizes the sharp identified set through the population criterion
	\begin{align*}
		T(\theta)
		\equiv
		-\inf_{u\in B^\infty}
		\mathbf E\left[h(Q_\theta(W^O),u)\right].
	\end{align*}
	Since \(0\in B^\infty\), \(T(\theta)\ge 0\), and Theorem~\ref{T:theorem1} implies $\Theta_I^*=\{\theta\in\Theta:T(\theta) = 0\}$. \\
	The criterion \(T(\theta)\) is not directly computable because it involves countably many moment coordinates and an infinite-dimensional set of directions. For estimation, we approximate the conditional moment restriction by retaining only a finite number of instruments. Let $\pi_K:\ell_1\to\mathbb R^{d_m \cdot K}$ denote the truncation map that selects the coordinates indexed by \((j,k)\), \(j=1,\ldots,d_m\), \(k=1,\ldots,K\). Let $\lambda_K:\mathbb R^{d_mK}\to\ell_1$ denote the zero-padding embedding that maps a finite vector into \(\ell_1\) by keeping its first \(d_mK\) coordinates and setting all remaining coordinates equal to zero. Define the truncated moment mappings (random vectors) and truncated random sets as\footnote{For comparisons with the infinite-dimensional set $Q_\theta(w^o)\subset\ell_1$, and to ensure that set distances, such as the Hausdorff distance, are well defined, we require the embedding
		\begin{align*}
			\widetilde Q_{\theta,K}(w^o) = \lambda_K Q_{\theta,K}(w^o) = \lambda_K\pi_K Q_\theta(w^o),
		\end{align*}
		where $\pi_K$ and $\lambda_K$ applied to sets denote their image under the corresponding point maps.}
	\begin{align*}
		M_{\theta,K}(y,x) &\equiv\pi_K M_\theta(y,x) \\
		Q_{\theta,K}(w^o) &\equiv \operatorname{cl}_{\mathbb R^{d_mK}}M_{\theta,K}(D(w^o))=\pi_KQ_\theta(w^o) 
	\end{align*}
	Let $B_K^\infty \equiv \{u\in\mathbb R^{d_mK}:\|u\|_\infty\le 1\}$. The finite-dimensional population criterion is defined as
	\begin{align*}
		T_K(\theta) &\equiv \vec{\rho}_{H}
		\left(0,\mathbb E[Q_{\theta,K}(W^O)] \right)  =
		-\inf_{u\in B_K^\infty} h\left( \mathbb E[Q_{\theta,K}(W^O)],u \right)  =
		-\inf_{u\in B_K^\infty} \mathbf{E}\left[ h(Q_{\theta,K}(W^O),u) \right].
	\end{align*}
	The corresponding finite-dimensional outer approximation to the identified set is $   \Theta_{I,K} \equiv \{\theta\in\Theta:T_K(\theta) = 0\}$.\\
	Since every admissible completion satisfying the conditional moment restriction also satisfies any finite collection of the unconditional moment restrictions, $\Theta_I^* \subseteq \Theta_{I,K}$ for every $K$. \\
	Given an i.i.d. sample \(\{W_i^O\}_{i=1}^n\), define the sample Aumann average\footnote{In contrast to the population Aumann expectation, the finite-sample Minkowski average need not be convex in the most general case. We therefore convexify it in defining its empirical analogue. This does not alter any support-function inequality because $h(A,u)=h(\operatorname{co}(A),u)$ for every $u$.}
	\begin{align*}
		\widehat{\mathbb E}_n[Q_{\theta,K}]
		\equiv
		\operatorname{co} \left(
		\frac{1}{n}
		\bigoplus_{i=1}^n
		Q_{\theta,K}(W_i^O) \right)
	\end{align*}
	The sample criterion is
	\begin{align*}
		\widehat T_{n,K}(\theta)
		&\equiv \vec{\rho}_{H} \left( 0, \widehat{\mathbb E}_n[Q_{\theta,K}] \right) 
		=
		-\inf_{u\in B_K^\infty} h\left( \widehat{\mathbb E}_n[Q_{\theta,K}],u \right) 
		=
		-\inf_{u\in B_K^\infty} \frac{1}{n} \sum_{i=1}^n h(Q_{\theta,K}(W_i^O),u).
	\end{align*}
	For a deterministic sequence $K(n)\to\infty$, we use $\widehat T_n(\theta) \equiv \widehat T_{n,K(n)}(\theta)$. Thus an estimator based on the criterion function of distance between $0$ and $\mathbb E[Q_{\theta,K(n)}(W^O)]$ is the contour set
	\begin{align*}
		\widehat\Theta_n  \equiv \left\{ \theta\in\Theta: a_n \widehat T_{n,K(n)}(\theta) \le \tau_n \right\}.
	\end{align*}
	With this we can state the result concerning consistency of the proposed estimator.
	
	\begin{theorem}
		\label{T:theorem2}
		Let Assumptions~\ref{A:assumption1}--\ref{A:thetalipschitz} hold. Suppose that \(\Theta_I^*\neq\emptyset\). Let \(K(n)\to\infty\), and let $a_n \equiv \sqrt{  \frac{n}{d_mK(n)+d_\theta} }$. Let $\frac{d_mK(n)}{n}\to0$, $\tau_n\to\infty$ and $\frac{\tau_n}{a_n}\to0$. Then $\rho_H(\widehat\Theta_n,\Theta_I^*) \overset p\to0$.
	\end{theorem}
	For the proof of Theorem~\ref{T:theorem2} we first establish a law of large numbers for random sets with growing dimension; see Appendix  \ref{sec:appendixC}. The estimator is built on the result of \citet{chernozhukov2007} for constructing set estimators using a criterion function approach. \\
	The sequence $K(n)$ controls the finite-dimensional approximation of the infinite collection of unconditional moment restrictions. Since each of the $K(n)$ retained instruments is interacted with the $d_m$ components of the moment function, the dimension of the finite random set is $d_mK(n)$. $K(n)\to\infty$ makes the deterministic truncation error vanish, while $d_mK(n)/n\to0$ ensures that the sampling error from estimating this growing-dimensional random set vanishes. Thus $K(n)$ grows to include more moment restrictions, but slowly enough to control sampling error. We note that since $d_{m}$ and $d_{\theta}$ are fixed, the rate conditions could be equivalently stated as $\frac{K(n)}{n} \to 0$ and $a_{n} \asymp \sqrt{\frac{n}{K(n)}}$. The estimator above is obtained by accepting parameter values for which the finite-sieve sample violation is small. 

	\subsection{Inference}
	\label{sec:inference}
	
	We construct confidence sets using the many-moment procedure of \citet{andrews2017}. The procedure is well suited to the present setting because the support-function inequalities are indexed by the uncountable direction class $B^\infty$; however, we note that alternative procedures exist, e.g., \citet{chernozhukov2017}, which might also be applied to our results.  The \citet{andrews2017} implementation requires a method for evaluating the supremum over $B^\infty$; the computational section discusses finite and truncated approximations to this problem.  We do not consider here the additional error caused by terminating a global optimization routine at an inexact solution, but such $\epsilon$-minimax solutions may be considered, as for example in \citet{aradillas2025}. \\
	For the closed infinite-dimensional unit ball under the $\Vert \cdot \Vert _\infty$ norm, write $u=(u_{jk})\in B^\infty$, with coordinates indexed in the same order as $M_\theta$. Define
	\begin{align*}
		\widetilde{m}(w^{o},\theta,u)
		&\equiv h(Q_\theta(w^o),u) =\sup_{(y,x)\in D(w^o)}
		\sum_{j=1}^{d_m}\sum_{k\geq1}
		u_{jk}g_k(x)m_j(y,x,\theta) \in \mathbb R^1
	\end{align*}
	Let
	\begin{align*}
		\mathcal U \equiv B^\infty =
		\left\{
		u=(u_{jk})_{j=1,\ldots,d_m,\ k\geq1}:
		\|u\|_\infty
		\leq1
		\right\}.
	\end{align*}
	Thus each $u\in\mathcal U$ is one direction for the complete stacked moment vector and indexes one scalar support-function inequality. The same feasible $(y,x)$ enters every term of the sum inside the supremum, as the latent completion must solve all moments simultaneously. We can slightly rewrite Theorem~\ref{T:theorem1} to bring it into the notation of \citet{andrews2017}:
	\begin{align*}
		\Theta_I=\Theta_I^*
		=\left\{\theta\in\Theta:
		\mathbf E_{P_{W^O}}
		[\widetilde{m}(W^O,\theta,u)]\geq0
		\quad\forall u \in \mathcal{U}
		\right\}.
	\end{align*} 
	The null hypothesis that $\theta \in \Theta_I^*$ can be stated as
	\begin{align*}
		H_0(\theta):\qquad
		\mathbf E_{P_{W^O}}[\widetilde{m}(W^O,\theta,u)]\geq0
		\quad\text{for every } \ u\in\mathcal U.
	\end{align*}
	Finite direction classes and their inferential interpretation are discussed below. The auxiliary conditioning-instrument class in \citet{andrews2017} is not needed here, as the functions $\tilde{m}$ already encode the original conditioning restriction of model \eqref{eq:model1} through Lemma~\ref{L:lemma1}, and the support direction $u$ indexes the resulting unconditional inequalities.\footnote{See the first paragraph of page 276 in \citet{andrews2017}.} \\
	Following \citet{andrews2017}, we use the scalar normalization $\sigma_F(\theta) = \widehat\sigma_n(\theta) \equiv 1$. Define the sample mean
	\begin{align*}
		\overline m_n(\theta,u)
		&\equiv
		\frac{1}{n}\sum_{i=1}^n
		\widetilde m(W_i^O,\theta,u),
	\end{align*}
	the corresponding sample variance
	\begin{align*}
		\widehat\Sigma_n(\theta,u)
		&\equiv
		\frac{1}{n}\sum_{i=1}^n
		\left(
		\widetilde m(W_i^O,\theta,u)-\overline m_n(\theta,u)
		\right)^2,
	\end{align*}
	and its regularized version
	\begin{align*}
		\overline\Sigma_n(\theta,u)
		&\equiv
		\widehat\Sigma_n(\theta,u)+\epsilon,
		\qquad \epsilon>0.
	\end{align*}
	
	The fixed ridge is the scalar specialization of the \citet{andrews2017} regularization. Since
	\(\sigma_F(\theta)=\widehat\sigma_n(\theta)\equiv1\), we use their benchmark choice $\epsilon=1/20$.\footnote{A direction-adaptive regularization would be an alternative, e.g., $\epsilon(u)=\epsilon_0\sum_{k\geq 1}|u_k|$. This addresses the concern that, for directions with very small support-function variance, a fixed ridge may dominate the empirical variance. The alternative is an open question, however, because the ridge can approach zero over the direction class. Its validity requires an additional argument.} \\
	For the single lower-bound moment inequality indexed by $u\in\mathcal U$, the $S_1$ function of
	\citet{andrews2017} reduces to
	\begin{align*}
		S_1(s,\Sigma) \equiv \left[ \frac{s}{\sqrt{\Sigma}} \right]_{-}^2, \qquad [x]_{-}\equiv\max\{-x,0\}.
	\end{align*}
	
	The negative part appears because our null hypothesis $H_0$ is that each moment should be weakly greater than zero. The corresponding test statistic is
	\begin{align*}
		T_n^{AS}(\theta)
		\equiv
		\sup_{u\in\mathcal U}
		S_1\!\left(
		\sqrt n\,\overline m_n(\theta,u),
		\overline\Sigma_n(\theta,u)
		\right).
	\end{align*}
	We use the generalized moment-selection critical value of
	\citet{andrews2010}, as implemented by \citet{andrews2017}.
	\begin{enumerate}[label=\textbf{Step \arabic*.},leftmargin=*,wide=0pt]
		\item Set
		\begin{align*}
			\kappa_n
			&\equiv(0.3\log n)^{1/2},
			&
			B_n
			&\equiv
			\left(\frac{0.4\log n}{\log\log n}\right)^{1/2},
		\end{align*}
		and define the scalar standardized slackness measure
		\begin{align*}
			\xi_n(\theta,u)
			&\equiv
			\kappa_n^{-1}
			\frac{\sqrt n \overline m_n(\theta,u)}
			{\sqrt{\overline\Sigma_n(\theta,u)}}
		\end{align*}
		The corresponding GMS adjustment is
		\begin{align*}
			\overline\varphi_n(\theta,u)
			&\equiv
			\sqrt{\overline\Sigma_n(\theta,u)} B_n
			\mathbf 1\{\xi_n(\theta,u)>1\}.
		\end{align*}
		
		\item Generate \(B_{\mathrm{boot}}\) nonparametric bootstrap samples \(\{W_{i,b}^{O,*}:i=1,\ldots,n\}\), \(b=1,\ldots,B_{\mathrm{boot}}\), by sampling with replacement from the empirical distribution. For every \(b\), calculate
		\begin{align*}
			\overline m_{n,b}^*(\theta,u)
			&\equiv
			\frac{1}{n}\sum_{i=1}^n
			\widetilde m(W_{i,b}^{O,*},\theta,u),\\
			\widehat\Sigma_{n,b}^*(\theta,u)
			&\equiv
			\frac{1}{n}\sum_{i=1}^n
			\left\{
			\widetilde m(W_{i,b}^{O,*},\theta,u)
			-\overline m_{n,b}^*(\theta,u)
			\right\}^2,\\
			\overline\Sigma_{n,b}^*(\theta,u)
			&\equiv
			\widehat\Sigma_{n,b}^*(\theta,u)+\epsilon.
		\end{align*}
		
		\item Form the recentered bootstrap statistic
		\begin{align*}
			T_{n,b}^{AS,*}(\theta)
			\equiv
			\sup_{u\in\mathcal U}
			\left[
			\frac{
				\sqrt n
				\left\{
				\overline m_{n,b}^*(\theta,u)
				-\overline m_n(\theta,u)
				\right\}
				+\overline\varphi_n(\theta,u)}
			{\sqrt{\overline\Sigma_{n,b}^*(\theta,u)}}
			\right]_-^2.
		\end{align*}
		
		\item Let \(q_{n,1-\alpha+\eta}^{AS,*}(\theta)\) be the empirical \((1-\alpha+\eta)\)-quantile of \(\{T_{n,b}^{AS,*}(\theta):b=1,\ldots,B_{\mathrm{boot}}\}\) and define
		\begin{align*}
			c_{n,1-\alpha}^{AS}(\theta)
			\equiv
			q_{n,1-\alpha+\eta}^{AS,*}(\theta)+\eta,
			\qquad 0<\eta<\alpha.
		\end{align*}
		The parameter \(\eta\) is part of the infinitesimal uniformity adjustment in \citet{andrews2017}. The theoretical results retain a fixed \(0<\eta<\alpha\). Following their implementation convention, \(\eta\) can be set equal to zero in computation.
	\end{enumerate}
	The confidence set obtained by test inversion is
	\begin{align*}
		CS_n^{AS}(1-\alpha)
		\equiv
		\left\{
		\theta\in\Theta:
		T_n^{AS}(\theta)
		\leq c_{n,1-\alpha}^{AS}(\theta)
		\right\}.
	\end{align*}
	
	The inference result requires a strengthening of the tail condition in Assumption~\ref{A:assumption2} (ii). Assumption~\ref{A:assumption2} (ii) is sufficient to ensure that the stacked moment map is \(\ell_1\)-valued and that finite truncations approximate the full moment vector uniformly. To control the complexity of the infinite-dimensional class of support directions, we additionally require a polynomial decay rate for the individual
	instrument coordinates. 
	\begin{assumption}
		\label{A:polynomial-decay}
		Define
		\begin{align*}
			a_k
			\equiv
			\sup_{x\in\mathcal X}|g_k(x)|,
			\qquad k\geq1.
		\end{align*}
		For some $\beta>3/2$, $C_a \equiv \sup_{k\geq1} k^\beta a_k <\infty$.
	\end{assumption}
	For many common applications of instrument classes, like those of indicator functions, the unweighted class has the property that $\sup_{x\in \mathcal X}|h_k(x)|\leq 1$.  If, additionally, $d_m=1$, then, for example, the researcher may define $g_k(x)\equiv k^{-\beta}h_k(x)$ which will satisfy Assumption \ref{A:polynomial-decay} when $\beta>3/2$. The polynomial decay condition above is stronger than Assumption~\ref{A:assumption2} (ii) in general and provides the entropy bound required for a technical requirement used to deal with an uncountable number of moment inequalities called \textit{manageability} \citep{pollard1990}. \\
	For the uniform coverage statement, we introduce additional notation.  Let $\mathcal P$ denote the class of observed-data laws on $\mathcal{W}^{O}$ over which inference is required to be valid.  For each $P\in\mathcal P$, let $\Theta_I^*(P)$ denote the identified set under $P$.  By Theorem~\ref{T:theorem1},
	\begin{align*}
		\Theta_I^*(P)
		=
		\left\{
		\theta\in\Theta:
		\mathbf E_P
		\left[
		\widetilde m(W^O,\theta,u)
		\right]
		\geq0
		\quad\forall u\in\mathcal U
		\right\}.
	\end{align*}
	Define $\mathcal F^+ \equiv \Theta\times\mathcal P$ and $\mathcal F \equiv \left\{ (\theta,P)\in\mathcal F^+: \theta\in\Theta_I^*(P) \right\}$.
	
	\begin{theorem}
		\label{T:theorem3}
		Let Assumptions~\ref{A:assumption1}--\ref{A:bounded-moments} and \ref{A:polynomial-decay} hold. Then, for every $\alpha\in(0,1)$,
		\begin{align*}
			\liminf_{n\to\infty}
			\inf_{(\theta,P) \in \mathcal{F}}
			P
			\left\{
			\theta\in CS_n^{AS}(1-\alpha)
			\right\}
			\geq
			1-\alpha.
		\end{align*}
		Moreover, for every fixed $P\in\mathcal P$ and every fixed $\theta\notin\Theta_I^*(P)$,
		$P\left\{ T_n^{AS}(\theta) > c_{n,1-\alpha}^{AS}(\theta) \right\} \to 1$.
	\end{theorem}

	The restriction $\beta>3/2$ ensures that the entropy integral for the infinite-dimensional direction class is finite. The proof is given in  \hyperref[sec:appendixD]{Appendix D}. Theorem~\ref{T:theorem3} gives uniformly valid marginal coverage over all null pairs $(\theta,P)\in\mathcal F$. It does not imply simultaneous coverage of the entire identified set $\Theta_I^*(P)$ or Hausdorff convergence of a fixed-level confidence set.

	\section{Implementation}
	\label{sec:implementation}
	
	This section discusses the computational implementation of the
	identification, estimation, and inference procedures developed above.
	We first show that, for an important class of regression models,
	the finite-dimensional estimation criterion admits an exact linear
	programming representation. We then discuss outer approximations
	obtained by restricting the direction class and present two practical
	strategies for constructing confidence sets using finite collections
	of directions.
	
	\subsection{Exact LP Representation for Monotone-Index Models}
	\label{sec:lp-representation}
	
	For scalar moment functions, the finite-dimensional estimation criterion
	$\widehat T_{n,K}(\theta)$ from Section~\ref{sec:finite_approximation}
	can be computed exactly by linear programming for an important class of
	models: monotone-index models, including linear, probit, and logit regressions. Formally, let $d_m=d_y=1$ and $m(y,x,\theta)=y-f(x,\theta)$ with $E[Y|X]=f(x,\theta_0)$ at the true parameter. For each fixed $\theta$, assume $f(\cdot,\theta)$ is continuous and coordinatewise monotone on the feasible rectangles considered below. \\
	Without loss of generality, the covariate domain is $[0,1]^{d_x}$. For our instrument class we use the
	weighted cube instruments of Appendix~\ref{sec:weigthedcubes}.
	Consider retaining $K$ weighted indicators $g_k(x)=\omega_k\mathbf 1\{x\in C_k\}$. Throughout this subsection, we work within the truncated problem $K<\infty$. As $\{C_k\}_{k=1}^{K}$ overlap, let $\{C_\ell^*\}_{\ell=1}^{L}$ be the common refinement of their
	partitions, such that $\forall x, x\in C_\ell^*$ for exactly one $\ell$.  Define $c_\ell\in\mathbb R^K$ as the mapping between $\{C_k\}$ and $\{C_\ell^*\}$ by
	\begin{align*}
		c_{\ell k}&\equiv\omega_k\mathbf 1\{C_\ell^*\subseteq C_k\},
		\qquad k=1,\ldots,K.
	\end{align*}
	Algebraically we have the observation $u^\top g(x)=c_\ell^\top u$ for each $x\in C_\ell^*$.
	For each observation, denote the domain of an observation $i$ within the set $C_\ell^*$ as
	\begin{align*}
		R_{i\ell}&\equiv\operatorname{cl}\bigl(D_i\cap
		(\mathbb R\times C_\ell^*)\bigr),
		\quad \ell\in S_i.
	\end{align*}
	Per our assumption of monotone-index models, $m(\cdot,\cdot,\theta)$ is continuous
	and coordinatewise monotone on each $R_{i\ell}$. Its extrema on
	$R_{i\ell}$ are then the corner values
	\begin{align*}
		m_{i\ell}^-(\theta)&\equiv
		\min_{z\in\operatorname{vert}(R_{i\ell})}m(z,\theta),
		&m_{i\ell}^+(\theta)&\equiv
		\max_{z\in\operatorname{vert}(R_{i\ell})}m(z,\theta).
	\end{align*}
	Because the instruments are constant on each refined cell, the
	observation-level support function satisfies
	\begin{align}
		\label{eq:lp-support-representation}
		h(Q_{\theta,K}(W_i^O),u)
		&=\max_{\ell\in S_i}\max_{s\in\{-,+\}}
		\bigl\{m_{i\ell}^s(\theta)c_\ell^\top u\bigr\},
		\qquad u\in B_K^\infty=[-1,1]^K.
	\end{align}
	Each term inside the maximum is linear in $u$. Consequently,
	$-\widehat T_{n,K}(\theta)$ is exactly the optimal value of the
	epigraph linear program
	\begin{align}
		\label{eq:lp-epigraph}
		\min_{u\in\mathbb R^K,\,t\in\mathbb R^n}\quad
		&\frac1n\sum_{i=1}^n t_i\notag\\
		\text{subject to}\quad
		&-1\le u_k\le1,
		&&k=1,\ldots,K,\notag\\
		&t_i\ge m_{i\ell}^s(\theta)c_\ell^\top u,
		&&i=1,\ldots,n,\quad\ell\text{ s.t } R_{i,\ell}\neq \emptyset,\quad s\in\{-,+\}.
	\end{align}
	The program has $K+n$ variables and $2\sum_i|S_i|$ epigraph inequalities,
	in addition to the box constraints. In particular, the constraints are
	imposed on the original direction $u$, not separately on the induced
	coefficients $c_\ell^\top u$. The representation is exact for each
	finite $K$; its derivation is given in Appendix~\ref{sec:appendix-lp}. The negative of the program's
	optimal value is the estimation criterion $\widehat T_{n,K}(\theta)$. The studentized statistic in Section \ref{sec:inference}
	is a different optimization problem because its variance denominator depends on the direction $u$. The GMS adjustment \citep{andrews2017} enters the bootstrap statistic used to construct its critical value. The program above does not directly compute either statistic. 

	\subsection{Construction of Outer Sets}
	\label{sec:abhcomparison}
	
	Theorem~\ref{T:theorem1} characterizes $\Theta_I^*$ through the inequalities $\mathbf E_{P_{W^O}}[\widetilde m(W^O,\theta,u)]\ge0$ for all directions $u\in\mathcal U=B^\infty$, where $\widetilde m(w^o,\theta,u)=h(Q_\theta(w^o),u)$. Any subcollection of these inequalities defines an \emph{outer set}, that is, a set containing $\Theta_I^*$. These population outer sets can be tightened by adding directions. Under the conditions of Corollary~\ref{C:as-truncations}, the corresponding tests control asymptotic size. Adding directions also changes the calibrated critical value, so the resulting confidence sets need not be nested in a given sample. For $\mathcal V\subseteq\mathcal U$, define
	\begin{align*}
		\Theta_O(\mathcal V)
		\equiv
		\bigl\{\theta\in\Theta:\
		\mathbf E_{P_{W^O}}[\widetilde m(W^O,\theta,u)]\ge0
		\ \ \forall u\in\mathcal V\bigr\}.
	\end{align*}
	Truncation of the instrument class is a special case where dropped instruments are not tested by the support function: $u_k=0$. Since support functions $h(\cdot,\cdot)$ must come from the same space, for $u\in\mathcal U$, let $\Pi_Ku\equiv\lambda_K\pi_Ku$ retain the first $K$ instrument coordinates and set the remaining coordinates to zero. Let $B_K\equiv\Pi_K\mathcal U$. Since $h(Q_\theta(w^o),\Pi_Ku)=h(Q_{\theta,K}(w^o),\pi_Ku)$, inequalities indexed by $\mathcal V\subseteq B_K$ involve only the finite-dimensional sets $Q_{\theta,K}$. In particular, $\Theta_O(B_K)$ coincides with the outer set $\Theta_{I,K}$ of Section~\ref{sec:finite_approximation}. We write $\Theta_O^K(\mathcal V)\equiv\Theta_O(\Pi_K\mathcal V)$.
	
	\begin{corollary}
		\label{C:outer-sets}
		For all $\mathcal V\subseteq\mathcal V'\subseteq\mathcal U$ and every $K$,
		\begin{align*}
			\Theta_I^*\subseteq\Theta_O(\mathcal V')\subseteq\Theta_O(\mathcal V)
			\qquad\text{and}\qquad
			\Theta_I^*\subseteq\Theta_O^K(\mathcal V).
		\end{align*}
		If, in addition, $\Pi_K\mathcal V\subseteq\mathcal V$, then
		$\Theta_O(\mathcal V)\subseteq\Theta_O^K(\mathcal V)$.
	\end{corollary}
	
	\begin{proof}
		By Theorem~\ref{T:theorem1}, the first claim is trivial. The second claim is a special case, since $\Pi_K\mathcal V\subseteq\mathcal U$. We note one caution in algebra. Even though $\Pi_K\mathcal U=\Pi_K B^\infty = B_K$, for $\mathcal V \subset \mathcal U$, it is not necessarily true that $\Pi_K\mathcal V \subset \mathcal V$, though again noting $\Pi_K\mathcal V \subset \mathcal U$.
	\end{proof}
	
	Examples of sets closed under truncation are $\mathcal U$, $B_K$, and the coordinate directions $\{\pm e_j\}$ used in Appendix~\ref{sec:appendixE}. Note that when truncating coordinate directions (instruments) $\Pi_Ke_j\in\{e_j,0\}$, where $h(A,0)=0$ adds no information.\\
	Under the decay condition of Theorem \ref{T:theorem3}, $(\mathcal U,\rho_a)$ is totally bounded. Thus it admits a finite $\epsilon$-net for each $\epsilon>0$. See Appendix \ref{sec:appendixD} and Lemma~\ref{L:D-manageability-revised} to construct a finite-dimensional grid $\mathcal U_\xi\subseteq\mathcal U$ such that $\min_{v\in\mathcal U_\xi}\rho_a(u,v)\le\xi/2$ for every $u\in\mathcal U$. Combined with the Lipschitz bound $|\widetilde m(w^o,\theta,u)-\widetilde m(w^o,\theta,v)|\le M_0\rho_a(u,v)$, which holds uniformly in $\theta$ and $P_{W^O}$, we note:
	\begin{align*}
		\inf_{u\in\mathcal U}\mathbf E[\widetilde m(W^O,\theta,u)]
		\ \ge\
		\min_{v\in\mathcal U_\xi}\mathbf E[\widetilde m(W^O,\theta,v)]
		-\frac{M_0\xi}{2}.
	\end{align*}
	For inference with a direction set $\mathcal V\subseteq\mathcal U$, let $CS_n^{AS}(1-\alpha;\mathcal V)$ be defined as $CS_n^{AS}(1-\alpha)$ in Section~\ref{sec:inference}, with every supremum over $\mathcal U$ replaced by a supremum over $\mathcal V$. Let $\Theta_O(\mathcal V)(P)$ denote $\Theta_O(\mathcal V)$ under the observed-data law $P$, and let $\mathcal F_{\mathcal V}\equiv\{(\theta,P)\in\mathcal F^+:
	\theta\in\Theta_O(\mathcal V)(P)\}$. By Corollary~\ref{C:outer-sets}, $\mathcal F\subseteq\mathcal F_{\mathcal V}$ and $\Theta_I\subseteq \Theta_O(\mathcal V)(P)$.
	
	\begin{corollary}
		\label{C:as-truncations}
		Let the conditions of Theorem~\ref{T:theorem3} hold.
		\begin{enumerate}[label=(\alph*),leftmargin=*]
			\item Coverage holds uniformly over direction sets and null pairs:
			\begin{align*}
				\liminf_{n\to\infty}\ \inf_{\mathcal V\subseteq\mathcal U}\
				\inf_{(\theta,P)\in\mathcal F_{\mathcal V}}
				P\bigl\{\theta\in CS_n^{AS}(1-\alpha;\mathcal V)\bigr\}
				\ge1-\alpha .
			\end{align*}
			In particular, since $\mathcal F\subseteq\mathcal F_{\mathcal V}$, the bound holds uniformly over $(\theta,P)\in\mathcal F$ and $\mathcal V\subseteq\mathcal U$.
			
			\item For fixed $\mathcal V\subseteq\mathcal U$, $P\in\mathcal P$ and $\theta\notin\Theta_O(\mathcal V)(P)$, $P\{\theta\notin CS_n^{AS}(1-\alpha;\mathcal V)\}\to1$.
			
			\item If $K(n)\to\infty$, then $CS_n^{AS}(1-\alpha;B_{K(n)})$ has coverage at least $1-\alpha$ uniformly over $\mathcal F$, by (a). It also excludes each fixed $\theta\notin\Theta_I^*(P)$ with probability approaching one. No rate restriction on $K(n)$ is required.
		\end{enumerate}
	\end{corollary}
	
	The proof can be found in Appendix \ref{sec:appendixD}. For a single fixed finite direction set, \textit{manageability}, a technical condition of \citet{andrews2017}, follows from its finite cardinality without Assumption \ref{A:polynomial-decay}. For uniformity over arbitrarily varying direction sets, we retain Assumption \ref{A:polynomial-decay} and the entropy bound of Appendix \ref{sec:appendixD}. In (c), the confidence level does not improve as $K(n)$ grows; it remains at least $1-\alpha$. What tightens is the collection of tested inequalities, so that every fixed $\theta\notin\Theta_I^*$ is eventually excluded. \\
	Finally, a particularly interesting connection to the related literature is given in Appendix~\ref{sec:appendixE}. It turns out that, up to minor implementation choices and the fact that they consider missing data instead of interval data, the two outer sets proposed in \citet{aucejo2017} are special cases of our estimator with particularly chosen direction sets $\mathcal V$. In particular, in the Appendix we show that taking $\mathcal{V}=\{\pm e_k\}_{k\leq K}$ consisting of basis vectors retrieves their first outer set construction. The intuition is simple: for their first outer set construction, \citet{aucejo2017} solve the identification problem by only checking $\forall k , \exists\eta \in \Gamma  \text{ s.t } E_\eta[m(\cdot,\theta)g_k(\cdot)]=0$. We note that $\Theta_I=\{\theta: \exists \eta \text { s.t } \forall k , E_\eta[m(\cdot,\theta)g_k(\cdot)]=0\}$ and for any proposition $\mathbf P$ one can write the logical proposition  $(\exists \eta \text { s.t } \forall k , \mathbf{P_k}) \Rightarrow (\forall k \exists \eta \text { s.t }  \mathbf{P_k})$. The second outer set instead integrates out the missing covariate. In Appendix \ref{sec:appendixE} we show that, when all covariates are integrated out, this corresponds to $\mathcal V =\{\pm\mathbf{1}_K\}$, which attaches equal weight to all cubes. \citet{aucejo2017} note that $E[m(\cdot,\theta)|X=x]=0\Rightarrow E[m(\cdot,\theta)]=0$, so instrument functions can receive equal weight everywhere. If some instruments are observed pointwise (non-interval) then these could be \textit{integrated out}. In general, it is interesting that the direction set $\mathcal V$ in the \textit{dual} space may offer some geometric intuition for support function constructions. \\
	In the simulations and the application, whenever we refer to \citet{aucejo2017}, we mean their estimator extended to the interval data case via our implementation and inference procedure.
	
	\subsection{Finite Sample Implementation}
	\label{sec:finite_implementation}
	
	The linear program in Section~\ref{sec:lp-representation} exactly minimizes the
	sample support function, whereas inference in
	Section~\ref{sec:inference} uses the studentized criterion
	$T_n^{AS}(\theta)$. Nevertheless, LP solutions to the unstudentized problem may provide a good direction bank $\mathcal V$ for assessing computationally convenient
	candidates for inference. More generally, approximating the
	infinite-dimensional direction class by finite subcollections is
	standard in the moment-inequality literature; see, for example,
	\citet{andrews2017}. The validity of inference with data-adaptively
	selected directions follows from Corollary~\ref{C:as-truncations} when the direction bank
	is selected on an independent subsample and the studentized statistic
	and bootstrap critical value are computed using that bank. We therefore
	consider two computational strategies: an adaptive procedure based
	on sample splitting and a fixed procedure using a prespecified
	direction bank. Their finite-sample performance is examined in
	Section~\ref{sec:simulation}. \\
	For the first implementation strategy, we use Corollary~\ref{C:as-truncations}(a) to select informative directions in a data-adaptive way while preserving validity. The procedure is as follows:
	\begin{enumerate}[label=\arabic*.,leftmargin=*]
		\item Randomly split the sample into two subsamples, denoted $A$ and $B$, with sizes $n_A,n_B$ of approximately equal size.
		\item Using the $n_A$ observations of $A$, find extremum directions $u_\theta$  at a small number of values of $\theta$. Specifically, determine $\widehat{\mathcal V}_A=\{u_\theta: u_\theta =\arg \min_{u\in B_K} \frac{1}{n_A} \sum_{i=1}^{n_A} h(Q_{\theta,K}(W_i^O),u),  \theta \in \{\theta_1,\ldots\}\subset \Theta\}$. For a particular $\theta$ and with $K<\infty$ this problem can be solved using the provided program (see Section \ref{sec:lp-representation}).  Theoretically, there is no guarantee that this is a good approximation to the true criterion $\hat{T}_{n_B}^{AS}(\theta)$; however, continuity of the criterion intuitively suggests it might be. We note that any direction bank that is a subset of $B_K$ will define an outer set and thus deliver correct size asymptotically.
		\item Compute the confidence set on $B$ with direction set $\widehat{\mathcal V}_A$. The test statistic $\hat{T}_{n_B}^{AS}(\theta)$ and the bootstrap values $\{ \hat{T}_{n_B}^{AS,b}(\theta)\}_{b=1}^{B_{boot}}$ are used to construct a confidence set $CS_{n_B}^{AS}(1-\alpha/2; \widehat{\mathcal V}_A)$, such that if $\theta\in \Theta_I$ or $\theta\in \Theta_O(\widehat{\mathcal V}_A)$ then $\lim_{n\rightarrow \infty}\Pr (\theta \in CS_{n_B}^{AS}(1-\alpha/2; \widehat{\mathcal V}_A))\geq 1-\alpha/2 $.  We elaborate on this result below. 
		\item Reverse the roles of $A$ and $B$, and intersect the two confidence sets, so that, if $\theta \in \Theta_I$, by a Bonferroni argument $\lim_{n\rightarrow \infty}\Pr (\theta \in \bigcap_{j\in \{A,B\}} CS_{n_j}^{AS}(1-\alpha/2; \widehat{\mathcal V}_{-j}))\geq 1-\alpha $.
		
	\end{enumerate}
	
	Because $\widehat{\mathcal V}_A$ is a function of $A$ alone, it is independent of $B$. Writing $n_B$ for the size of $B$, conditioning on $A$ gives, for every $(\theta,P)\in\mathcal F$,
	\begin{align*}
		P\bigl\{\theta\in CS_{n_B}^{AS}(1-\alpha/2;\widehat{\mathcal V}_A)\bigr\}
		&=\mathbf E\Bigl[P\bigl\{\theta\in CS_{n_B}^{AS}(1-\alpha/2;
		\widehat{\mathcal V}_A)\,\big|\,A\bigr\}\Bigr] \\
		&\ge
		\inf_{\mathcal V\subseteq\mathcal U}
		P^{\,n_B}\bigl\{\theta\in CS_{n_B}^{AS}(1-\alpha/2;\mathcal V)\bigr\}.
	\end{align*}
	By Corollary~\ref{C:as-truncations}(a), the limit inferior of the right-hand side is at least $1-\alpha/2$, uniformly over $\mathcal F$. This is where uniformity over $\mathcal V$ matters: a guarantee for each fixed direction set separately would not cover a bank that changes with the sample. We believe this procedure avoids treating directions selected from the same observations as fixed, while retaining most of the computational advantage of concentrating the bootstrap on a small set of informative directions. A potential shortcoming of this approach is that splitting the sample might affect inference in finite samples. \\
	The second complementary strategy for choosing a direction bank consists of considering an extended collection of ex-ante prespecified directions. \\
	Specifically, we consider an extended fixed direction bank consisting of the global directions $\pm\mathbf{1}_J$, the coordinate directions $\pm e_{rj}$, contiguous block directions $\pm\sum_{j=a}^{b}e_{rj}$, and single-sign-change directions $\pm(\sum_{j=1}^{k}e_{rj}-\sum_{j=k+1}^{K_r}e_{rj})$. The latter two classes are constructed separately within each resolution $r$, where $K_r$ denotes its number of cells. Exact duplicates are removed. Since the directions are prespecified, inference uses the full sample without sample splitting. \\
	For the remainder of this paper we will refer to the split-sample procedure as the \textit{adaptive} implementation, as it chooses the direction set specifically for the data and model studied, whereas the prespecified direction implementation is referred to as the \textit{fixed} implementation.

	\section{Monte Carlo Simulation}
	\label{sec:simulation}
	
	We study the finite-sample performance of the proposed confidence sets
	in the linear regression model
	\begin{align*}
		Y_i=\theta_0+\theta_1X_i+\varepsilon_i,
		\qquad X_i\sim U[0,1],
		\qquad (\theta_0,\theta_1)=(0,1),
	\end{align*}
	where $\varepsilon_i$ is a centered Gaussian draw clipped to a large compact
	support. Outcomes are always observed, while covariates may be observed
	only through intervals
	\begin{align*}
		[X_{i,l},X_{i,u}]
		=[\max\{0,X_i-ws_i\},X_i],
		\qquad s_i\sim U[0.5,1.5].
	\end{align*}
	We first examine coverage and confidence-set geometry under three
	designs with varying degrees of interval observation, before turning
	to a point-identified design to study power against alternatives.
	
	The first design features random interval observation, independent of
	$(X_i,Y_i)$, with probability $p=0.10$ and interval-width parameter
	$w=0.1$. In the second design, all covariates are interval-observed
	($p=1$) and $w=0.1$. The third design allows outcome-dependent interval observation with $p=0.30$ and $w=0.25$: the probability of interval observation is $1.6p$ for $Y_i\leq0.5$ and $0.4p$ otherwise. By symmetry, the unconditional probability remains $p$. \\
	For each design, we conduct 250 Monte Carlo replications with
	$n\in\{500,1000,5000\}$ observations and construct 90\% confidence sets using
	399 bootstrap draws. We compare our (Adaptive) and (Fixed) implementations, defined in Section \ref{sec:finite_implementation}, with an extension of the outer-set procedure of \citet{aucejo2017} (ABH) to the interval case as well as conventional OLS confidence ellipses based on midpoint imputation (OLS MID) or complete cases (OLS CC). We report empirical coverage and summarize confidence-set geometry
	using Minkowski averages across replications.
	
	\begin{figure}[H]
		\centering
		\includegraphics[width=\textwidth]{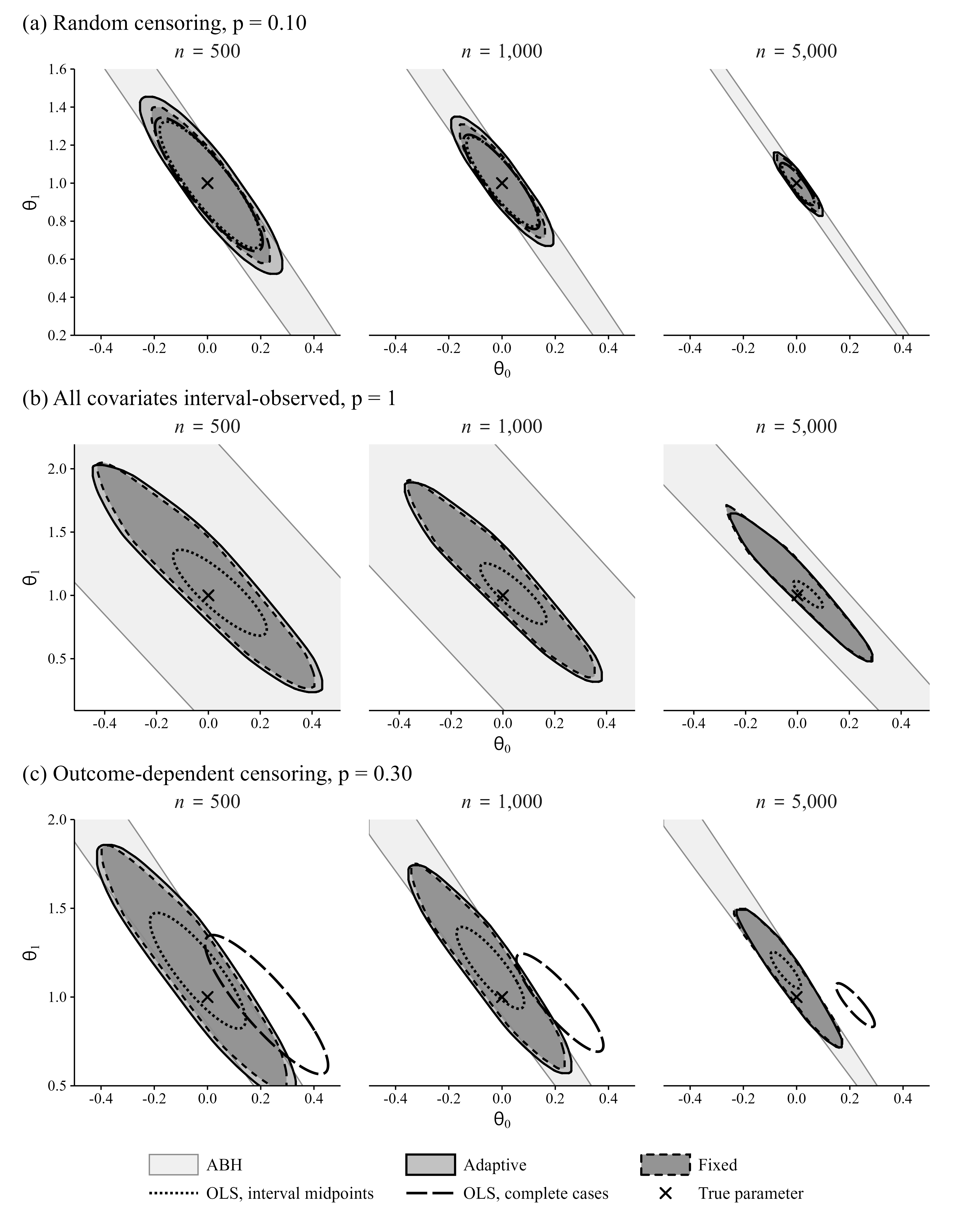}
		\caption{Minkowski-Averaged Confidence Sets across Simulations}
		\label{fig:fig1}
		\par\vspace{0.2cm}
		\begin{minipage}{\textwidth}
			\scriptsize
			\textit{Note:} Minkowski averages of the joint 90\% confidence sets for
			$(\theta_0,\theta_1)$ across the Monte Carlo replications of
			Table~\ref{tab:tab1}. The ABH
			sets extend beyond the plotted range.
		\end{minipage}
	\end{figure}
	
	Figure~\ref{fig:fig1} displays the Minkowski-averaged 90\% confidence sets across
	the three simulation designs, while Table~\ref{tab:tab1} reports
	empirical coverage. Under random censoring with 10\% of
	covariates interval-observed (a), our adaptive and fixed
	procedures perform similarly to conventional OLS, with confidence
	sets contracting rapidly as the sample size increases. The ABH
	outer-set procedure also becomes more informative, but its
	confidence regions retain an elongated, band-like shape,
	reflecting substantially weaker joint restrictions on the
	intercept and slope. By exploiting these restrictions jointly,
	our procedure produces considerably tighter confidence sets
	that better distinguish the two coefficients. \\
	The difference becomes more pronounced when all covariates
	are interval-observed (b). The ABH confidence regions
	remain relatively wide, whereas our procedure continues to provide informative joint restrictions on the regression parameters. Nevertheless, our confidence sets remain substantially wider than the midpoint-imputation OLS ellipses, even for $n=5{,}000$. This reflects the fact that increasing
	the sample size reduces sampling uncertainty but does not eliminate the ambiguity inherent in interval measurement. Midpoint OLS, by contrast, becomes increasingly concentrated around the parameter implied by its particular imputation
	rule. Since this imputation is incorrect in the simulated design, its empirical coverage deteriorates from 75.0\% at $n=500$ to only 9.9\% at $n=5{,}000$, as reported in Table~\ref{tab:tab1}. \\
	Finally, outcome-dependent censoring (c) illustrates
	the potential failure of complete-case estimation. Because
	the probability of observing a covariate exactly depends
	on the outcome, complete-case OLS is subject to selection
	bias and its confidence regions become increasingly
	concentrated away from the true parameter. Its empirical
	coverage is only 3.6\% at $n=500$ and falls to zero for
	larger samples. Midpoint OLS also performs poorly, with
	coverage declining to 2.7\% at $n=5{,}000$. In contrast,
	our adaptive and fixed procedures retain 95.6\% coverage
	at this sample size, despite the more severe and endogenous
	interval observation. Taken together, the simulations
	illustrate that conventional estimators can become
	increasingly precise without becoming more reliable,
	whereas our procedure accounts for the uncertainty
	generated by incomplete data.
	
	\begin{table}[H]
		\centering
		\caption{Empirical Coverage by DGP and Method}
		\vspace{0.1cm}
		\label{tab:tab1}
		\begingroup
		\setlength{\tabcolsep}{14pt}
		\begin{tabular}{lrrrrr}
\hline
$n$ & Adaptive & Fixed & ABH & OLS MID & OLS CC \\
\hline
\multicolumn{6}{l}{\textit{(a) Random censoring, $p = 0.10$}} \\
500 & 0.920 & 0.964 & 0.936 & 0.920 & 0.896 \\
1,000 & 0.912 & 0.960 & 0.908 & 0.920 & 0.916 \\
5,000 & 0.960 & 0.980 & 0.904 & 0.872 & 0.900 \\[3pt]
\multicolumn{6}{l}{\textit{(b) All covariates interval-observed, $p = 1$}} \\
500 & 0.980 & 0.980 & 0.940 & 0.750 & --- \\
1,000 & 0.990 & 1.000 & 0.960 & 0.580 & --- \\
5,000 & 0.956 & 0.945 & 0.945 & 0.099 & --- \\[3pt]
\multicolumn{6}{l}{\textit{(c) Outcome-dependent censoring, $p = 0.30$}} \\
500 & 0.976 & 0.984 & 0.948 & 0.760 & 0.036 \\
1,000 & 0.978 & 0.991 & 0.964 & 0.476 & 0.000 \\
5,000 & 0.956 & 0.956 & 0.920 & 0.027 & 0.000 \\
\hline
\end{tabular}
		\endgroup
		\par\vspace{0.3cm}
		\begin{minipage}{\textwidth}
			\scriptsize
			\textit{Note:} Share of Monte Carlo replications in which the joint 90\%
			confidence set contains the true parameter $(0,1)$. Adaptive and Fixed are our procedures (Section~\ref{sec:finite_implementation}), with 399 bootstrap draws; ABH is the outer set of \citet{aucejo2017} extended to interval
			data; OLS MID and OLS CC are joint 90\% $F$-based confidence ellipses from
			OLS with midpoint imputation and complete cases. OLS CC is not available in
			design (b), where every covariate is interval-observed.
		\end{minipage}
	\end{table}
	
\begin{figure}[H]
		\centering
		\includegraphics[width=0.9\linewidth]{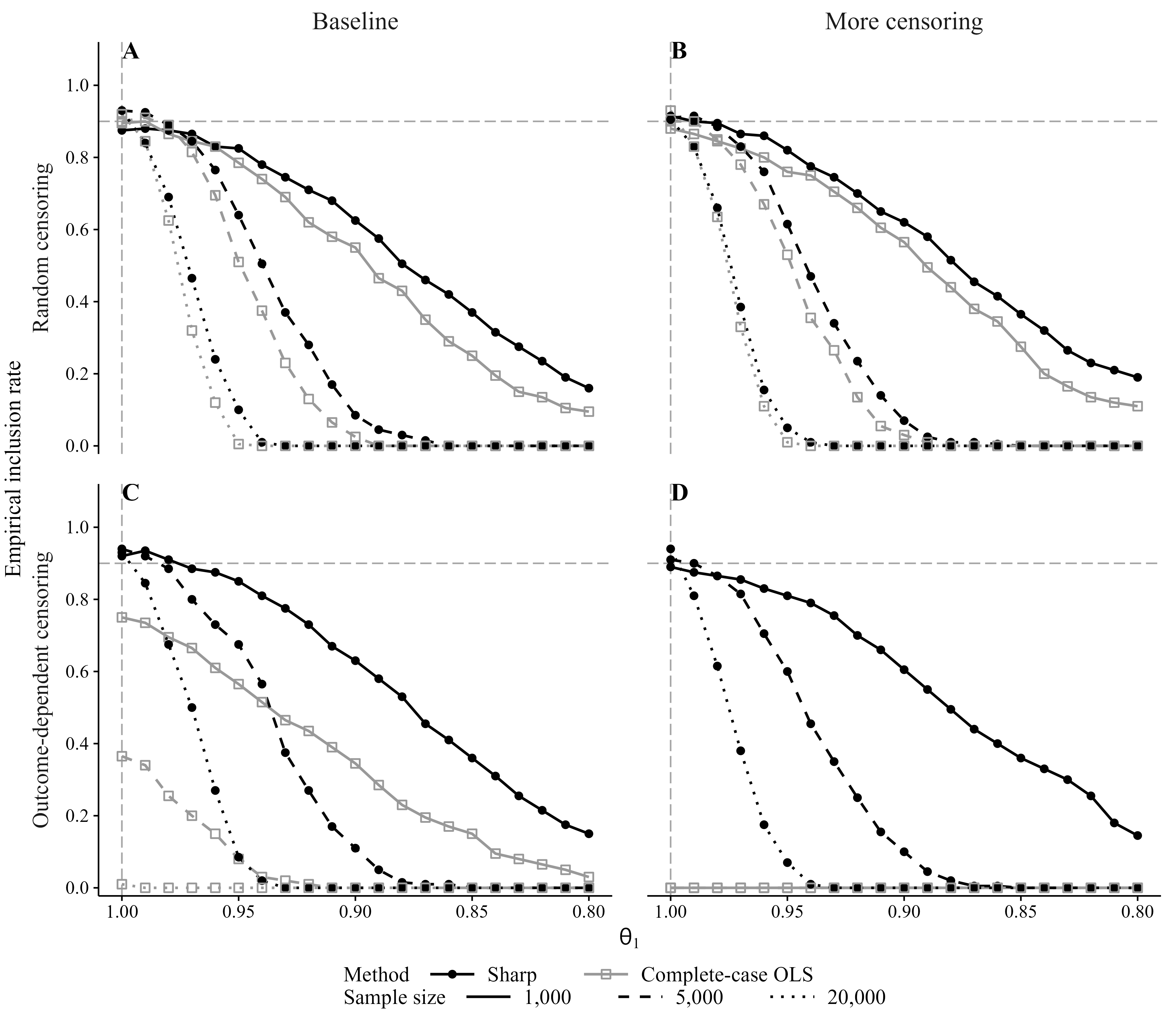}
		\vspace{0.2cm}
		\caption{Empirical Inclusion Frequencies along the Slope Path}
		\label{fig:pointid_power}
		\vspace{0.2cm}
		\begin{minipage}{\textwidth}
			\scriptsize
			\textit{Note:} Share of 200 Monte Carlo replications in which $(0,\theta_1)$
			lies in the 90\% confidence set, for the adaptive procedure (\emph{Sharp})
			and the joint Wald confidence set from complete-case OLS. The true parameter
			is $(0,1)$, so the inclusion rate at $\theta_1=1$ is the coverage and the
			rates at $\theta_1<1$ are one minus the power. The horizontal dashed line
			marks the nominal level of 0.90. Baseline: $p=0.05$; more censoring:
			$p=0.25$.
		\end{minipage}
	\end{figure}

	The preceding simulations illustrate the failure of conventional approaches and the robustness of our method. However, the true shape of the identified set is not known analytically, and the true parameter may lie in its interior. To provide a more stringent assessment of coverage and examine power against fixed local alternatives, we therefore consider a closely related design in which the regression parameter is point identified despite interval observation of some covariates. \\
	We retain the same regression model and interval construction but
	restrict censoring to the lower half of the support of $X_i$.
	Let $C_i=1$ indicate that $X_i$ is interval-observed.
	Censoring can occur only when $X_i\leq 1/2$, while $X_i$
	is always observed exactly for $X_i>1/2$. Per Equation~\eqref{eq:model1}, $E[m(\cdot,\theta)|X=x]=E[m(\cdot,\theta)|X> 1/2]=0$. Since $X_i$ is observed exactly on the nondegenerate interval $(1/2,1]$, the model implies point identification or a rejection of the model. Since the model is simulated with $\theta_0=(0,1)$ we therefore know that $\Theta_I^*=\{(0,1)\}$. The true parameter is therefore a known boundary point of the identified set. This allows us to evaluate finite-sample coverage at the boundary and power against nearby fixed alternatives. \\
	We consider both random and outcome-dependent censoring as well as two levels of the overall frequency of missing data $p=P(C_i=1)$. For censored observations, we use $w=0.1$. Under random censoring,
	$\Pr(C_i=1\mid X_i\leq 1/2)=2p$ ($\Pr(C_i=1)=p$). For outcome-dependent censoring we set $ \Pr(C_i=1\mid X_i\leq 1/2,Y_i\leq 0.25)=3.2p, \space \text{and}
	\Pr(C_i=1\mid X_i\leq 1/2,Y_i>0.25)=0.8p$, implying $P(C_i=1)=p$. We examine both $p\in \{0.05,0.25\}$. \\
	Figure~\ref{fig:pointid_power} reports empirical inclusion frequencies over
	200 Monte Carlo replications along the slope path
	$(0,\theta_1)$, with $\theta_1\in[0.8,1]$, for
	$n\in\{1{,}000,5{,}000,20{,}000\}$. To avoid cluttering the graph, we compare our adaptive implementation with complete-case OLS; the fixed implementation yields similar curves. In Appendix~\ref{sec:appendixE} we compare to \citet{aucejo2017} as well.  The columns compare the baseline and more-censoring designs, while the rows distinguish random from outcome-dependent censoring.
	
	Figure~\ref{fig:pointid_power} illustrates the trade-off between
	robustness and power. Under random censoring, complete-case OLS
	is valid and generally more powerful than our procedure, since
	discarding interval-observed observations does not induce
	selection bias. The lower power of the adaptive procedure in
	this benign design is therefore the price paid for remaining
	agnostic about the censoring mechanism. The power gap tends to
	narrow as the sample size increases and, in this design, under
	more severe censoring. The latter reflects the relative
	informational advantage of our approach, which retains
	interval-observed observations that complete-case OLS discards. \\
	Importantly, our procedure continues to deliver approximately
	nominal coverage at the true parameter under both random and
	outcome-dependent censoring, while its power against fixed
	alternatives increases substantially with sample size. Under
	outcome-dependent censoring, however, more severe censoring
	makes complete-case OLS increasingly unreliable: its inclusion
	frequency at the true parameter deteriorates sharply, whereas
	our procedure retains coverage and comparable power to OLS.

	\section{Application}
	\label{sec:application}
	
	As an empirical illustration, we revisit \citet{castillo2018}, who elicit children’s time preferences using a multiple price list (20 binary choices between \$49 in one month and a larger amount in seven months) and study how the resulting discount-rate measure predicts on-time high school graduation. In a multiple price list, respondents face a sequence of binary intertemporal choices in which one option is held fixed while the alternative (here, the later payment) varies monotonically. Under standard rationality assumptions, choices should exhibit a single cutoff: the respondent selects the sooner option for lower later amounts and then switches once to the later option when the offered premium becomes sufficiently large. \\
	The setting is attractive for our purposes for three reasons. First, multiple price lists are widely used in experimental and applied microeconomics, making the example relevant to a broad class of studies. The design inherently generates interval-valued covariates: even with a single switch, because only a finite number of questions are asked and researchers want to use rounded dollar values, the discount rate is only known to lie within a bracket determined by two adjacent rows. Second, respondents sometimes answer inconsistently, switching
	multiple times between the earlier and later payments. Such respondents have
	been found to differ systematically in observable characteristics (see \citet{yu2021} for a discussion and potential design-based solutions). Treating these responses as random error or using the interval between the first and last switch generates heterogeneous
	interval lengths that may be correlated with unobservables, raising concerns for identification strategies relying on mean independence under interval measurement \citep{manski2002}. Third, graduation outcomes are missing for roughly 30\% of participants due to attrition or departure from the district. This motivated inverse-probability weighting in the original analysis and provides a natural setting for methods that accommodate missing outcomes alongside interval measurement.\\
	In their original analysis, to deal with observing multiple switches, \citet{castillo2018} use an algorithm that maps inconsistent responses into consistent single-switch patterns, which we call their corrected switching pattern. Because the correction need not be unique, for robustness they consider several
	alternatives. The resulting discount-rate intervals are then represented by their midpoints in the regression. \\
	Our approach instead accounts explicitly for the uncertainty within these intervals. We consider three progressively more permissive specifications assuming a probit model of graduation on an intercept and the potentially interval-observed discount rate. First, taking the corrected switching patterns as given,
	we allow discount rates to vary within the survey-implied intervals of width $0.0735$ (\emph{MPL intervals}). Second, for respondents with inconsistent choices, we allow discount rates to lie anywhere between their first and last switches
	(\emph{+ Inconsistent responses}). Finally, we also include respondents with missing graduation outcomes for whom discount-rate information is available and who were dropped from the original analysis (\emph{+ Missing outcomes}). \\
	Table~\ref{tab:tab2} reports results for two specifications. Panel A considers the baseline probit model with the discount rate $X$ as the only covariate. Panel B examines robustness to including the additional covariates $W$ used by \citet{castillo2018} and applying inverse-propensity weighting as in their original analysis.
	For this specification, we retain instruments that depend only on the discount rate $X$, rather than the full covariate vector $(X,W)$. Since this restricts the class of moment inequalities, it yields
	an outer set to the identified set of the fully conditioned model.\footnote{\label{fn:offset}For computational tractability, we do not search over a grid of nuisance parameters $\gamma$ associated with $W$. Instead, we fix $\hat{\gamma}$ at the coefficients
		estimated from a conventional probit using midpoint-imputed discount rates, so that $W^\top\hat{\gamma}$ enters as an offset. The estimated propensity weights are likewise treated as fixed. Consequently, the reported confidence sets do not account for first-stage estimation uncertainty, and the formal coverage
		guarantees of Corollary \ref{C:as-truncations} do not extend to Panel B.} Although theoretically a richer instrument class could incorporate $W$, a fully
	interacted partition of its 16 covariates would generate up to $2^{16}=65{,}536$ cells even if all were binary, before accounting for the partition of $X$. With only $n=547$ observations, most cells
	would contain few or no observations, leading to imprecisely estimated moment inequalities and potentially unstable inference.   \\
	Table~\ref{tab:tab2} reports 95\% confidence sets for the marginal
	effect of the discount rate on the probability of on-time graduation,
	evaluated at a common reference value. In Panel A, the conventional
	probit yields a marginal effect of $-0.086$ with a confidence interval
	of $[-0.152,-0.020]$, excluding zero. To distinguish the effect of
	interval measurement from that of the inference procedure, the
	\emph{Midpoint $X$} specification treats discount rates as exactly
	observed at their imputed midpoints but applies our projection-based
	inference procedure using \citet{andrews2017}. The resulting
	confidence sets are already wider than the conventional probit
	interval, reflecting the change in inference method rather than
	uncertainty about the latent discount rates. Relative to this
	benchmark, accounting for the MPL intervals increases confidence-set
	width by approximately 27\%, yielding $[-0.226,0.020]$ for the
	adaptive and $[-0.216,0.018]$ for the fixed implementation. Both
	now include zero. Allowing for inconsistent responses widens the
	sets considerably further, while incorporating respondents with
	missing graduation outcomes increases the sample from 547 to 790
	but introduces additional ambiguity, resulting in substantially
	wider confidence sets. \\
	The comparison with \citet{aucejo2017} highlights the gains from
	exploiting the joint moment restrictions underlying sharp
	identification. For the basic MPL intervals, the ABH outer-set
	procedure yields $[-0.396,0.403]$, substantially wider than either
	of our confidence sets, with even larger differences when
	inconsistent responses and missing outcomes are incorporated.
	The adaptive and fixed implementations, by contrast, deliver
	broadly similar results. Panel B shows that the qualitative
	pattern persists when additional covariates and inverse-propensity
	weighting are included: conventional probit inference excludes
	zero, while accounting for interval measurement produces wider
	sets containing zero. 
	
	\begin{table}[H]
		\footnotesize
		\caption{Marginal Effects at a Common Reference Discount Rate}
		\vspace{0.2cm}
		\centering
		\resizebox{\textwidth}{!}{
			\begin{tabular}{@{}lrccc@{}}
\toprule
Specification & $n$ & Adaptive & Fixed & ABH \\
\midrule
\multicolumn{5}{@{}l}{\textit{Panel A: Without controls}} \\
\addlinespace[2pt]
Conventional probit & 547 & \multicolumn{3}{c}{$-0.086$ \quad $[-0.152,\,-0.020]$} \\
\cmidrule(l){3-5}
Midpoint $X$ & 547 & $[-0.194,\,0.000]\phantom{^{*}}$ & $[-0.185,\,-0.003]\phantom{^{*}}$ & $[-0.223,\,0.039]\phantom{^{*}}$ \\
MPL intervals & 547 & $[-0.226,\,0.020]\phantom{^{*}}$ & $[-0.216,\,0.018]\phantom{^{*}}$ & $[-0.396,\,0.403]^{*}$ \\
\quad + Inconsistent responses & 547 & $[-0.446,\,0.160]\phantom{^{*}}$ & $[-0.465,\,0.211]\phantom{^{*}}$ & $[-0.638,\,0.587]^{*}$ \\
\quad + Missing outcomes & 790 & $[-0.471,\,0.439]\phantom{^{*}}$ & $[-0.518,\,0.465]\phantom{^{*}}$ & $[-8.872,\,3.446]^{*}$ \\
\addlinespace
\multicolumn{5}{@{}l}{\textit{Panel B: With controls}} \\
\addlinespace[2pt]
Conventional probit & 547 & \multicolumn{3}{c}{$-0.091$ \quad $[-0.174,\,-0.008]$} \\
\cmidrule(l){3-5}
Midpoint $X$ & 547 & $[-0.269,\,0.045]\phantom{^{*}}$ & $[-0.221,\,0.008]\phantom{^{*}}$ & $[-0.237,\,0.044]\phantom{^{*}}$ \\
MPL intervals & 547 & $[-0.283,\,0.097]\phantom{^{*}}$ & $[-0.258,\,0.051]\phantom{^{*}}$ & $[-0.733,\,1.096]^{*}$ \\
\quad + Inconsistent responses & 547 & $[-0.528,\,0.355]\phantom{^{*}}$ & $[-0.446,\,0.328]\phantom{^{*}}$ & $[-7.059,\,4.991]^{*}$ \\
\quad + Missing outcomes & 790 & $[-0.407,\,0.557]\phantom{^{*}}$ & $[-0.458,\,0.581]\phantom{^{*}}$ & $[-8.872,\,3.446]^{*}$ \\
\bottomrule
\end{tabular}
		}
		\label{tab:tab2}
		\vspace{0.2cm}
		\begin{minipage}{\textwidth}
			\scriptsize
			\textit{Note:} Entries are projections of joint 95\% confidence sets for
			$(\theta_0,\theta_1)$ onto the marginal effect
			$\phi(\theta_0+\theta_1\bar{x}+\bar{c})\,\theta_1$. Following
			\citet{castillo2018}, the marginal effect is evaluated at the sample mean of
			the discount rate: $\bar{x}=0.799$ in Panel~A and the IPW-weighted mean
			$\bar{x}=0.823$ in Panel~B for the 547 observations with observed graduation
			status, and the unweighted mean $\bar{x}=0.797$ over all 790 observations in
			the \emph{+ Missing outcomes} rows. Controls enter as an offset centered at their
			(weighted) sample means, so $\bar{c}=0$. Because the parameter is
			set-identified, no point estimate is reported. The \emph{Conventional probit}
			row is the exception: it reports the estimated marginal effect from a standard
			probit with midpoint discount rates (Panel~A: unweighted, no controls;
			Panel~B: IPW with the controls of \citet{castillo2018}) and its marginal
			95\% delta-method confidence interval based on heteroskedasticity-robust
			standard errors. Projection intervals are conservative for the scalar marginal
			effect relative to such marginal intervals. In Panel~B, the control
			coefficients and propensity weights are estimated in a first step and held
			fixed, so these confidence sets do not account for first-step uncertainty and
			the coverage guarantee of Corollary~\ref{C:as-truncations} does not apply
			(see footnote~\ref{fn:offset}).
			$^*$The ABH confidence set reaches the boundary of the parameter grid, so the
			reported interval is truncated.
		\end{minipage}
	\end{table}

\begin{figure}[H]
		\centering
		\includegraphics[width=\linewidth]{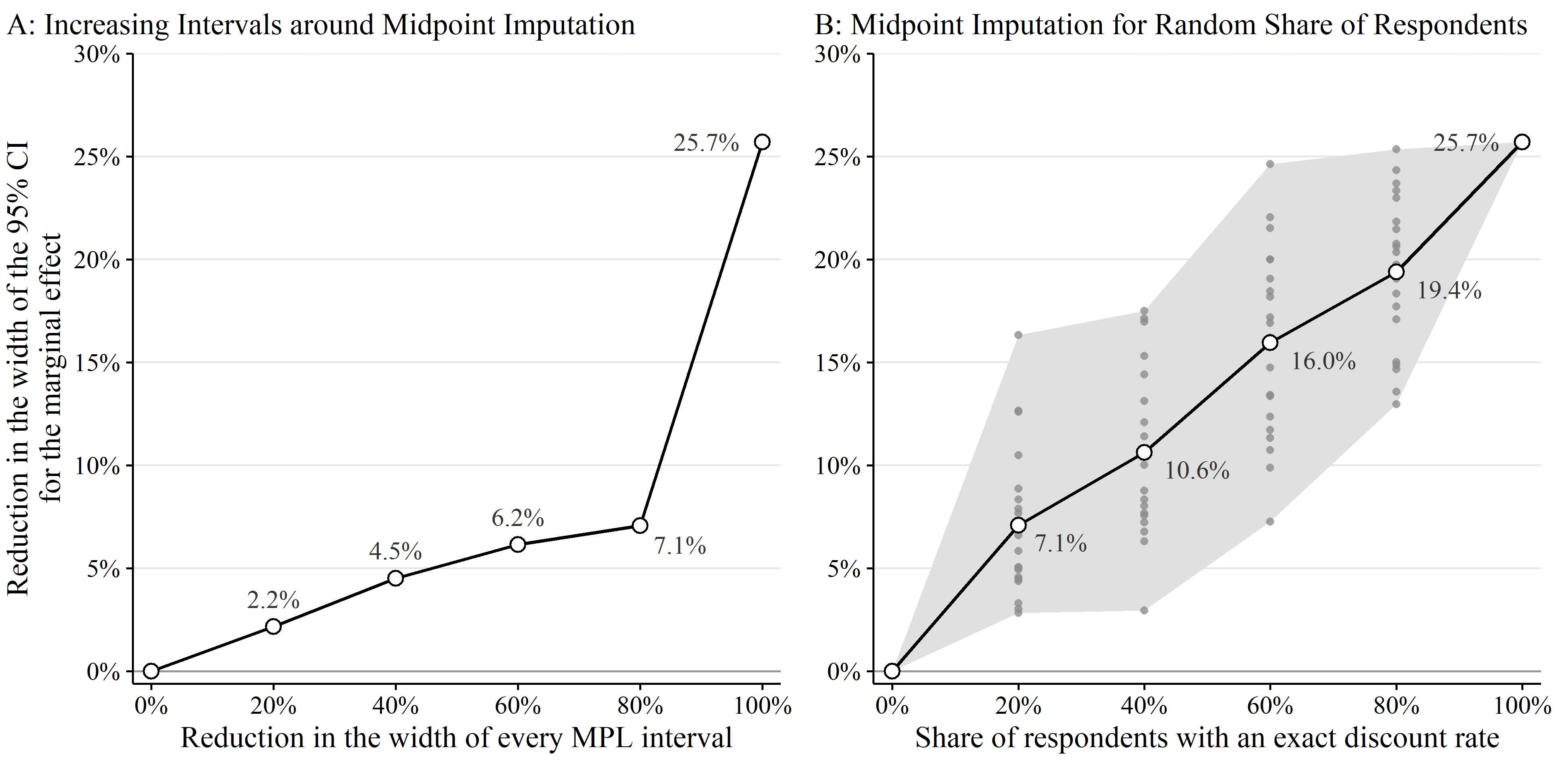}
		\vspace{0.1cm}
		\caption{The Value of More Precise Covariate Measurement}
		\label{fig:intervalreduction}
		\vspace{0.1cm}
		\begin{minipage}{\textwidth}
			\scriptsize
			\textit{Note:} Reduction in the width of the projected 95\% confidence
			interval for the marginal effect, relative to the \emph{MPL intervals}
			specification in Panel~B of Table~\ref{tab:tab2} (fixed direction bank).
			Panel~B: the line is the mean over 20 random draws of the respondents
			who become exactly observed, the dots are the individual draws, and the
			shaded area is the minimum--maximum range across draws.
		\end{minipage}
	\end{figure}

	The application illustrates how even modest uncertainty about
	incompletely observed covariates, here a binning of 7.35 percentage points, can affect empirical conclusions.
	Relative to the midpoint benchmark, accounting for the intervals
	inherent to the MPL design increases the width of the adaptive
	95\% confidence set by approximately 27\%. 
	An additional question for applied researchers is what improvements
	in data quality and/or additional assumptions would be needed to obtain more informative confidence sets. Figure~\ref{fig:intervalreduction} explores this by progressively reducing the ambiguity in the observed discount rates.
	The horizontal axes of both panels represent the same aggregate
	reduction in interval width. Panel A narrows every MPL interval
	symmetrically around its midpoint, whereas Panel B replaces the
	intervals of a randomly selected share of respondents by their
	midpoints, treating these discount rates as exactly observed. These exercises represent either a shrinking of the intervals or having fewer interval observations---both a weakening of the midpoint-imputation assumption. 

	The results reveal substantial differences between these two
	ways of reducing measurement uncertainty. Narrowing every
	interval by 80\% reduces the width of the confidence interval
	by only 7.1\%, whereas replacing 80\% of the intervals by
	their midpoints yields a reduction of 19.4\%. Both approaches
	converge to the same 25.7\% reduction when all interval
	uncertainty is eliminated. The relationship is notably
	nonlinear in Panel A, where moving from point
	observations to 20\% length (equivalently 1.47 p.p.) produces a pronounced decrease in
	precision. The variation across
	random subsamples in Panel B also illustrates that the
	gains from more precise measurement depend on which
	observations become more informative. Such sensitivity exercises can inform the design of future
	surveys by comparing the potential benefits of more precise
	measurement for all respondents with those of obtaining
	more detailed information for a subset. 

	\section{Conclusion}
	\label{sec:conclusion}
	
	Conditional moment equality models are central to empirical economics, but the variables on which they condition are often not fully observed. This paper studies this problem without replacing the latent variables by imputations, conditioning on the observed intervals, or imposing missing-at-random or mean-independence assumptions. We characterize the sharp identified set: the set of all parameter values consistent with the maintained conditional moment model and the incomplete data alone. The resulting object provides a worst-case benchmark against which stronger identifying assumptions and imputation-based procedures can be assessed. \\
	The main identification result rewrites the conditional moment restriction as a countable collection of unconditional moments and embeds these restrictions in an infinite-dimensional random set. A support-function characterization then converts the existence of an admissible completion of the latent data into a collection of moment inequalities. This extends the random-set approach to a setting in which incomplete continuous covariates generate infinitely many restrictions. We also show how the characterization can be operationalized through finite-dimensional sieve approximations, yielding a Hausdorff-consistent estimator and uniformly valid confidence regions. For monotone-index models, including linear, probit, and logit regressions, the finite approximations are computationally tractable. \\
	The simulations illustrate both the benefits and costs of robustness
	to incomplete observation. Under outcome-dependent censoring,
	conventional methods can suffer severe coverage failures, whereas
	our procedure retains approximately nominal coverage. In a
	point-identified design, our procedure also gains power against
	fixed alternatives as the sample size increases, although
	robustness entails a power cost when complete-case analysis is
	valid. \\
	The empirical application demonstrates that even modest
	data limitations can affect substantive conclusions. In the
	multiple-price-list setting, accounting for the intervals
	inherent in the survey increases confidence-set width by
	approximately 27\% relative to the midpoint benchmark,
	rendering a previously significant association inconclusive.
	Inconsistent responses and missing outcomes introduce further
	uncertainty, while our sharp approach yields substantially
	tighter confidence sets than non-sharp outer approximations.
	Finally, the sensitivity exercises illustrate how the gains
	from more precise measurement depend on whether uncertainty
	is reduced for all respondents or eliminated for a subset.
	Such exercises can help guide future survey design and data
	collection by quantifying the potential value of more
	informative observations. \\
	Several extensions remain open. The exact linear-programming representation relies on scalar moments with a monotone index; vector-valued moment functions require a different computational approach. Further, the linear program evaluates the unstudentized criterion and is used here to select directions at finitely many parameter values; it could potentially be replaced by an optimization routine that maximizes the studentized statistic over directions directly, at the expense of computational efficiency. Confidence sets for scalar functionals such as marginal effects are obtained here by projecting joint confidence sets, which is conservative; methods for subvector inference, such as \citet{bugni2017} and \citet{kaido2019}, could deliver shorter intervals. Finally, the choice of instruments and directions affects finite-sample power, and data-driven rules for this choice are left for future work.

	\bibliographystyle{ecta}
	\bibliography{ref}
	
	\appendix

	\newpage 
	
	\section{Proofs of Section 2}
	\label{sec:appendixA}
	
	\subsection{Proof of Lemma~\ref{L:lemma1}}
	
	For compact notation, let $\mathsf C(\theta,\eta)$ denote the statement that
	\begin{align*}
		\mathbf E_\eta[m(Y,X,\theta)\mid X]=0
		\qquad\eta_X\text{-a.s.},
	\end{align*}
	and let $\mathsf U(\theta,\eta)$ denote the statement that
	\begin{align*}
		\int m_j(y,x,\theta)g_k(x)\,
		\eta(dy,dx,dw^o)=0
		\qquad\forall j,\ k.
	\end{align*}
	
	\begin{proof}
		Fix $(\theta,\eta)\in\Theta\times\Gamma$. \\
		Suppose $\mathsf C(\theta,\eta)$ holds and set
		\begin{align*}
			\mu_j(x)
			\equiv
			\mathbf E_\eta[m_j(Y,X,\theta)\mid X=x].
		\end{align*}
		Iterated expectations give
		\begin{align*}
			\int m_j(y,x,\theta)g_k(x)\,d\eta
			=
			\int\mu_j(x)g_k(x)\,\eta_X(dx)
			=0
		\end{align*}
		for every $(j,k)$. Thus $\mathsf U(\theta,\eta)$ holds. \\
		Suppose $\mathsf U(\theta,\eta)$ holds and define
		$\mu_j(X)=\mathbf E_\eta[m_j(Y,X,\theta)\mid X]$. By linearity, the equality holds for every
		$v\in\operatorname{span}(\mathcal G)$. Fix
		$f\in C(\mathcal X)$. Assumption~\ref{A:assumption2}(i) supplies $v_r\in\operatorname{span}(\mathcal G)$ with
		$\|f-v_r\|_\infty\to0$. For all sufficiently large $r$,
		\begin{align*}
			|\mu_jv_r|
			\leq(\|f\|_\infty+1)|\mu_j|.
		\end{align*}
		Dominated convergence therefore gives
		\begin{align*}
			\int\mu_j(x)f(x)\,\eta_X(dx)
			=
			\lim_{r\to\infty}\int\mu_j(x)v_r(x)\,\eta_X(dx)
			=0
			\qquad\forall f\in C(\mathcal X).
		\end{align*}
		Let $\nu_j$ be the finite signed measure with density $\mu_j$ relative to
		$\eta_X$. Since $\mathcal X$ is compact, uniqueness in the Riesz
		representation theorem implies from the preceding display that $\nu_j=0$.
		Thus $\mu_j=0$ $\eta_X$-almost surely for every $j$, and
		$\mathsf C(\theta,\eta)$ holds.
		
		We have proved the fixed-pair equivalence
		\begin{align}
			\tag{FP}
			\label{eq:fixed-pair-equivalence}
			\mathsf C(\theta,\eta)
			\quad\Longleftrightarrow\quad
			\mathsf U(\theta,\eta).
		\end{align}
		Both identified sets quantify over the same $\eta\in\Gamma$, with the
		appropriate integrability included in $\mathsf C$ and $\mathsf U$. Therefore
		$\Theta_I=\Theta_I^*$.
	\end{proof}

	\subsection{Class of Weighted Cubes}
	\label{sec:weigthedcubes}
	
	The following proposition introduces the instrument class used and establishes that the class satisfies Assumption~\ref{A:assumption2}. The choice of weighted cubes allows the exact linear program epigraph formulation described in the main text. For this reason, it is the selected class in our implementations; however, we note that many other classes would also fulfill the conditions needed for our theoretical results.
	
	\begin{proposition}
		\label{P:instrument-classes}
		For the cube construction, let $\mathcal X=[0,1]^{d_x}$, following
		\citet{andrews2013}. Let $r_0\geq1$ be an integer. For each $r\geq r_0$
		let $N_r\equiv\sum_{s=r_0}^r(2s)^{d_x}$, with $N_{r_0-1}=0$,
		and for $a=(a_1,\ldots,a_{d_x})\in\{1,\ldots,2r\}^{d_x}$, define
		\begin{align*}
			I_{a_j,r}
			&\equiv
			\begin{cases}
				[0,(2r)^{-1}],&a_j=1,\\
				((a_j-1)/(2r),a_j/(2r)],&a_j=2,\ldots,2r,
			\end{cases}
			&
			C_{a,r}&\equiv\prod_{j=1}^{d_x}I_{a_j,r},\\
			h_{a,r}(x)&\equiv\mathbf 1\{x\in C_{a,r}\},
			&
			\omega_{a,r}&\equiv w(r)=(c_0+N_r^\beta)^{-1},\\
			g_{a,r}(x)&\equiv\omega_{a,r}h_{a,r}(x).
		\end{align*}
		If $c_0>0$, $\beta>1$, and the pairs $(a,r)$ are enumerated by increasing
		resolution, the resulting classes $\mathcal H=\{h_{a,r}\}$ and $\mathcal G=\{g_{a,r}\}$ satisfy
		Assumption~\ref{A:assumption2}.
	\end{proposition}
	
	\begin{proof}
		At every resolution $r$, the cubes
		$\{C_{a,r}:a\in\{1,\ldots,2r\}^{d_x}\}$ form a partition of
		$[0,1]^{d_x}$ under the stated boundary convention. At each $x$, exactly one
		indicator $h_{a,r}$ is active per resolution, so
		\begin{align*}
			\sup_{x\in[0,1]^{d_x}}
			\sum_{r=r_0}^\infty\sum_a
			|g_{a,r}(x)|
			=
			\sum_{r=r_0}^\infty(c_0+N_r^\beta)^{-1}
			<\infty.
		\end{align*}
		The series converges because $N_r\geq r-r_0+1$ and $\beta>1$. Let $R(K)$ be the largest resolution
		completed among the first $K$ instruments, with $R(K)=r_0-1$ if none is complete. Enumeration by increasing
		resolution implies $R(K)\to\infty$. At any $x$, at most one of the remaining
		indicators is active at each unresolved resolution; hence the tail is bounded
		uniformly by
		\begin{align*}
			\sum_{r>R(K)}(c_0+N_r^\beta)^{-1}
			\longrightarrow0.
		\end{align*}
		This proves Assumption~\ref{A:assumption2}(ii).
		
		For Part~(i), fix $f\in C([0,1]^{d_x})$ and $\varepsilon>0$. Uniform
		continuity supplies a $\delta>0$ such that
		$|f(x)-f(x')|<\varepsilon$ whenever $\|x-x'\|_2<\delta$. Choose $r$ with
		$\sqrt{d_x}/(2r)<\delta$; this quantity bounds the diameter of every
		$C_{a,r}$. Select $x_{a,r}\in C_{a,r}$ and define
		\begin{align*}
			v_r(x)
			\equiv
			\sum_a f(x_{a,r})h_{a,r}(x).
		\end{align*}
		The sum is finite, so $v_r\in\operatorname{span}(\mathcal H)=\operatorname{span}(\mathcal G)$, and the
		diameter bound gives $\|f-v_r\|_\infty<\varepsilon$.
	\end{proof}
	
	The positive weights do not alter this finite-span approximation, because
	$g_{a,r}=\omega_{a,r}h_{a,r}$ with $\omega_{a,r}>0$ implies
	$\operatorname{span}(\mathcal H)=\operatorname{span}(\mathcal G)$. Their role
	is to place the infinite stack in $\ell_1$ and control its tail.
	
	For inference, the same weights also satisfy
	Assumption~\ref{A:polynomial-decay} when $\beta>3/2$. To see this, let $(a(k),r(k))$ be the cell and resolution of instrument $k$
	in this enumeration, so that
	$(h_k,\omega_k,g_k)=(h_{a(k),r(k)},\omega_{a(k),r(k)},g_{a(k),r(k)})$ and
	$N_{r(k)-1}<k\leq N_{r(k)}$. Recall
	$a_k\equiv\sup_{x\in\mathcal X}|g_k(x)|$ from Assumption~\ref{A:polynomial-decay}.
	Since each cube indicator $h_k$ has supremum one,
	\begin{align*}
		a_k=\omega_k
		=(c_0+N_{r(k)}^\beta)^{-1}
		\leq N_{r(k)}^{-\beta}\leq k^{-\beta},
	\end{align*}
	so that $C_a\leq1$.
	
	For truncation through resolution $R$, let $K=N_R$ and let
	$\Pi_Kv$ retain the first $K$ instruments in each primitive-moment block
	and zero-pad the rest. Uniformly over $w^o$, $\theta$, and $v\in\mathcal U=B^\infty$,
	\begin{align*}
		|\widetilde m(w^o,\theta,v)-\widetilde m(w^o,\theta,\Pi_Kv)|
		&\leq d_mC_1\sup_{x\in\mathcal X}
		\sum_{k>K}|g_k(x)|\\
		&=d_mC_1\sum_{r>R}(c_0+N_r^\beta)^{-1}.
	\end{align*}
	Exactly one cube indicator is active at each resolution. Since
	$N_r\asymp r^{d_x+1}$, the last expression is
	$O(R^{1-\beta(d_x+1)})$.

	\subsection{Exact feasibility and closure-based identification}
	\label{sec:schennach-definition}
	
	This subsection compares the exact set $\Theta_I$ with a distance-based
	formulation and shows that they coincide under the paper's assumptions. The
	point is not that $I_\theta$ must be closed; it need not be. Rather, any
	sequence of admissible moment vectors converging to zero yields an admissible
	law that attains zero.
	
	In the notation of this paper, the distance-based formulation associated with
	\citet[Equation~(3)]{schennach2014} is
	\begin{align}
		\label{eq:schennach-set}
		\Theta_I^{\mathrm{Sch}}
		\equiv
		\left\{
		\theta\in\Theta:
		\inf_{\eta\in\Gamma}
		\left\|
		\int M_\theta(y,x)\,\eta(dy,dx,dw^o)
		\right\|_1=0
		\right\}.
	\end{align}
	By \eqref{eq:attainable-moments},
	\begin{align*}
		\inf_{\eta\in\Gamma}
		\left\|\int M_\theta\,d\eta\right\|_1
		=
		\operatorname{dist}_{\ell_1}(0,I_\theta).
	\end{align*}
	Distance to a set equals distance to its closure, so
	\begin{align}
		\label{eq:schennach-closure}
		\Theta_I^{\mathrm{Sch}}
		=
		\{\theta\in\Theta:0\in\operatorname{cl}_{\ell_1}I_\theta\}.
	\end{align}
	Because $\Gamma$ is nonempty,
	\begin{align*}
		\Theta_I\subseteq\Theta_I^{\mathrm{Sch}},
	\end{align*}
	so that
	\begin{align*}
		\Theta_I^{\mathrm{Sch}}\setminus\Theta_I
		=\left\{\theta\in\Theta:
		0\in\operatorname{cl}_{\ell_1}I_\theta
		\text{ and } 0\notin I_\theta\right\}.
	\end{align*}
	
	\begin{proposition}[Equivalence in the present model]
		\label{P:schennach-equivalence}
		Under Assumptions~\ref{A:assumption1}--\ref{A:bounded-moments},
		\begin{align*}
			\Theta_I=\Theta_I^{\mathrm{Sch}}.
		\end{align*}
	\end{proposition}
	
	By \eqref{eq:schennach-closure},
	$\Theta_I^{\mathrm{Sch}}=\{\theta\in\Theta:0\in\operatorname{cl}_{\ell_1}I_\theta\}$,
	and Appendix Lemma~\ref{L:origin-attainment} gives $0\in I_\theta$ if and only
	if $0\in\operatorname{cl}_{\ell_1}I_\theta$. The two sets therefore coincide.
	Lemma~\ref{L:origin-attainment} is proved in Appendix~\ref{sec:proof-theorem1}.
	
	For comparison, the following general observation gives a familiar sufficient
	condition under which exact and distance-based definitions agree.
	
	\begin{proposition}[A general compactness condition]
		\label{P:compact-attainment}
		Let $\Gamma'$ be a nonempty compact topological space and suppose that
		$b_\theta:\Gamma'\to\ell_1$, or $\mathbb R^L$ for finite $L$, is well defined
		for every $\theta\in\Theta$.
		Define
		\begin{align*}
			\Theta^{\mathrm{exact}}
			&\equiv
			\{\theta:b_\theta(\gamma)=\mathbf 0\text{ for some }\gamma\in\Gamma'\},\\
			\Theta^{\mathrm{dist}}
			&\equiv
			\{\theta:\inf_{\gamma\in\Gamma'}\|b_\theta(\gamma)\|_1=0\}.
		\end{align*}
		If $\gamma\mapsto\|b_\theta(\gamma)\|_1$ is lower semicontinuous on
		$\Gamma'$ for every $\theta$, then
		$\Theta^{\mathrm{exact}}=\Theta^{\mathrm{dist}}$.
	\end{proposition}
	
	\begin{proof}
		The inclusion $\Theta^{\mathrm{exact}}\subseteq\Theta^{\mathrm{dist}}$ is
		immediate. Fix $\theta\in\Theta^{\mathrm{dist}}$ and for each $n\geq1$ define
		\begin{align*}
			L_n(\theta)
			\equiv
			\left\{\gamma\in\Gamma':\|b_\theta(\gamma)\|_1\leq n^{-1}\right\}.
		\end{align*}
		Because $\inf_{\gamma\in\Gamma'}\|b_\theta(\gamma)\|_1=0$, every $L_n(\theta)$
		is nonempty. Lower semicontinuity makes each $L_n(\theta)$ closed in $\Gamma'$
		\citep{beer1993},
		and $L_{n+1}(\theta)\subseteq L_n(\theta)$. Compactness and the
		finite-intersection property therefore give
		$\bigcap_{n\geq1}L_n(\theta)\neq\varnothing$. For any $\gamma^*$ in this
		intersection, $0\leq\|b_\theta(\gamma^*)\|_1\leq n^{-1}$ for every $n$, so
		$b_\theta(\gamma^*)=\mathbf 0$ and $\theta\in\Theta^{\mathrm{exact}}$.
	\end{proof}
	
	In the present model $b_\theta(\eta)=\int M_\theta\,d\eta$. If every
	instrument $g_k$ is continuous, then $M_\theta$ is a uniform $\ell_1$-limit of
	continuous truncations and is therefore continuous, so
	$\eta\mapsto\|\int M_\theta\,d\eta\|_1$ is continuous and satisfies the
	lower-semicontinuity condition of Proposition~\ref{P:compact-attainment}. This
	is a sufficient condition, not the route used for
	Theorem~\ref{T:theorem1}. Continuity of the instruments would substantially simplify some parts of the following identification proof; see the remarks in the respective section. For the general theory developed in this paper we do not rely on continuity. \\
	The multiresolution cube geometry follows \citet{andrews2013}; the $\ell_1$ coordinate weights are imposed here for Assumption~\ref{A:assumption2}(ii) and Theorem~\ref{T:theorem3}. The offset $c_0$ is a positive
	normalization or tuning constant, not an identifying restriction. Setting
	$c_0=100$ and $\beta=2$ retains the tuning constants in
	\citet{andrews2013}; here they enter $w(r)=(c_0+N_r^\beta)^{-1}$
	to satisfy the enumerated decay condition. No restriction is imposed on the probability
	that $X$ lies on a cube boundary.

	\section{Proof of Theorem~\ref{T:theorem1}}
	\label{sec:proof-theorem1}
	
	\subsection{Roadmap and probability-space convention}
	
	The proof has five steps. Lemma~\ref{L:random-compact} establishes that the moment image is a measurable integrably bounded random compact set. Lemma~\ref{L:lawselection} identifies admissible laws with laws of feasible latent selections. Lemma~\ref{L:gamma-compact} records compactness and convexity of the admissible-law set. Lemma~\ref{L:aumann-closure} connects the Aumann expectation to the attainable mean-moment set $I_\theta$. Lemma~\ref{L:origin-attainment} shows that taking the closure does not alter membership at zero. The main proof then applies the convexity and support-function results of \citet{molchanov2017}, similar to \citet{beresteanu2011}. \\
	The probability space carrying the true latent variables $(\Omega,\mathcal{A},F_{0})$ is only a representation of their joint law. An element $\eta \in \Gamma$ should not be understood as a perturbation of the true law $\eta_{0} = \mathcal{L}_{F_{0}}(W,W^{O})$ on that particular space. Rather, it represents an alternative complete data-generating process having the observed
	marginal $P_{W^O}$ and assigning probability one to latent values compatible with the observed set. Because the maintained restrictions and the definition of $\Gamma$ depend only on joint laws, we may, without changing the model or its identified set, work throughout this appendix on the completed enrichment of the original probability space obtained by adjoining an independent $V\sim\operatorname{Unif}[0,1]$. All original random variables are identified with their lifts to the enriched space, which preserves their joint laws and therefore also preserves the conditional moment restriction in \eqref{eq:model1}. Selections and Aumann expectations below are understood on this enriched space, while the original notation is retained for simplicity. The uniform variable allows every conditional probability kernel represented by an element of $\Gamma$ to be realized jointly with $W^O$, as established in Lemma~\ref{L:lawselection}.\footnote{This enrichment follows the construction in the proof of \citet[Theorem~1.4.8, pp.~82--83]{molchanov2017}; see also \citet[Definition~1.4.7 and Example~1.4.9, pp.~82--83]{molchanov2017}.} The enriched probability space is nonatomic, so the Aumann expectations used below are convex by \citet[Theorem~2.1.30]{molchanov2017}. Thus their population convexity does not rely on nonatomicity of the original probability space or on the i.i.d.\ sampling assumption; the latter remains in force for the sampling and asymptotic results. \\
	Recall that $\pi_K$ is the truncation map and $\lambda_K$ is the corresponding zero-padding embedding defined in Section~\ref{sec:estimation}.

	\subsection{Auxiliary lemmas}
	
	\begin{lemma}
		\label{L:random-compact}
		The map $M_\theta:\mathcal W\to\ell_1$ is bounded and Borel measurable.
		Moreover,
		\begin{align*}
			Q_\theta(W^O)
			=
			\operatorname{cl}_{\ell_1}M_\theta(D(W^O))
		\end{align*}
		is an integrably bounded, almost surely nonempty random compact set in
		$\ell_1$. For every finite $K$, $Q_{\theta,K}\equiv\pi_KQ_\theta$ is a random compact set in $\mathbb R^{d_mK}$.
	\end{lemma}
	
	\begin{proof}
		Index the coordinates by $(j,k)$ and write
		\begin{align*}
			M_\theta(y,x)_{j,k}
			=
			m_j(y,x,\theta)g_k(x).
		\end{align*}
		Assumptions~\ref{A:assumption2} and~\ref{A:bounded-moments} give
		\begin{align}
			\label{eq:M-envelope}
			\|M_\theta(y,x)\|_1
			&\leq
			\sum_{j=1}^{d_m}\sum_{k\geq1}
			|m_j(y,x,\theta)||g_k(x)|\notag\\
			&\leq
			d_mC_1
			\sup_{x'\in\mathcal X}
			\sum_{k\geq1}|g_k(x')|
			<\infty,
		\end{align}
		uniformly in $(y,x)$. Thus $M_\theta$ is bounded and $\ell_1$-valued. \\
		Let $M_{\theta,K}^0=\lambda_K\pi_KM_\theta$. The finite-coordinate map
		$\pi_KM_\theta:\mathcal W\to\mathbb R^{d_mK}$ is Borel because its
		coordinates are products of continuous and Borel functions; composing it
		with the continuous embedding $\lambda_K$ shows that
		$M_{\theta,K}^0$ is Borel. Moreover,
		\begin{align}
			\label{eq:M-tail}
			\sup_{(y,x)\in\mathcal W}
			\|M_\theta(y,x)-M_{\theta,K}^0(y,x)\|_1
			\leq
			d_mC_1
			\sup_{x\in\mathcal X}
			\sum_{k>K}|g_k(x)|
			\longrightarrow0.
		\end{align}
		A pointwise limit of Borel maps into a metric space is Borel, so
		$M_\theta$ is Borel. \\
		The same tail bound makes the range totally bounded. Given
		$\varepsilon>0$, choose $K$ so that the bound in \eqref{eq:M-tail} is less than $\varepsilon/2$. The bounded set $M_{\theta,K}^0(\mathcal W)$ lies in a finite-dimensional subspace and has a
		finite $\varepsilon/2$-net. That net is an $\varepsilon$-net for
		$M_\theta(\mathcal W)$. Since $\ell_1$ is complete,
		$\operatorname{cl}_{\ell_1}M_\theta(\mathcal W)$ is compact. Each
		$Q_\theta(w^o)$ is a closed subset of this common compact set and is nonempty because $D(w^o)$ is almost surely nonempty under Assumption~\ref{A:assumption1}. Thus its values are almost surely compact and nonempty. \\
		It remains to verify random-set measurability on $\Omega$. Let
		$O\subseteq\ell_1$ be open. Then $B=M_\theta^{-1}(O)$ is Borel, and
		\begin{align*}
			H_B
			&\equiv
			\{w^o:D(w^o)\cap B\neq\varnothing\}\\
			&=
			\operatorname{proj}_{\mathcal W^O}
			\left\{(w,w^o)\in\operatorname{Gr}(D):w\in B\right\}.
		\end{align*}
		The set inside the projection is Borel in a product of Polish spaces, so
		$H_B$ is analytic and therefore $P_{W^O}$-universally measurable. Closure does
		not change whether a set meets an open set, and hence
		\begin{align*}
			\{\omega:Q_\theta(W^O(\omega))\cap O\neq\varnothing\}
			=
			\{\omega:W^O(\omega)\in H_B\}.
		\end{align*}
		The right-hand side belongs to $\mathcal A$ because the probability space is
		complete. This proves Effros measurability of
		$\omega\mapsto Q_\theta(W^O(\omega))$. The envelope
		\eqref{eq:M-envelope} gives integrable boundedness. Finally, the continuous
		image of a compact random set under $\pi_K$ is a compact random set, which
		establishes the finite-dimensional claim.
	\end{proof}
	
	\begin{lemma}
		\label{L:lawselection}
		Let $\mathcal S\equiv\mathcal Y\times\mathcal X$ and let $\mathcal W_o$ denote the state space of $W_o$. Suppose $\mathcal S\times\mathcal W_o$ is Polish and $D:\mathcal W_o\rightrightarrows \mathcal S$ is nonempty compact-valued with closed graph. Let $\Gamma$ be as defined in Section~\ref{sec:identification}. Equivalently, for $\operatorname{Gr}(D)\equiv\{(s,w)\in\mathcal S\times\mathcal W_o:s\in D(w)\}$,
		we have
		\begin{align*}
			\Gamma
			=
			\left\{
			\eta\in\Delta(\mathcal S\times\mathcal W_o):
			\eta_{W_o}=P_{W_o},\ 
			\eta(\operatorname{Gr}(D))=1
			\right\}.
		\end{align*}
		
		Under the enriched probability-space convention described above, any measurable selection $S'=(Y',X')$ of $D(W_o)$ induces an element $\eta\equiv\mathcal L(S',W_o)\in\Gamma$. Conversely, for every $\eta\in\Gamma$, there exists a measurable selection $S^\eta=(Y^\eta,X^\eta)$ of $D(W_o)$ such that $\mathcal L(S^\eta,W_o)=\eta$. In other words, $\Gamma=\{\mathcal{L}(S,W_o):S(\omega)\in D(W_o(\omega))\ \text{a.s.}\}$.
	\end{lemma}
	
	\begin{proof}
		First let $S'=(Y',X')$ be a measurable selection of $D(W_o)$ and define $\eta\equiv\mathcal L(S',W_o)$. Then $\eta_{W_o}=P_{W_o}$ by construction. Moreover, $F_0(S'\in D(W_o))=1$, so $\eta(\operatorname{Gr}(D))=1$. Thus $\eta\in\Gamma$. \\
		Conversely, fix $\eta\in\Gamma$. Define
		\begin{align*}
			\mathbb D(\omega):=D(W_o(\omega))\times\{W_o(\omega)\}
			\subseteq \mathcal S\times\mathcal W_o.
		\end{align*}
		By Assumption~\ref{A:assumption1}, $\mathbb D$ is a nonempty random compact set. We verify Artstein's inequalities. Let $F\subseteq\mathcal S\times\mathcal W_o$ be closed. Since $\eta(\operatorname{Gr}(D))=1$,
		$\eta(F)=\eta(F\cap\operatorname{Gr}(D))$. If $(s,w_o)\in F\cap\operatorname{Gr}(D)$, then $s\in D(w_o)$ and
		$(s,w_o)\in F$. Hence $F\cap(D(w_o)\times\{w_o\})\neq\varnothing$. Therefore, 
		\begin{align*}
			\mathbb{I}\{(s,w_o)\in F\cap\operatorname{Gr}(D)\}
			\leq
			\mathbb{I}\{F\cap(D(w_o)\times\{w_o\})\neq\varnothing\}.
		\end{align*}
		Integrating with respect to $\eta$ gives
		\begin{align*}
			\eta(F)
			\leq
			\int
			\mathbb{I}\{F\cap(D(w)\times\{w\})\neq\varnothing\}\,
			\eta_{W_o}(dw).
		\end{align*}
		Because $\eta_{W_o}=P_{W_o}$,
		\begin{align*}
			\eta(F)
			\leq
			F_0\left(F\cap(D(W_o)\times\{W_o\})\neq\varnothing\right)
			=
			F_0\left(\mathbb D\cap F\neq\varnothing\right).
		\end{align*}
		Thus $\eta$ satisfies Artstein's inequalities for $\mathbb D$. By Artstein's selectionability theorem
		\citep[Theorem~1.4.8, pp.~82--83]{molchanov2017}, $\eta$ is selectionable with respect to the law of $\mathbb D$. More specifically, the construction in the proof of that theorem yields a measurable map $f_\eta$ such that, using the independent $V\sim\operatorname{Unif}[0,1]$ supplied by the enriched probability space, the random element $Z^\eta\equiv f_\eta(V,\mathbb D)$ satisfies $Z^\eta\in\mathbb D$ almost surely and $\mathcal L(Z^\eta)=\eta$. Since $\mathbb D=D(W_o)\times\{W_o\}$, the last component of $Z^\eta$ equals $W_o$ almost surely. Writing $Z^\eta=(Y^\eta,X^\eta,W_o)$ therefore gives $(Y^\eta,X^\eta)\in D(W_o)$ $F_0$-a.s. and $\mathcal L(Y^\eta,X^\eta,W_o)=\eta$.
	\end{proof}
	
	This law-selection bridge is the formal random-set foundation for the sharpness statement in the main text. It plays a role analogous to that of Assumption~2.4 in \citet{beresteanu2011}, but here the existence of the relevant selection is derived from the definition of $\Gamma$ and the enriched probability-space convention rather than imposed as an additional model assumption.
	
	\begin{lemma}
		\label{L:gamma-compact}
		Under Assumption \ref{A:assumption1}, the set $\Gamma$ is nonempty, convex, compact, and metrizable under weak convergence.
	\end{lemma}
	
	\begin{proof}
		Nonemptiness follows from
		$\eta_0=\mathcal L_{F_0}(W,W^O)\in\Gamma$. The ambient space
		$\mathcal W\times\mathcal W^O$ is a compact metric space under Assumption \ref{A:assumption1}, so its space of probability
		measures is compact and metrizable under weak convergence
		\citep[Section~8.6]{bogachev2007}. It remains to show that $\Gamma$ is
		closed.
		
		Let $\eta_n\in\Gamma$ and $\eta_n\Rightarrow\eta$. Continuity of the
		coordinate projection gives $(\eta_n)_{W^O}\Rightarrow\eta_{W^O}$, so
		$\eta_{W^O}=P_{W^O}$. Since $\operatorname{Gr}(D)$ is closed, the Portmanteau theorem gives
		\begin{align*}
			1
			=
			\limsup_{n\to\infty}\eta_n(\operatorname{Gr}(D))
			\leq
			\eta(\operatorname{Gr}(D))
			\leq1.
		\end{align*}
		Thus $\eta\in\Gamma$, proving closedness and hence compactness. Finally, the
		fixed-marginal and probability-one support restrictions defining $\Gamma$
		are affine. Therefore every mixture of two elements of $\Gamma$ also belongs
		to $\Gamma$, so $\Gamma$ is convex.
	\end{proof}
	
	The preceding lemmata are sufficient to establish that $Q_\theta$ is well defined and to connect the admissible completion laws in $\Gamma$ to feasible latent selections. It remains to relate the attainable mean-moment set $I_\theta$ generated by $\Gamma$ to the Aumann expectation of $Q_\theta(W^O)$. Two closure operations must be handled carefully. First, because $M_\theta$ need only be Borel measurable, $M_\theta(D(w^o))$ need not be closed, so $Q_\theta(w^o)$ is defined as its closure. Second, the Aumann expectation is the closure of the set of selection means. Lemma~\ref{L:aumann-closure} shows that the combined effect of these operations is $\mathbb E[Q_\theta(W^O)]=\operatorname{cl}_{\ell_1}I_\theta$, while Lemma~\ref{L:origin-attainment} shows that this closure does not affect
	identification because it does not alter membership at zero. \\
	Let
	\begin{align*}
		\mathcal A_\theta
		\equiv
		\left\{
		\mathbf E[q]:
		q\text{ is an integrable selection of }Q_\theta(W^O)
		\right\}
	\end{align*}
	denote the unclosed set of selection means. Recall that $I_\theta$ was defined
	in \eqref{eq:attainable-moments} and that
	$\mathbb E[Q_\theta(W^O)]=\operatorname{cl}_{\ell_1}\mathcal A_\theta$.

	\begin{lemma}
		\label{L:aumann-closure}
		For every $\theta\in\Theta$, $I_\theta \subseteq \mathcal A_\theta \subseteq \operatorname{cl}_{\ell_1} I_\theta$, and consequently $\mathbb E[Q_\theta(W^O)] = \operatorname{cl}_{\ell_1}\mathcal A_\theta = \operatorname{cl}_{\ell_1}I_\theta$.
	\end{lemma}
	
	\begin{proof}
		For the first inclusion, let $z\in I_\theta$. Then there exists
		$\eta\in\Gamma$ such that
		\begin{align*}
			z
			=
			\int M_\theta(w) \eta(dw,dw^o).
		\end{align*}
		By Lemma~\ref{L:lawselection}, there exists a measurable
		$S^\eta\in D(W^O)$ whose joint law with $W^O$ is $\eta$. Hence
		$q^\eta\equiv M_\theta(S^\eta)$ is an integrable selection of
		$Q_\theta(W^O)$ and
		\begin{align*}
			\mathbf E[q^\eta]
			=
			\int M_\theta(w) \eta(dw,dw^o)
			=
			z.
		\end{align*}
		Thus $I_\theta\subseteq\mathcal A_\theta$. \\
		For $\mathcal A_\theta \subseteq \operatorname{cl}_{\ell_1} I_\theta$, let $q$ be an arbitrary integrable
		selection of $Q_\theta(W^O)$. By definition of $Q_\theta$,
		\begin{align*}
			q(\omega) \in \operatorname{cl}_{\ell_1} M_\theta(D(W^O(\omega)))
			\qquad F_0\text{-a.s.}
		\end{align*}
		The closure is relevant because $M_\theta$ is only Borel. Hence
		$q(\omega)$ need not admit an exact preimage in $D(W^O(\omega))$,
		although it admits admissible preimages whose moment vectors are
		arbitrarily close. Fix $r\geq1$ and define the relation
		\begin{align*}
			\mathcal R_r \equiv
			\left\{
			((w^o,z),w)\in (\mathcal W^O\times\ell_1)\times\mathcal W :
			w\in D(w^o),\,
			||M_\theta(w)-z||_1 < r^{-1}
			\right\}.
		\end{align*}
		For $e=(w^o,z)$, let $\mathcal R_r^\circ(e) \equiv \{w:(e,w)\in\mathcal R_r\}$. Thus $\mathcal R_r^\circ(w^o,z)$ is the set of admissible latent values
		$w\in D(w^o)$ whose moment vectors approximate the target $z$ within
		$r^{-1}$. \\
		The relation $\mathcal R_r$ is Borel, because $\{(w^o,w):w\in D(w^o)\} = \operatorname{Gr}(D)$ is closed, while $(w,z)
		\mapsto ||M_\theta(w)-z||_1$ is Borel because $M_\theta$ is Borel by
		Lemma~\ref{L:random-compact} and the norm is continuous. \\
		Next define
		\begin{align*}
			\mathsf{E} \equiv \mathcal W^O\times\ell_1, \quad
			C_r \equiv \operatorname{proj}_{\mathsf E}\mathcal R_r, \quad
			Z_q(\omega) \equiv (W^O(\omega),q(\omega)), \quad 
			\mu \equiv \mathcal L_{F_0}(Z_q).
		\end{align*}
		Since $q$ is an integrable $\ell_1$-valued selection, it is measurable, and hence $Z_q$ is an $\mathsf E$-valued measurable random element. Moreover, $\mathsf E$ and $\mathcal W$ are standard Borel spaces. \\
		Since $\mathcal R_r$ is Borel in $\mathsf E\times\mathcal W$, its projection $C_r$ is analytic. Hence $C_r$ is universally measurable and therefore belongs to the $\mu$-completion $\overline{\mathcal B(\mathsf E)}^{\mu}$. \\
		By $q(\omega)\in \operatorname{cl}_{\ell_1}M_\theta(D(W^O(\omega)))$ $F_0$-a.s., for every $r$ there exists an admissible $w\in D(W^O(\omega))$ such that $|M_\theta(w)-q(\omega)|_1<r^{-1}$. Hence $\mathcal R_r^\circ(W^O(\omega),q(\omega))\neq\varnothing$ $F_0$-a.s. Equivalently, $Z_q(\omega)\in C_r$ $F_0$-a.s. Thus the random pair $Z_q=(W^O,q)$ lies almost surely in the part of the relation on which an approximate inverse exists, and therefore $\mu(C_r)=1$. \\
		By the Jankov--von Neumann uniformization theorem
		\citep[Theorem~18.1]{kechris1995}, there exists a universally
		measurable map $\widetilde\sigma_r:C_r\to\mathcal W$ such that
		\begin{align*}
			\widetilde\sigma_r(e)
			\in
			\mathcal R_r^\circ(e)
			\qquad
			\text{for every }e\in C_r.
		\end{align*}
		Fix $w_0\in\mathcal W$ and extend $\widetilde\sigma_r$ to all of
		$\mathsf E$ by setting $\widetilde\sigma_r(e)=w_0$ for
		$e\notin C_r$. Since $C_r$ and the inverse images of Borel sets under the selector are universally measurable, this extension remains universally measurable. In particular, it is measurable with respect to the $\mu$-completion of $\mathcal B(\mathsf E)$. \\
		Define
		\begin{align*}
			S_{q,r}(\omega) \equiv \widetilde\sigma_r(Z_q(\omega)) = \widetilde\sigma_r(W^O(\omega),q(\omega)).
		\end{align*}
		Then $S_{q,r}$ is $\mathcal A$-measurable. Indeed, for every
		$B\in\mathcal B(\mathcal W)$, $\widetilde\sigma_r^{-1}(B)$ belongs to the $\mu$-completion of $\mathcal B(\mathsf E)$. Hence it differs from a Borel subset of $\mathsf E$ only by a $\mu$-null set. Pulling back through $Z_q$ therefore gives a set that differs from an $\mathcal A$-measurable set only by an $F_0$-null set. Completeness of $(\Omega,\mathcal A,F_0)$ yields measurability. \\
		Since $Z_q\in C_r$ almost surely, the defining property of the selector gives
		\begin{align*}
			S_{q,r} \in D(W^O),\
			||M_\theta(S_{q,r})-q||_1 <r^{-1}
			\qquad F_0\text{-a.s.}
		\end{align*}
		Thus $S_{q,r}$ is an admissible latent completion whose moment vector approximates the selected vector $q$ pointwise. \\
		Let $\eta_r \equiv \mathcal L_{F_0}(S_{q,r},W^O)$. By Lemma~\ref{L:lawselection}, $\eta_r\in\Gamma$. Moreover, the
		uniform $\ell_1$ bound on $M_\theta$ implies that
		$M_\theta(S_{q,r})$ is Bochner integrable. Therefore
		\begin{align*}
			z_r \equiv \int M_\theta(w)\,\eta_r(dw,dw^o) = \mathbf{E}[M_\theta(S_{q,r})] \in I_\theta.
		\end{align*}
		Using the above and the Bochner triangle inequality,
		\begin{align*}
			|z_r-\mathbf E[q]|_1 = \left| \mathbf E[M_\theta(S_{q,r})-q] \right|_1\
			\leq \mathbf E\left[ |M_\theta(S_{q,r})-q|_1 \right]\ < r^{-1}.
		\end{align*}
		Hence $z_r\in I_\theta$ and $z_r \to \mathbf E[q]$ in $\ell_1$. Therefore $\mathbf E[q] \in \operatorname{cl}_{\ell_1}I_\theta$. Since $q$ was an arbitrary integrable selection of $Q_\theta(W^O)$, $\mathcal A_\theta \subseteq \operatorname{cl}_{\ell_1}I_\theta$. Together with the first inclusion, this proves the first part of the lemma. \\
		Finally, taking norm closures yields $\operatorname{cl}_{\ell_1}\mathcal A_\theta = \operatorname{cl}_{\ell_1}I_\theta$. By definition of the Aumann expectation, $\mathbb E[Q_\theta(W^O)] = \operatorname{cl}_{\ell_1}\mathcal A_\theta$, which establishes the second claim of the lemma.
	\end{proof}

	\begin{lemma}
		\label{L:origin-attainment}
		Under Assumptions~\ref{A:assumption1}--\ref{A:bounded-moments},
		\begin{align}
			\label{eq:origin-attainment}
			0\in\operatorname{cl}_{\ell_1}I_\theta
			\qquad\Longleftrightarrow\qquad
			0\in I_\theta.
		\end{align}
	\end{lemma}
	
	\begin{proof}
		Only the forward implication requires proof. Suppose
		$0\in\operatorname{cl}_{\ell_1}I_\theta$. There are
		$\eta_n\in\Gamma$ such that
		\begin{align*}
			z_n
			\equiv
			\int M_\theta(w)\,\eta_n(dw,dw^o),
			\qquad
			\|z_n\|_1\to0.
		\end{align*}
		By Lemma~\ref{L:gamma-compact}, pass to a subsequence with
		$\eta_n\Rightarrow\eta\in\Gamma$.
		
		Fix $j\in\{1,\ldots,d_m\}$, $f\in C(\mathcal X)$, and
		$\varepsilon>0$. Assumption~\ref{A:assumption2}(i) gives
		\begin{align*}
			v_\varepsilon(x)
			=
			\sum_{k\in K_\varepsilon}a_kg_k(x),
			\qquad
			K_\varepsilon\subset\mathbb N\text{ finite},
			\qquad
			\|f-v_\varepsilon\|_\infty<\varepsilon.
		\end{align*}
		Let
		$B_\varepsilon=\sum_{k\in K_\varepsilon}|a_k|<\infty$.
		Because the $(j,k)$ coordinate of $z_n$ is
		$\int m_jg_k\,d\eta_n$,
		\begin{align}
			\label{eq:finite-coordinate-bound}
			\left|
			\int m_j(y,x,\theta)v_\varepsilon(x)\,d\eta_n
			\right|
			&=
			\left|
			\sum_{k\in K_\varepsilon}
			a_k
			\int m_j(y,x,\theta)g_k(x)\,d\eta_n
			\right| \notag\\
			&\leq
			B_\varepsilon\|z_n\|_1
			\longrightarrow0.
		\end{align}
		Assumption~\ref{A:bounded-moments} then gives
		\begin{align}
			\label{eq:continuous-test-bound}
			\left|
			\int m_j(y,x,\theta)f(x)\,d\eta_n
			\right|
			\leq
			B_\varepsilon\|z_n\|_1+C_1\varepsilon.
		\end{align}
		The function
		$(y,x,w^o)\mapsto m_j(y,x,\theta)f(x)$ is bounded and continuous. Weak
		convergence therefore permits passage to the limit in its integral. From
		\eqref{eq:continuous-test-bound},
		\begin{align*}
			\left|
			\int m_j(y,x,\theta)f(x)\,d\eta
			\right|
			\leq C_1\varepsilon.
		\end{align*}
		Letting $\varepsilon\downarrow0$ gives
		\begin{align}
			\label{eq:all-continuous-tests}
			\int m_j(y,x,\theta)f(x)\,d\eta=0
			\qquad
			\forall f\in C(\mathcal X),\
			\forall j.
		\end{align}
		
		Let
		$\mu_j(X)=\mathbf E_\eta[m_j(Y,X,\theta)\mid X]$. By iterated expectations,
		the finite signed measure with density $\mu_j$ relative to $\eta_X$
		integrates every $f\in C(\mathcal X)$ to zero. The Riesz representation
		theorem therefore implies $\mu_j=0$ $\eta_X$-almost surely for every $j$.
		The fixed-pair equivalence \eqref{eq:fixed-pair-equivalence}, together with
		the coordinate identity for Bochner integrals, gives
		\begin{align*}
			\int M_\theta(w)\,\eta(dw,dw^o)=0.
		\end{align*}
		Because $\eta\in\Gamma$, this vector belongs to $I_\theta$. Thus
		$0\in I_\theta$. \\
		Notice that the proof passes to the weak limit only through continuous test functions. It never requires the limiting law to assign zero probability to the discontinuity boundaries of the cube instruments.
	\end{proof}
	
	\begin{remark}
		A different and simpler proof strategy could be employed under a continuous moment class, effectively making Lemma~\ref{L:aumann-closure} and Lemma~\ref{L:origin-attainment} superfluous. Suppose $M_\theta:\mathcal W\to\ell_1$ is continuous. Then $Q_\theta(W^O)=M_\theta(D(W^O))$ $F_0\text{-a.s.}$ and $\mathcal A_\theta= I_\theta =\mathbb E[Q_\theta(W^O)]$. Moreover, $I_\theta$ is compact in $\ell_1$.
		
		\begin{proof}
			Because $D(W^O)$ is compact and $M_\theta$ is continuous,
			$M_\theta(D(W^O))$ is compact and already closed. Thus the closure defining
			$Q_\theta$ is redundant.
			
			Let $q$ be an integrable selection of $Q_\theta(W^O)$ and define
			\begin{align*}
				H_q(\omega)
				\equiv
				\left\{
				w\in D(W^O(\omega)):
				M_\theta(w)=q(\omega)
				\right\}.
			\end{align*}
			The correspondence is nonempty and random closed by the random-inverse
			theorem \citep[Theorem~1.3.27]{molchanov2017}. The Fundamental Selection
			Theorem \citep[Theorem~1.4.1]{molchanov2017} supplies a measurable
			$S_q\in H_q$. Lemma~\ref{L:lawselection} implies
			$\eta_q=\mathcal L_{F_0}(S_q,W^O)\in\Gamma$, and hence
			\begin{align*}
				\mathbf E[q]
				=
				\int M_\theta(w)\,\eta_q(dw,dw^o)
				\in I_\theta.
			\end{align*}
			Therefore $\mathcal A_\theta\subseteq I_\theta$; the first inclusion in
			Lemma~\ref{L:aumann-closure} gives the reverse inclusion.
			
			To show closedness, define
			\begin{align*}
				T_\theta:\Gamma&\to\ell_1,&
				T_\theta(\eta)&=\int M_\theta(w)\,\eta(dw,dw^o).
			\end{align*}
			If $\eta_n\Rightarrow\eta$, the first $d_mK$ coordinates converge for every
			fixed $K$ because they are bounded and continuous. The tail bound
			\eqref{eq:M-tail} holds uniformly over $\Gamma$. Finite-coordinate
			convergence followed by the uniform tail bound gives
			$\|T_\theta(\eta_n)-T_\theta(\eta)\|_1\to0$. Thus $T_\theta$ is continuous.
			Lemma~\ref{L:gamma-compact} makes $\Gamma$ compact, so
			$I_\theta=T_\theta(\Gamma)$ is compact and closed. Since
			$\mathbb E[Q_\theta(W^O)]=\operatorname{cl}_{\ell_1}\mathcal A_\theta$, all claimed
			equalities follow.
		\end{proof}
		
	\end{remark}
	
	\subsection{Main Proof of Theorem~\ref{T:theorem1}}
	
	\begin{proof}
		Lemma~\ref{L:random-compact} establishes the asserted random-set properties.
		The Fundamental Selection Theorem
		\citep[Theorem~1.4.1]{molchanov2017} supplies a measurable selection of
		$Q_\theta(W^O)$; the common envelope in \eqref{eq:M-envelope} makes that
		selection integrable. Hence $\mathbb E[Q_\theta(W^O)]$ is nonempty.
		By the coordinate identity for Bochner integrals and the fixed-pair
		equivalence \eqref{eq:fixed-pair-equivalence},
		\begin{align*}
			\theta\in\Theta_I^*
			\quad\Longleftrightarrow\quad
			\theta\in\Theta_I
			\quad\Longleftrightarrow\quad
			0\in I_\theta.
		\end{align*}
		Lemmas~\ref{L:aumann-closure} and~\ref{L:origin-attainment} give
		\begin{align*}
			0\in\mathbb E[Q_\theta(W^O)]
			\quad\Longleftrightarrow\quad
			0\in\operatorname{cl}_{\ell_1}I_\theta
			\quad\Longleftrightarrow\quad
			0\in I_\theta.
		\end{align*}
		This proves \eqref{eq:aumann-identification} and the first line of
		\eqref{eq:support-identification}. \\
		The set $\mathbb E[Q_\theta(W^O)]$ is closed by
		\citet[Definition~2.1.21]{molchanov2017}. It is convex because the extended probability space is nonatomic and \citet[Theorem~2.1.30]{molchanov2017} applies to the integrably bounded random closed set $Q_\theta(W^O)$ in the separable Banach space $\ell_1$.
		Equivalently, after Lemma~\ref{L:aumann-closure}, convexity also follows from convexity of $\Gamma$ and affinity of $\eta\mapsto\int M_\theta\,d\eta$.
		
		By \citet[Theorem~2.1.35]{molchanov2017},
		\begin{align*}
			h(\mathbb E[Q_\theta(W^O)],u)
			=
			\mathbb E_{P_{W^O}}[h(Q_\theta(W^O),u)]
			\qquad\forall u\in\ell_\infty.
		\end{align*}
		For a nonempty closed convex subset $C$ of $\ell_1$, the Hahn--Banach
		separation theorem gives
		\begin{align*}
			0\in C
			\quad\Longleftrightarrow\quad
			h(C,u)\geq0
			\qquad\forall u\in(\ell_1)^*=\ell_\infty .
		\end{align*}
		Positive homogeneity of $u\mapsto h(C,u)$ permits restriction to
		$B^\infty$. Combining
		the last two displays proves the final equality in
		\eqref{eq:support-identification}.
	\end{proof}
	
	\section{Proof of Theorem~\ref{T:theorem2}}
	\label{sec:appendixC}
	
	This section first proves a triangular law of large numbers for random sets. We then proceed by stating and proving several auxiliary lemmata, which are needed to verify the high-level conditions in \citet{chernozhukov2007} for our case and thus establish consistency of the proposed estimator.
	
	\subsection{Triangular Law of Large Numbers for Random Sets}
	Before stating the consistency result for the proposed estimator, we introduce an auxiliary lemma establishing a law of large numbers for random sets growing at rate $p_{n}$, which might be of independent interest and is a crucial part of the proof.
	\begin{lemma}
		Let $p_n$ denote the ambient dimension of the $n$-th row of a triangular array,
		with $p_n\to\infty$. For each $n$, let
		$X_{n,1},\ldots,X_{n,n}$ be i.i.d. nonempty compact random subsets of
		$\mathbb R^{p_n}$, and let $X_n$ denote a generic copy from the $n$-th row and suppose they are uniformly bounded by a deterministic constant $b$. Then
		\begin{align*}
			\rho_H\left(
			\operatorname{co}\left[\frac1n\bigoplus_{i=1}^n X_{n,i}\right],
			\mathbb E[\operatorname{co}X_n]
			\right)
			=
			O_p\left(\sqrt{\frac{p_n}{n}}\right).
		\end{align*}
		If, in addition, the Aumann expectation satisfies $\mathbb E[X_n]=\mathbb E[\operatorname{co}X_n]$, then
		\begin{align*}
			\rho_H\left(
			\frac1n\bigoplus_{i=1}^n X_{n,i},
			\mathbb E[X_n]
			\right)
			=
			O_p\left(
			\sqrt{\frac{p_n}{n}}
			+
			\frac{p_n}{n}
			\right).
		\end{align*}
		Consequently, if $p_n/n\to0$, then
		\begin{align*}
			\rho_H\left(
			\frac1n\bigoplus_{i=1}^n X_{n,i},
			\mathbb E[X_n]
			\right)
			=
			O_p\left(\sqrt{\frac{p_n}{n}}\right).
		\end{align*}
	\end{lemma}
	
	\begin{proof}
		Let $\hat{X}_{n} \equiv \frac{1}{n} \bigoplus_{i=1}^n X_{n,i}$. First, let us control the convexified average. Note that
		$\operatorname{co}(\hat{X}_{n}) = \frac1n\bigoplus_{i=1}^n \operatorname{co}(X_{n,i})$. Using
		$h(\operatorname{co}(\hat{X}_{n}), u) = h(\hat X_n,u)=\frac{1}{n} \sum_{i=1}^{n} h(X_{n,i},u)$ and
		$h(\mathbb{E}[\operatorname{co}(X_{n})],u) = \mathbf{E}[h(\operatorname{co}(X_{n}),u)]=\mathbf{E}[h(X_{n},u)]$, we can rewrite the Hausdorff distance between the convexified average and the Aumann expectation as the empirical process supremum:
		\begin{align*}
			\rho_{H}\left(
			\operatorname{co}(\hat X_n),
			\mathbb{E}[\operatorname{co}(X_n)]
			\right)
			= \sup\limits_{u \in \mathbb B_n^*} \Big| \frac{1}{n} \sum_{i=1}^{n} h(X_{n,i},u) - \mathbf{E}[h(X_{n},u)] \Big|
		\end{align*}
		where $\mathbb B_n^*$ denotes the finite-dimensional dual unit ball corresponding to the ambient space $\mathbb R^{p_n}$. \\
		Define the support function class as $\mathcal{H}_{n} \equiv \{f_{u}: \ f_{u}(X) = h(X,u) , u \in \mathbb B_n^*\}$, thus the above is akin to $\sup\limits_{h \in \mathcal{H}_{n}} |(P_{n} - P_{W^O})h|$, which is controlled by the entropy of the function class $\mathcal{H}_{n}$. From Lemma 4.10 in \citet{pisier1989} we have $N(\mathbb B_n^*,\delta,||\cdot||_{\infty}) \leq (1+\frac{2}{\delta})^{p_{n}}$, which we combine with Lipschitz continuity of the support function, i.e. $|f_u(X)-f_v(X)| = |h(X,u)-h(X,v)| \le \sup_{x\in X}|\langle x,u-v\rangle| \le \sup_{x\in X}\|x\|_1\|u-v\|_\infty \le b\|u-v\|_\infty$, where $b$ is the uniform deterministic bound from the statement of the lemma. Hence $||f_{u}-f_{v}||_{L_{2}(P_{W^O})} \leq b ||u-v||_{\infty}$ and we have $N(\epsilon, \mathcal{H}_{n}, L_{2}(P_{W^O})) \leq N(\frac{\epsilon}{b}, \mathbb B_n^*, ||\cdot||_{\infty})$, thus
		\begin{align*}
			\log N(\epsilon, \mathcal{H}_{n}, L_{2}(P_{W^O})) \leq p_{n} \log \left(1+\frac{2b}{\epsilon} \right) \leq p_{n} \log \left(\frac{Cb}{\epsilon} \right)
		\end{align*}
		By Dudley's entropy theorem \citep{dudley1967}, we have
		\begin{align*}
			\mathbb{E} \left[ \sup\limits_{u \in \mathbb B_n^*} \Big| \frac{1}{n} \sum_{i=1}^{n} h(X_{n,i},u) - \mathbf{E}[h(X_{n},u)] \Big| \right ] &\leq \frac{C}{\sqrt{n}} \int_{0}^{b} \sqrt{p_{n} \log \left( \frac{Cb}{\epsilon}\right)} d\epsilon \\
			&=
			C\sqrt{\frac{p_{n}}{n}} b \int_{0}^{1} \sqrt{\log(C/t)} dt \\
			&\leq C b \sqrt{\frac{p_{n}}{n}}
		\end{align*}
		where the equality uses a change of variable and the last inequality uses $\int_{0}^{1} \sqrt{\log(C/t)} dt < \infty$. Using Markov's inequality it follows that
		\begin{align*}
			\sup\limits_{u \in \mathbb B_n^*} \Big| \frac{1}{n} \sum_{i=1}^{n} h(X_{n,i},u) - \mathbf{E}[h(X_{n},u)] \Big| = O_{p}\left( \sqrt{\frac{p_{n}}{n}} \right).
		\end{align*}
		Hence,
		\begin{align*}
			\rho_H\left(
			\operatorname{co}(\hat X_n),
			\mathbb{E}[\operatorname{co}(X_n)]
			\right)
			=
			O_{p}\left( \sqrt{\frac{p_{n}}{n}} \right).
		\end{align*}
		
		It remains to pass from the convexified average back to the original, potentially non-convex, Minkowski average. For each $n$, the sets $X_{n,i}$ are compact subsets of the finite-dimensional space $\mathbb R^{p_n}$. Therefore, applying the finite-dimensional Shapley--Folkman--Starr theorem row by row gives
		\begin{align*}
			\rho_H\left(
			\hat X_n,
			\operatorname{co}(\hat X_n)
			\right)
			\leq
			\frac{p_n}{n}
			\max_{1\leq i\leq n}\operatorname{diam}_1(X_{n,i})
			\leq
			2 b \frac{p_n}{n}.
		\end{align*}
		If $\mathbb{E}[X_n]=\mathbb{E}[\operatorname{co}(X_n)]$, then by the triangle inequality,
		\begin{align*}
			\rho_H\left(
			\hat X_n,
			\mathbb{E}[X_n]
			\right)
			&\leq
			\rho_H\left(
			\hat X_n,
			\operatorname{co}(\hat X_n)
			\right)
			+
			\rho_H\left(
			\operatorname{co}(\hat X_n),
			\mathbb{E}[\operatorname{co}(X_n)]
			\right) \\
			&=
			O\left(\frac{p_n}{n}\right)
			+
			O_p\left( \sqrt{\frac{p_n}{n}} \right).
		\end{align*}
		Thus,
		\begin{align*}
			\rho_H\left(
			\frac{1}{n} \bigoplus_{i=1}^n X_{n,i},
			\mathbb{E}[X_n]
			\right)
			=
			O_p\left( \sqrt{\frac{p_n}{n}}+\frac{p_n}{n} \right).
		\end{align*}
		If $p_n/n \to 0$, then $p_n/n=o(\sqrt{p_n/n})$, and therefore
		\begin{align*}
			\rho_H\left(
			\frac{1}{n} \bigoplus_{i=1}^n X_{n,i},
			\mathbb{E}[X_n]
			\right)
			=
			O_p\left( \sqrt{\frac{p_n}{n}} \right).
		\end{align*}
	\end{proof}

	\subsection{Auxiliary Lemmata}
	
	The first auxiliary lemma establishes a uniform version of the triangular law of large numbers for the random sets established above.
	
	\begin{lemma}
		\label{L:uniform_rs_lln}
		Let Assumptions~\ref{A:assumption1}--\ref{A:thetalipschitz} hold. Let \(K_n\to\infty\), and let
		\(Q_{\theta,K_n}\subset\mathbb R^{d_mK_n}\) denote the truncation of
		\(Q_\theta\) to the first \(K_n\) instruments \(g_1,\ldots,g_{K_n}\).
		For \(i=1,\ldots,n\), write $Q_{\theta,K_n,i} \equiv Q_{\theta,K_n}(W_i^O)$.
		Suppose that $\frac{d_mK_n}{n}\to0$. Then
		\begin{align*}
			\sup_{\theta\in\Theta}
			\rho_H\left(
			\frac1n\bigoplus_{i=1}^n Q_{\theta,K_n,i},
			\mathbb E[Q_{\theta,K_n}]
			\right)
			=
			O_p\left(
			\sqrt{\frac{d_mK_n+d_\theta}{n}}
			\right).
		\end{align*}
		Since \(d_m\) and \(d_\theta\) are fixed, this is equivalently
		\begin{align*}
			\sup_{\theta\in\Theta}
			\rho_H\left(
			\frac1n\bigoplus_{i=1}^n Q_{\theta,K_n,i},
			\mathbb E[Q_{\theta,K_n}]
			\right)
			=
			O_p\left(\sqrt{\frac{K_n}{n}}\right).
		\end{align*}
	\end{lemma}
	
	\begin{proof}
		Let
		\[
		\hat Q_{n,\theta,K_n}
		\equiv
		\frac1n\bigoplus_{i=1}^n Q_{\theta,K_n,i}.
		\]
		First, control the convexified average. Since
		\(Q_{\theta,K_n}\subset\mathbb R^{d_mK_n}\), the support-function representation
		of the Hausdorff distance for compact convex sets gives
		\begin{align*}
			\rho_H\left(
			\operatorname{co}\left(\hat Q_{n,\theta,K_n}\right),
			\mathbb E[\operatorname{co}(Q_{\theta,K_n})]
			\right)
			=
			\sup_{u\in B_{K_n}}
			\left|
			\frac1n\sum_{i=1}^n h(Q_{\theta,K_n,i},u)
			-
			\mathbb E[h(Q_{\theta,K_n},u)]
			\right|,
		\end{align*}
		where
		\[
		B_{K_n}
		\equiv
		\{u\in\mathbb R^{d_mK_n}:\|u\|_\infty\le1\}.
		\]
		Here we used
		\(h(\operatorname{co}(A),u)=h(A,u)\),
		\(h(n^{-1}\oplus_i A_i,u)=n^{-1}\sum_i h(A_i,u)\), and
		\[
		h(\mathbb E[\operatorname{co}(Q_{\theta,K_n})],u)
		=
		\mathbb E[h(Q_{\theta,K_n},u)].
		\]
		
		Define the support-function class
		\[
		\mathcal H_n^\Theta
		\equiv
		\left\{
		f_{\theta,u}(W_i^O)
		=
		h(Q_{\theta,K_n}(W_i^O),u)
		:
		\theta\in\Theta,\ u\in B_{K_n}
		\right\}.
		\]
		By Assumptions~\ref{A:assumption2} and~\ref{A:bounded-moments}, there exists \(b<\infty\), independent of \(n\), such that
		\[
		\sup_{\theta\in\Theta}
		\sup_{q\in Q_{\theta,K_n}}
		\|q\|_1
		\le b .
		\]
		Indeed, one may take
		\[
		b
		=
		d_m C_1
		\sup_{x\in\mathcal X}
		\sum_{k=1}^\infty |g_k(x)|
		<\infty.
		\]
		Hence the class has envelope \(b\). Moreover, for \(u,v\in B_{K_n}\),
		\begin{align*}
			|h(Q_{\theta,K_n},u)-h(Q_{\theta,K_n},v)|
			\le
			b\|u-v\|_\infty .
		\end{align*}
		Assumption~\ref{A:thetalipschitz} and the summability condition in Assumption~\ref{A:assumption2} imply that, for some
		constant \(\bar L<\infty\) independent of \(n\),
		\begin{align*}
			|h(Q_{\theta,K_n},u)-h(Q_{\theta',K_n},u)|
			\le
			\bar L\|\theta-\theta'\|.
		\end{align*}
		Therefore, covering both \(B_{K_n}\) and \(\Theta\) gives
		\begin{align*}
			\log N(\epsilon,\mathcal H_n^\Theta,L_2(P_{W^O}))
			&\le
			d_mK_n \log\left(\frac{C b}{\epsilon}\right)
			+
			d_\theta \log\left(\frac{C \bar L}{\epsilon}\right).
		\end{align*}
		By Dudley's entropy theorem,
		\begin{align*}
			\mathbb E\left[
			\sup_{\theta\in\Theta}
			\sup_{u\in B_{K_n}}
			\left|
			\frac1n\sum_{i=1}^n
			\left\{
			h(Q_{\theta,K_n,i},u)
			-
			\mathbb E[h(Q_{\theta,K_n},u)]
			\right\}
			\right|
			\right]
			&\le
			\frac{C}{\sqrt n}
			\int_0^b
			\sqrt{
				d_mK_n\log\left(\frac{C b}{\epsilon}\right)
				+
				d_\theta\log\left(\frac{C \bar L}{\epsilon}\right)
			}
			d\epsilon \\
			&\le
			C\sqrt{\frac{d_mK_n+d_\theta}{n}} .
		\end{align*}
		Hence, by Markov's inequality,
		\begin{align*}
			\sup_{\theta\in\Theta}
			\rho_H\left(
			\operatorname{co}\left(\hat Q_{n,\theta,K_n}\right),
			\mathbb E[\operatorname{co}(Q_{\theta,K_n})]
			\right)
			=
			O_p\left(\sqrt{\frac{d_mK_n+d_\theta}{n}}\right).
		\end{align*}
		
		It remains to remove convexification. For each \(n\) and each \(\theta\),
		\(Q_{\theta,K_n,i}\) is a compact subset of the finite-dimensional space
		\(\mathbb R^{d_mK_n}\). Thus the finite-dimensional Shapley--Folkman--Starr theorem gives,
		uniformly in \(\theta\),
		\begin{align*}
			\rho_H\left(
			\hat Q_{n,\theta,K_n},
			\operatorname{co}\left(\hat Q_{n,\theta,K_n}\right)
			\right)
			\le
			\frac{d_mK_n}{n}
			\max_{1\le i\le n}\operatorname{diam}_1(Q_{\theta,K_n,i})
			\le
			2b\frac{d_mK_n}{n}.
		\end{align*}
		By the convexification identity for the Aumann expectation,
		\[
		\mathbb E[Q_{\theta,K_n}]
		=
		\mathbb E[\operatorname{co}(Q_{\theta,K_n})].
		\]
		Therefore, by the triangle inequality,
		\begin{align*}
			\sup_{\theta\in\Theta}
			\rho_H\left(
			\hat Q_{n,\theta,K_n},
			\mathbb E[Q_{\theta,K_n}]
			\right)
			&\le
			\sup_{\theta\in\Theta}
			\rho_H\left(
			\hat Q_{n,\theta,K_n},
			\operatorname{co}\left(\hat Q_{n,\theta,K_n}\right)
			\right) \\
			&\quad+
			\sup_{\theta\in\Theta}
			\rho_H\left(
			\operatorname{co}\left(\hat Q_{n,\theta,K_n}\right),
			\mathbb E[\operatorname{co}(Q_{\theta,K_n})]
			\right) \\
			&=
			O\left(\frac{d_mK_n}{n}\right)
			+
			O_p\left(\sqrt{\frac{d_mK_n+d_\theta}{n}}\right).
		\end{align*}
		Hence
		\begin{align*}
			\sup_{\theta\in\Theta}
			\rho_H\left(
			\frac1n\bigoplus_{i=1}^n Q_{\theta,K_n,i},
			\mathbb E[Q_{\theta,K_n}]
			\right)
			=
			O_p\left(
			\sqrt{\frac{d_mK_n+d_\theta}{n}}
			+
			\frac{d_mK_n}{n}
			\right).
		\end{align*}
		If \(d_mK_n/n\to0\), the second term is absorbed into the first. Since \(d_m\) and
		\(d_\theta\) are fixed, this yields
		\begin{align*}
			\sup_{\theta\in\Theta}
			\rho_H\left(
			\frac1n\bigoplus_{i=1}^n Q_{\theta,K_n,i},
			\mathbb E[Q_{\theta,K_n}]
			\right)
			=
			O_p\left(\sqrt{\frac{K_n}{n}}\right).
		\end{align*}
	\end{proof}
	
	The following lemma links the deviation of directed Hausdorff distances to the undirected Hausdorff distances between two sets. 
	
	\begin{lemma}
		\label{L:hausdorffdist}
		Let $A$ and $B$ be nonempty closed bounded subsets of a common normed space. Then
		\begin{align*}
			\big|\vec{\rho}_{H}(0,A)-\vec{\rho}_{H}(0,B)\big|
			\leq
			\rho_H(A,B).
		\end{align*}
	\end{lemma}
	
	\begin{proof}
		By the triangle inequality for hemi-metrics, for all $q \in B$, $\vec{\rho}_{H}(0,A) \le \vec{\rho}_{H}(0,q) + \vec{\rho}_{H}(q,A)$ \citep[see][Lemma~6.2.16]{goubaultlarrecq2013}. Therefore,
		\begin{align*}
			\vec{\rho}_{H}(0,A)
			&\leq \min_{q \in B}\Big( \vec{\rho}_{H}(0,q) + \vec{\rho}_{H}(q,A) \Big) \leq \min_{q \in B}\Big( \rho_H(0,q) + \rho_H(q,A) \Big) \\
			&\leq \vec{\rho}_{H}(0,B) + \rho_H(A,B).
		\end{align*}
		Hence $\vec{\rho}_{H}(0,A) - \vec{\rho}_{H}(0,B) \leq \rho_H(A,B)$. Reversing the inequality, for all $q \in A$ we similarly obtain $\vec{\rho}_{H}(0,B) \leq \vec{\rho}_{H}(0,q) + \vec{\rho}_{H}(q,B)$ so that
		\begin{align*}
			\vec{\rho}_{H}(0,B)
			&\leq \min_{q \in A}\Big( \vec{\rho}_{H}(0,q) + \vec{\rho}_{H}(q,B) \Big) \leq \min_{q \in A}\Big( \rho_{H}(0,q) + \rho_H(q,B) \Big) \\
			&\leq \vec{\rho}_{H}(0,A) + \rho_H(A,B).
		\end{align*}
		Hence $\vec{\rho}_{H}(0,B) - \vec{\rho}_{H}(0,A) \leq \rho_H(A,B)$.
		Combining both inequalities gives the desired result.
	\end{proof}
	
	Next we bound the deterministic approximation error from retaining only the first $K$ instruments $(g_1,\ldots,g_K)$ instead of the full infinite instrument class.
	
	\begin{lemma}
		\label{L:truncation_error}
		Let Assumptions~\ref{A:assumption1}--\ref{A:bounded-moments} hold. Let \(Q_{\theta,K}\subset\mathbb R^{d_mK}\) denote the truncation of
		\(Q_\theta\) to the first \(K\) instruments \(g_1,\ldots,g_K\), and let
		\(\widetilde Q_{\theta,K}\subset\ell_1\) denote its zero-padded embedding, $\widetilde Q_{\theta,K}(w^o) = \lambda_K Q_{\theta,K}(w^o) = \lambda_K\pi_K Q_\theta(w^o)$. Then
		\begin{align*}
			\sup_{\theta\in\Theta}
			\left|
			\vec{\rho}_{H}
			\bigl(0,\mathbb E[Q_\theta]\bigr)
			-
			\vec{\rho}_{H}
			\bigl(0,\mathbb E[\widetilde Q_{\theta,K}]\bigr)
			\right|
			\to 0
			\qquad\text{as }K\to\infty.
		\end{align*}
		Moreover, since $\vec{\rho}_{H} \bigl(0,\mathbb E[\widetilde Q_{\theta,K}]\bigr) = \vec{\rho}_{H} \bigl(0,\mathbb E[Q_{\theta,K}]\bigr)$, the finite-dimensional criterion satisfies
		\begin{align*}
			\sup_{\theta\in\Theta} |T(\theta)-T_K(\theta)| \le d_m C_1 \sup_{x\in\mathcal X} \sum_{k>K} |g_k(x)| \to0.
		\end{align*}
	\end{lemma}
	
	\begin{proof}
		Fix $\theta\in\Theta$. By Lemma~\ref{L:hausdorffdist}, $\left| \vec{\rho}_{H} \bigl(0,\mathbb{E}[Q_\theta]\bigr) - \vec{\rho}_{H} \bigl(0,\mathbb{E}[\widetilde Q_{\theta,K}]\bigr) \right| \le \rho_H \bigl( \mathbb{E}[Q_\theta], \mathbb{E}[\widetilde Q_{\theta,K}] \bigr)$. The right-hand side can be bounded using the support-function representation of the Hausdorff distance for closed convex bounded sets and the support-function representation of the Aumann expectation:
		\begin{align*}
			\rho_H
			\bigl(
			\mathbb E[Q_\theta],
			\mathbb E[\widetilde Q_{\theta,K}]
			\bigr)
			&=
			\sup_{u\in B^\infty}
			\left|
			h(\mathbb E[Q_\theta],u)
			-
			h(\mathbb E[\widetilde Q_{\theta,K}],u)
			\right| \\
			&=
			\sup_{u\in B^\infty}
			\left|
			\mathbb E_{P_{W^O}}[h(Q_\theta,u)]
			-
			\mathbb E_{P_{W^O}}[h(\widetilde Q_{\theta,K},u)]
			\right| \\
			&\le
			\mathbf E\left[
			\sup_{u\in B^\infty}
			\left|
			h(Q_\theta,u)
			-
			h(\widetilde Q_{\theta,K},u)
			\right|
			\right].
		\end{align*}
		The last inequality follows from $\sup_u |\mathbf E[Z_u]| \le \mathbf E\left[\sup_u |Z_u|\right]$. By Lipschitz continuity of support functions and \(\|u\|_\infty\le1\),
		\begin{align*}
			\sup_{u\in B^\infty}
			\left|
			h(Q_\theta,u)
			-
			h(\widetilde Q_{\theta,K},u)
			\right|
			\le
			\rho_H
			\bigl(
			Q_\theta,
			\widetilde Q_{\theta,K}
			\bigr).
		\end{align*}
		Since \(Q_\theta(W^O)\) and \(\widetilde Q_{\theta,K}(W^O)\) are generated from the
		same feasible set \(D(W^O)\), we have pathwise
		\begin{align*}
			\rho_H
			\bigl(
			Q_\theta(W^O),
			\widetilde Q_{\theta,K}(W^O)
			\bigr)
			&\le
			\sup_{(y,x)\in D(W^O)}
			\left\|
			M_\theta(y,x)
			-
			\lambda_K\pi_K M_\theta(y,x)
			\right\|_1 \\
			&=
			\sup_{(y,x)\in D(W^O)}
			\sum_{j=1}^{d_m}
			\sum_{k>K}
			|m_j(y,x,\theta)|
			|g_k(x)| \\
			&\le
			d_m C_1
			\sup_{x\in\mathcal X}
			\sum_{k>K} |g_k(x)|,
		\end{align*}
		where the last inequality follows from Assumption~\ref{A:bounded-moments}. \\
		Combining the preceding displays yields, uniformly in \(\theta\),
		\begin{align*}
			\left|
			\vec{\rho}_{H}
			\bigl(0,\mathbb E[Q_\theta]\bigr)
			-
			\vec{\rho}_{H}
			\bigl(0,\mathbb E[\widetilde Q_{\theta,K}]\bigr)
			\right|
			\le
			d_m C_1
			\sup_{x\in\mathcal X}
			\sum_{k>K} |g_k(x)|.
		\end{align*}
		Taking the supremum over \(\theta\in\Theta\) and using the weighted-tail condition in
		Assumption~\ref{A:assumption2} gives
		\begin{align*}
			\sup_{\theta\in\Theta}
			\left|
			\vec{\rho}_{H}
			\bigl(0,\mathbb E[Q_\theta]\bigr)
			-
			\vec{\rho}_{H}
			\bigl(0,\mathbb E[\widetilde Q_{\theta,K}]\bigr)
			\right|
			\to0.
		\end{align*}
		Finally, under the zero-padding identification, $\vec{\rho}_{H} \bigl(0,\mathbb E[\widetilde Q_{\theta,K}]\bigr)  = \vec{\rho}_{H} \bigl(0,\mathbb E[Q_{\theta,K}]\bigr)$ so the displayed bound also gives
		\begin{align*}
			\sup_{\theta\in\Theta}|T(\theta)-T_K(\theta)| \le d_m C_1 \sup_{x\in\mathcal X} \sum_{k>K} |g_k(x)| \to 0
		\end{align*}
	\end{proof}

	\subsection{Main Proof of Theorem~\ref{T:theorem2}}
	
	We want to show $\rho_H(\widehat\Theta_n,\Theta_I^*) \overset p\to0$ by applying Theorem 3.1 in \citet{chernozhukov2007}. Their result rests on condition C.1 of the same paper, which we later verify for our setting. Define
	\begin{align*}
		a_n = \sqrt{\frac{n}{d_mK(n)+d_\theta}}, \qquad \delta_{K} = \sup_{\theta\in\Theta}|T(\theta)-T_K(\theta)|,
	\end{align*}
	and, for the verification of \citet{chernozhukov2007} condition C.1(d), define
	\begin{align*}
		b_n
		=
		\left(
		\delta_{K(n)}
		+
		\sqrt{\frac{d_mK(n)+d_\theta}{n}}
		\right)^{-1}.
	\end{align*}
	By Lemma \ref{L:truncation_error} and Assumption \ref{A:assumption2},
	\(\delta_{K(n)}\to0\). Since \(d_mK(n)/n\to0\), it follows that \(b_n\to\infty\). Recall,
	\begin{align*}
		T(\theta)
		&\equiv
		\vec{\rho}_{H}\left(0,\mathbb E[Q_\theta]\right), \\
		T_{K(n)}(\theta)
		&\equiv
		\vec{\rho}_{H}\left(0,\mathbb E[Q_{\theta,K(n)}]\right), \\
		\widehat T_n(\theta)
		&\equiv
		-\inf_{u\in B_{K(n)}}
		\frac1n\sum_{i=1}^n h(Q_{\theta,K(n),i},u).
	\end{align*}
	Equivalently,
	\begin{align*}
		\widehat T_n(\theta) = \vec{\rho}_{H}\left( 0, \operatorname{co}\left[ \frac1n\bigoplus_{i=1}^n Q_{\theta,K(n),i} \right] \right).
	\end{align*}
	We restate the estimator defined in the main text:
	\begin{align*}
		\widehat\Theta_n
		&=
		\left\{
		\theta\in\Theta:
		-\inf_{u\in B_{K(n)}}
		\frac1n\sum_{i=1}^n h(Q_{\theta,K(n),i},u)
		\le
		\frac{\tau_n}{a_n}
		\right\} \\
		&=
		\{\theta\in\Theta:a_n\widehat T_n(\theta)\le\tau_n\}.
	\end{align*}
	with $K(n)\to\infty$, $d_mK(n)/n\to0$, $\tau_n\to\infty$, and $\tau_n/a_n\to0$. \\
	With these we are equipped to check conditions (a)--(e) in condition C.1 of \citet{chernozhukov2007}:
	\begin{enumerate}[label=\alph*)]
		\item $\Theta$ is a compact subset of $\mathbb{R}^{d_\theta}$. \\
		Holds by Assumption \ref{A:compact-theta}.
		\item $T(\theta)$ is continuous and $\inf\limits_{\theta \in \Theta} T(\theta) = 0$ \\
		By Lemma~\ref{L:hausdorffdist}, $|T(\theta)-T(\theta')| \le  \rho_H(\mathbb E[Q_\theta],\mathbb E[Q_{\theta'}])$. For any feasible \((y,x)\), let \(q_\theta=M_\theta(y,x)\) and
		\(q_{\theta'}=M_{\theta'}(y,x)\). Then
		\begin{align*}
			\|q_\theta-q_{\theta'}\|_1
			&\le
			\sum_{j=1}^{d_m}\sum_{k\ge1}
			|m_j(y,x,\theta)-m_j(y,x,\theta')|
			|g_k(x)| \\
			&\le
			\left[
			\sup_{x\in\mathcal X}
			\sum_{k\ge1}|g_k(x)|
			\right]
			\sup_{(y,x)\in\mathcal W}
			\sum_{j=1}^{d_m}
			|m_j(y,x,\theta)-m_j(y,x,\theta')|.
		\end{align*}
		By Assumption~\ref{A:assumption2}, the bracketed term is finite, and by Assumption \ref{A:thetalipschitz} the second
		factor converges to zero as \(\theta'\to\theta\), uniformly on the compact domain.
		Thus
		\begin{align*}
			\rho_H\bigl(\mathbb E[Q_\theta],\mathbb E[Q_{\theta'}]\bigr)\to0
			\qquad\text{as }\theta'\to\theta,
		\end{align*}
		and hence \(T\) is continuous.
		
		\item $\hat{T}_{n}(\theta)$ is jointly measurable. \\
		Not strictly necessary according to \citet{chernozhukov2007}. Nevertheless, it follows from the measurability of $Q_{\theta,K(n)}$, using the same arguments as in the proof of Theorem \ref{T:theorem1}.
		
		\item The sample criterion function is uniformly no smaller than
		the population function in large samples:
		\begin{align*}
			\sup_{\theta\in\Theta} \bigl(T(\theta)-\widehat T_n(\theta)\bigr)_+
			&\leq \sup_{\theta\in\Theta} \left| T(\theta)-\widehat T_n(\theta) \right| \\
			&=\sup_{\theta\in\Theta} \left| T(\theta)-T_{K(n)}(\theta) + T_{K(n)}(\theta)-\widehat T_n(\theta) \right| \\
			&\leq \sup_{\theta\in\Theta} \left| T(\theta)-T_{K(n)}(\theta) \right| + \sup_{\theta\in\Theta} \left| T_{K(n)}(\theta)-\widehat T_n(\theta) \right|.
		\end{align*}
		The first term is the deterministic truncation error. By Lemma
		\ref{L:truncation_error},
		\begin{align*}
			\sup_{\theta\in\Theta} \left| T(\theta)-T_{K(n)}(\theta) \right|
			&\leq d_m C_1 \sup_{x\in\mathcal X} \sum_{k>K(n)} |g_k(x)| \equiv \delta_{K(n)} \to 0.
		\end{align*}
		For the second term, recall that
		\begin{align*}
			T_{K(n)}(\theta) =
			\vec{\rho}_{H}
			\bigl(0,\mathbb E[Q_{\theta,K(n)}]\bigr),
			\qquad
			\widehat T_n(\theta) =
			\vec{\rho}_{H}\left(0, \operatorname{co}\left[ \frac1n \mathop{\oplus}\limits_{i=1}^n Q_{\theta,K(n),i} \right] \right).
		\end{align*}
		Hence, by Lemma \ref{L:uniform_rs_lln} and Lemma \ref{L:hausdorffdist},
		\begin{align*}
			\sup_{\theta\in\Theta}
			\left|
			T_{K(n)}(\theta)-\widehat T_n(\theta) \right|
			&\leq \sup_{\theta\in\Theta} \rho_H\left( \mathbb E[Q_{\theta,K(n)}], \operatorname{co}\left[ \frac1n \mathop{\oplus}\limits_{i=1}^n Q_{\theta,K(n),i} \right] \right) \\
			&= O_p\left( \sqrt{ \frac{d_mK(n)+d_\theta}{n} } \right).
		\end{align*}
		Therefore,
		\begin{align*}
			\sup_{\theta\in\Theta}
			\bigl(T(\theta)-\widehat T_n(\theta)\bigr)_+
			&=
			O_p\left(
			\delta_{K(n)}
			+
			\sqrt{
				\frac{d_mK(n)+d_\theta}{n}
			}
			\right) \\
			&=
			O_p(1/b_n).
		\end{align*}
		By Assumption \ref{A:assumption2}, Lemma \ref{L:truncation_error}, and
		\(d_mK(n)/n\to0\), we have \(b_n\to\infty\). The term
		\(\delta_{K(n)}\) captures the deterministic approximation error from using
		finitely many instruments instead of the full infinite instrument class, while
		the second term captures the sampling error of the growing-dimensional random
		set estimator.
		
		\item The sample criterion converges to the limit criterion function over the
		identified set \(\Theta_I^*\) at the \(1/a_n\) rate.
		\begin{align*}
			\sup_{\theta\in\Theta_I^*} \widehat T_n(\theta)
			&=
			\sup_{\theta\in\Theta_I^*}
			\vec{\rho}_{H}\left(
			0,
			\operatorname{co}\left[
			\frac{1}{n}
			\mathop{\oplus}_{i=1}^n
			Q_{\theta,K(n),i}
			\right]
			\right) \\
			&\leq
			\sup_{\theta\in\Theta_I^*}
			\vec{\rho}_{H}
			\bigl(
			0,
			\mathbb E[Q_{\theta,K(n)}]
			\bigr) \\
			&\quad+
			\sup_{\theta\in\Theta_I^*}
			\rho_H\left(
			\mathbb E[Q_{\theta,K(n)}],
			\operatorname{co}\left[
			\frac{1}{n}
			\mathop{\oplus}_{i=1}^n
			Q_{\theta,K(n),i}
			\right]
			\right) \\
			&=
			\sup_{\theta\in\Theta_I^*}
			\rho_H\left(
			\mathbb E[Q_{\theta,K(n)}],
			\operatorname{co}\left[
			\frac{1}{n}
			\mathop{\oplus}_{i=1}^n
			Q_{\theta,K(n),i}
			\right]
			\right) \\
			&=
			O_p(1/a_n).
		\end{align*}
		The first line is definitional. The second line follows from the inequality $\vec\rho_H(0,B) \le \vec\rho_H(0,A)+\rho_H(A,B)$, applied with $A=\mathbb E[Q_{\theta,K(n)}]$ and $B= \operatorname{co}\left[ \frac{1}{n} \mathop{\oplus}_{i=1}^n Q_{\theta,K(n),i} \right]$. The third line uses the fact that if $\theta\in\Theta_I^*$, then
		$0\in\mathbb E[Q_\theta]$. Lemmas~\ref{L:aumann-closure}
		and~\ref{L:origin-attainment} yield an $\eta\in\Gamma$ with
		$\int M_\theta\,d\eta=0$, and Lemma~\ref{L:lawselection} then produces an
		integrable
		selection $q_\theta(W^O)\in Q_\theta(W^O)$ satisfying
		$\mathbb E[q_\theta(W^O)]=0$.
		Applying the coordinate projection \(\pi_{K(n)}\) gives an integrable selection $\pi_{K(n)}q_\theta(W^O) \in Q_{\theta,K(n)}(W^O)$ satisfying $\mathbb E[\pi_{K(n)}q_\theta(W^O)] = \pi_{K(n)}\mathbb E[q_\theta(W^O)] = 0$.
		Therefore $0\in\mathbb E[Q_{\theta,K(n)}]$ and $\vec\rho_H \bigl(0, \mathbb E[Q_{\theta,K(n)}]\bigr) = 0$. \\
		Finally, the last rate follows from the convexified part of the uniform
		triangular-array law of large numbers and the definition
		\[
		a_n
		=
		\sqrt{
			\frac{n}{d_mK(n)+d_\theta}
		}.
		\]
	\end{enumerate}
	
	To complete the proof, after verifying condition C.1 in
	\citet{chernozhukov2007}, the result follows from Theorem 3.1(i) in
	\citet{chernozhukov2007}. Indeed, by part (e), $\mathcal C_n \equiv \sup_{\theta\in\Theta_I^*} a_n\widehat T_n(\theta) = O_p(1)$. Since $\tau_n\to\infty$, we have $\tau_n\geq \mathcal C_n$ with probability approaching one. Moreover, since $\tau_n/a_n\to0$, the
	unscaled contour level vanishes. Therefore Theorem 3.1(i) of
	\citet{chernozhukov2007} implies $\rho_H(\widehat\Theta_n,\Theta_I^*) \overset p\to0$.
	\qed
	
	\newcommand{\Binf}{B^\infty}

	\section{Proof of Theorem~\ref{T:theorem3}}
	\label{sec:appendixD}
	
	This appendix verifies the high-level conditions of \citet{andrews2017}, applies their Theorem 6.1, and adapts the proof of their Theorem 5.1 to the full class of null pairs. The principal technical step is to verify manageability of the process indexed by $\mathcal{U}=B^\infty$.

	\subsection{Manageability of the support-function process}
	
	Recall the uniform coordinate envelopes
	\begin{align*}
		a_k
		\equiv \sup_{x\in\mathcal X}|g_k(x)|,
		\qquad
		A\equiv\sum_{k=1}^{\infty}a_k.
	\end{align*}
	The polynomial decay condition, Assumption \ref{A:polynomial-decay}, used in Theorem~\ref{T:theorem3} gives $A<\infty$ and an
	explicit rate for its tail. Equip $B^\infty$ with
	the normalized weighted semimetric
	\begin{align*}
		\rho_a(v,v')
		\equiv
		\frac{1}{d_mA}\sum_{\ell=1}^{d_m}\sum_{k=1}^{\infty}
		a_k|v_{\ell k}-v'_{\ell k}|.
	\end{align*}
	Assumption \ref{A:bounded-moments} and the elementary inequality
	$|\sup_s f(s)-\sup_s g(s)|\leq\sup_s|f(s)-g(s)|$ imply, uniformly over
	$w^o\in\mathcal W^O$, $\theta\in\Theta$, and $u,u'\in\mathcal U$,
	\begin{align*}
		\left| \widetilde m(w^o,\theta,u) -\widetilde m(w^o,\theta,u') \right| &\leq C_1 \sum_{\ell=1}^{d_m}\sum_{k=1}^{\infty} a_k|u_{\ell k}-u'_{\ell k}| \\
		&= M_0\rho_a(u,u')
	\end{align*}
	with $M_0 \equiv d_mC_1A$. Moreover, for every $u\in\mathcal U$,
	\begin{align*}
		\left| \widetilde m(w^o,\theta,u) \right| &\leq C_1 \sum_{\ell=1}^{d_m}\sum_{k=1}^{\infty} a_k|u_{\ell k}| \\
		&\leq d_mC_1A = M_0.
	\end{align*}
	Thus $M_0$ is a uniform scalar envelope for the support-function moment class. \\
	Fix an arbitrary sequence
	$\{(\theta_n,P_n)\in\mathcal F^+:n\geq1\}$, and let $\{W_{n,i}^O:i\leq n\}$ be a row-wise i.i.d. triangular array with $W_{n,i}^O\sim P_n$.  For a realization $\omega$, define
	\begin{align*}
		\mathcal M_{n,\omega}(\theta_n)
		\equiv
		\Bigl\{
		\bigl(
		\widetilde m(W_{n,1}^O(\omega),\theta_n,u),
		\ldots,
		\widetilde m(W_{n,n}^O(\omega),\theta_n,u)
		\bigr)': \ u\in\mathcal{U}
		\Bigr\}
		\subset\mathbb R^n.
	\end{align*}
	For a subset $V$ of a metric space, let $N(\xi,V)$ and $D(\xi,V)$ denote its
	covering and packing numbers.  Appendix B of \citet{andrews2017} records
	the standard inequalities
	\begin{align*}
		N(\xi,V)
		\leq D(\xi,V)
		\leq N(\xi/2,V).
	\end{align*}
	The same reference restates Definition 7.9 of \citet{pollard1990}: the
	triangular array is manageable with respect to an envelope array if there is a
	deterministic $\lambda:(0,1]\to\mathbb R$ such that
	\begin{align*}
		\int_0^1\sqrt{\log\lambda(\xi)}\,d\xi<\infty
	\end{align*}
	and, for every $\xi\in(0,1]$, every realization $\omega$, every $n$, and every
	nonnegative $n$-vector $\alpha$,
	\begin{align*}
		D\!\left(
		\xi\|\alpha\odot M_n\|_2,
		\alpha\odot\mathcal M_{n,\omega}(\theta_n)
		\right)
		\leq\lambda(\xi),
		\qquad
		M_n\equiv(M_0,\ldots,M_0)'.
	\end{align*}
	
	The first lemma determines the number of instrument coordinates needed to control the coordinate-envelope tail.
	
	\begin{lemma}
		\label{L:D-tail-revised}
		Suppose $C_a\equiv\sup_{k\geq1}k^\beta a_k<\infty$ for some $\beta>1$.
		Equivalently, $a_k\leq C_a k^{-\beta}$ for every $k$.  For
		$\xi\in(0,1]$, let
		\begin{align*}
			K_\xi
			\equiv
			\left\lceil
			\left(
			\frac{4C_a}{(\beta-1)A\xi}
			\right)^{1/(\beta-1)}
			\right\rceil.
		\end{align*}
		Then
		\begin{align*}
			\sum_{k>K_\xi}a_k
			\leq\frac{A\xi}{4}
		\end{align*}
		and $K_\xi=O\bigl(\xi^{-1/(\beta-1)}\bigr)$ as $\xi\downarrow0$.
	\end{lemma}

	\begin{proof}
		By the definition of $C_a$,
		\begin{align*}
			a_k\leq C_a k^{-\beta}
			\qquad\text{for every }k\geq1.
		\end{align*}
		Hence, for every integer $K\geq1$, the integral comparison for the
		decreasing function $x\mapsto x^{-\beta}$ gives
		\begin{align*}
			\sum_{k>K}a_k
			&\leq
			C_a\sum_{k>K}k^{-\beta} \\
			&\leq
			C_a\int_K^\infty x^{-\beta}\,dx \\
			&=
			\frac{C_a}{\beta-1}K^{1-\beta},
		\end{align*}
		where the last expression is finite because $\beta>1$. \\
		By definition,
		\begin{align*}
			K_\xi
			\geq
			\left(
			\frac{4C_a}{(\beta-1)A\xi}
			\right)^{1/(\beta-1)}.
		\end{align*}
		Since $1-\beta<0$, it follows that
		\begin{align*}
			K_\xi^{1-\beta}
			&\leq
			\left(
			\frac{4C_a}{(\beta-1)A\xi}
			\right)^{-1} \\
			&=
			\frac{(\beta-1)A\xi}{4C_a}.
		\end{align*}
		Therefore,
		\begin{align*}
			\sum_{k>K_\xi}a_k
			&\leq
			\frac{C_a}{\beta-1}K_\xi^{1-\beta} \\
			&\leq
			\frac{A\xi}{4}.
		\end{align*}
		
		Finally, using $\lceil x\rceil\leq x+1$,
		\begin{align*}
			K_\xi
			&\leq
			1+
			\left(
			\frac{4C_a}{(\beta-1)A}
			\right)^{1/(\beta-1)}
			\xi^{-1/(\beta-1)}.
		\end{align*}
		Hence
		\begin{align*}
			K_\xi
			=
			O\left(\xi^{-1/(\beta-1)}\right)
			\qquad\text{as }\xi\downarrow0.
		\end{align*}
	\end{proof}
	
	The next lemma verifies the entropy-integrability requirement.

	\begin{lemma}
		\label{L:D-entropy-revised}
		Let $K_\xi$ be defined as in Lemma~\ref{L:D-tail-revised}.  For a sufficiently
		large constant $C_0>1$, define
		\begin{align*}
			\lambda(\xi)
			\equiv
			\left(\frac{C_0}{\xi}\right)^{d_mK_\xi},
			\qquad \xi\in(0,1].
		\end{align*}
		If $\beta>3/2$, then
		\begin{align*}
			\int_0^1\sqrt{\log\lambda(\xi)}\,d\xi<\infty.
		\end{align*}
	\end{lemma}
	
	\begin{proof}
		Let $q\equiv1/(\beta-1)$. Lemma~\ref{L:D-tail-revised} implies that, for some finite constant $C$,
		\begin{align*}
			\log\lambda(\xi)
			&=d_mK_\xi\log(C_0/\xi) \\
			&\leq C\xi^{-q}\log(C_0/\xi)
		\end{align*}
		for all $\xi\in(0,1]$.  Consequently,
		\begin{align*}
			\int_0^1\sqrt{\log\lambda(\xi)}\,d\xi
			\leq
			\sqrt C\int_0^1
			\xi^{-q/2}\sqrt{\log(C_0/\xi)}\,d\xi.
		\end{align*}
		Set $t=\log(1/\xi)$, so that $\xi=e^{-t}$ and $d\xi=-e^{-t}dt$.  The right-hand side becomes
		\begin{align*}
			\sqrt C\int_0^{\infty}
			e^{-ct}\sqrt{t+\log C_0}\,dt,
			\qquad
			c\equiv1-\frac{q}{2}
			=\frac{2\beta-3}{2(\beta-1)}.
		\end{align*}
		Because $C_0>1$,
		\begin{align*}
			\sqrt{t+\log C_0}
			&\leq
			\sqrt t+\sqrt{\log C_0} \\
			&\leq
			1+t+\sqrt{\log C_0}.
		\end{align*}
		Hence, if $c>0$,
		\begin{align*}
			\int_0^{\infty}
			e^{-ct}\sqrt{t+\log C_0}\,dt
			&\leq
			\int_0^{\infty}
			e^{-ct}
			\left(1+t+\sqrt{\log C_0}\right)\,dt \\
			&=
			\frac{1+\sqrt{\log C_0}}{c}
			+\frac{1}{c^2}
			<\infty.
		\end{align*}
		It follows that
		\begin{align*}
			\int_0^1\sqrt{\log\lambda(\xi)}\,d\xi<\infty.
		\end{align*}
		Finally, $c>0$ is equivalent to $\beta>3/2$.
	\end{proof}

	\begin{lemma}
		\label{L:D-manageability-revised}
		Assumption PS2 of \citet{andrews2017} holds in the present setting with
		\begin{align*}
			\lambda(\xi)
			=\left(\frac{C_0}{\xi}\right)^{d_mK_\xi},
		\end{align*}
		where $K_\xi$ is the integer chosen in
		Lemma~\ref{L:D-tail-revised} so that the coordinate-envelope tail is at most
		$A\xi/4$.
	\end{lemma}
	
	\begin{proof}
		Fix $\xi\in(0,1]$.  Let $\Pi_{K_\xi}u$ retain the first $K_\xi$ instrument
		coordinates in each primitive-moment block of $u$ and set the remaining
		coordinates equal to zero.  Cover $[-1,1]^{d_mK_\xi}$ in the sup norm by a grid $\mathcal U_\xi$ of radius
		$\xi/4$.  For a constant $C_0>1$ independent of $\xi$, the grid may be chosen
		so that
		\begin{align*}
			|\mathcal U_\xi|
			\leq
			\left(\frac{C_0}{\xi}\right)^{d_mK_\xi}.
		\end{align*}
		For every $u\in\mathcal U$, choose
		$v\in\mathcal U_\xi$ within $\xi/4$ of $\Pi_{K_\xi}u$ in the sup norm
		and extend $v$ by zeros.  Then
		\begin{align*}
			\rho_a(u,v)
			&\leq
			\frac{1}{d_mA}\sum_{\ell=1}^{d_m}
			\left\{\sum_{k\leq K_\xi}a_k|u_{\ell k}-v_{\ell k}|
			+\sum_{k>K_\xi}a_k|u_{\ell k}|\right\} \\
			&\leq\frac{\xi}{4}+\frac{\xi}{4}
			=\frac{\xi}{2}.
		\end{align*}
		Thus $\mathcal U_\xi$ is a $\xi/2$ cover of $B^\infty$ under $\rho_a$. \\
		Let $\alpha=(\alpha_1,\ldots,\alpha_n)'$ be nonnegative.  The Lipschitz bound above implies
		\begin{align*}
			&\left\|
			\alpha\odot
			\left(
			\bigl(\widetilde m(W_{n,i}^O,\theta_n,u)\bigr)_{i=1}^n
			-
			\bigl(\widetilde m(W_{n,i}^O,\theta_n,v)\bigr)_{i=1}^n
			\right)
			\right\|_2 \\
			&\hspace{4cm}
			\leq
			\rho_a(u,v)\|\alpha\odot M_n\|_2.
		\end{align*}
		The same direction grid therefore covers the sample paths.  The packing--covering inequality from \citet{andrews2017} then gives
		\begin{align*}
			D\!\left(
			\xi\|\alpha\odot M_n\|_2,
			\alpha\odot\mathcal M_{n,\omega}(\theta_n)
			\right)
			\leq
			\left(\frac{C_0}{\xi}\right)^{d_mK_\xi}
			=\lambda(\xi).
		\end{align*}
		Lemma~\ref{L:D-entropy-revised} verifies the required entropy integral.
		Hence the support-function process is manageable in the sense of Definition 7.9 of
		\citet{pollard1990}.
	\end{proof}
	
	Since the preceding bounds are deterministic and do not depend on the sequence $\{P_n\}$, the same manageability function $\lambda$ applies to every sequence $\{(\theta_n,P_n)\in\mathcal F^+\}$.
	
	\subsection{Verification of the Andrews--Shi assumptions}
	
	We apply \citet{andrews2017} to the scalar process
	$\{\widetilde m(\cdot,\theta,u):u\in\mathcal U\}$. In their notation there is
	one moment inequality, their index $\tau$ is $u\in\mathcal U$, the
	conditioning variable is degenerate, the instrument class is the singleton
	containing the constant function, and the weight function is a point mass on
	it. Their test statistic, GMS function and bootstrap critical value then
	reduce to $T_n^{AS}(\theta)$, $\overline\varphi_n(\theta,u)$ and
	$c^{AS}_{n,1-\alpha}(\theta)$. Recall the envelope and Lipschitz bounds
	established above, which hold uniformly in $(w^o,\theta)$:
	\begin{align}
		\label{eq:env-lip}
		|\widetilde m(w^o,\theta,u)|\le M_0,
		\qquad
		|\widetilde m(w^o,\theta,u)-\widetilde m(w^o,\theta,u')|
		\le M_0\rho_a(u,u').
	\end{align}
	For $(\theta,P)\in\mathcal F^+$ and $u,u'\in\mathcal U$ define
	\begin{align*}
		\nu_{n,P}(\theta,u)
		&\equiv
		\frac{1}{\sqrt n}\sum_{i=1}^n
		\bigl\{\widetilde m(W_i^O,\theta,u)
		-\mathbf E_P[\widetilde m(W_i^O,\theta,u)]\bigr\},
		\\
		h_{1,n,P}(\theta,u)
		&\equiv
		\sqrt n\,\mathbf E_P[\widetilde m(W_i^O,\theta,u)],
		\\
		h_{2,P}(\theta;u,u')
		&\equiv
		\operatorname{Cov}_P\bigl(\widetilde m(W_i^O,\theta,u),
		\widetilde m(W_i^O,\theta,u')\bigr).
	\end{align*}
	Let $\widehat h_{2,n}(\theta;u,u')$ denote the sample analogue of
	$h_{2,P}(\theta;u,u')$. Then
	$\sqrt n\,\overline m_n(\theta,u)=\nu_{n,P}(\theta,u)+h_{1,n,P}(\theta,u)$
	and $\overline\Sigma_n(\theta,u)=\widehat h_{2,n}(\theta;u,u)+\epsilon$.
	Moreover, $h_{1,n,P}(\theta,\cdot)\ge0$ whenever $(\theta,P)\in\mathcal F$.
	Kernels on $\mathcal U\times\mathcal U$ are compared in the uniform metric
	$d(h,h')\equiv\sup_{u,u'\in\mathcal U}|h(u,u')-h'(u,u')|$.
	
	\begin{enumerate}[label=(\roman*),leftmargin=*]
		\item \textbf{PS1.} Under every $(\theta,P)\in\mathcal F^+$ the
		observations are i.i.d. and $\sigma_P(\theta)=1$. By
		\eqref{eq:env-lip}, $M_0$ is a deterministic envelope, so
		$\sup_{P\in\mathcal P}\mathbf E_P[M_0^{2+\delta}]<\infty$ for every
		$\delta>0$.
		
		\item \textbf{PS2.} This is Lemma~\ref{L:D-manageability-revised}, whose
		bound holds for every sequence in $\mathcal F^+$ with row-wise i.i.d.
		arrays. The map $w^o\mapsto\widetilde m(w^o,\theta,u)=h(Q_\theta(w^o),u)$
		is measurable, being the support function of the random compact set of
		Lemma~\ref{L:random-compact}. By \eqref{eq:env-lip}, suprema over
		$\mathcal U$ coincide with suprema over a countable $\rho_a$-dense
		subset.
		
		\item \textbf{PS3.} With a degenerate conditioning variable, the moment
		inequality reads $\mathbf E_P[\widetilde m(W_i^O,\theta,u)]\ge0$ for all
		$u\in\mathcal U$. By Theorem~\ref{T:theorem1}, this holds if and only if
		$\theta\in\Theta_I^*(P)$.
		
		\item \textbf{SIG1 and SIG2.} Both hold identically, because
		$\widehat\sigma_n(\theta)=\sigma_P(\theta)=1$.
		
		\item \textbf{MFA.} If $\theta\notin\Theta_I^*(P)$,
		Theorem~\ref{T:theorem1} gives some $u_{\theta,P}\in\mathcal U$ with
		$\mathbf E_P[\widetilde m(W_i^O,\theta,u_{\theta,P})]<0$. With a
		degenerate conditioning variable, this is a violation on a set of
		probability one.
		
		\item \textbf{PS4 and PS5.} The constant instrument satisfies Assumption
		M of \citet{andrews2017}, and $S_1$ satisfies their Assumptions S1
		and S2 by Lemma 1 of \citet{andrews2013}. Their Lemma D.2 derives PS4
		and PS5 from these assumptions together with PS1 and PS2. As formulated
		there, PS4 and PS5 apply to every sequence $(n_r,\theta_r,P_r)$ with
		$(n_r)$ strictly increasing, $(\theta_r,P_r)\in\mathcal F^+$ and
		$d(h_{2,P_r}(\theta_r;\cdot,\cdot),h_{2,0})\to0$, for \emph{some}
		covariance kernel $h_{2,0}$, whether or not it is attained.
		\begin{itemize}
			\item PS4 gives $\nu_{n_r,P_r}(\theta_r,\cdot)\Rightarrow\nu_{h_{2,0}}$
			and $d(\widehat h_{2,n_r}(\theta_r),h_{2,0})\overset{p}{\to}0$.
			\item PS5 gives the same two conclusions for the bootstrap process
			$\sqrt{n_r}(\overline m^*_{n_r}-\overline m_{n_r})(\theta_r,\cdot)$ and
			the bootstrap kernel $\widehat h^*_{2,n_r}(\theta_r)$. These hold
			conditionally on almost every sample path along which
			$d(\widehat h_{2,n_r}(\theta_r),h_{2,0})\to0$.
		\end{itemize}
	\end{enumerate}
	
	Next let us show relative compactness of the covariance kernels. Let $\mathcal H_{2,\mathcal P}\equiv\{h_{2,P}(\theta;\cdot,\cdot):
	(\theta,P)\in\mathcal F\}$. Identify directions at zero $\rho_a$-distance. By
	\eqref{eq:env-lip}, such directions generate identical moment functions, so
	all kernels are defined on the quotient. The quotient is compact. Given
	$\{u^{(r)}\}\subseteq\mathcal U$, a diagonal argument yields a subsequence
	converging coordinatewise to some $u\in\mathcal U$. For every $K$, this
	subsequence satisfies
	$\limsup_r\rho_a(u^{(r)},u)\le(2/A)\sum_{k>K}a_k$, which vanishes as
	$K\to\infty$ because $A<\infty$. Write $f_u\equiv\widetilde m(W_i^O,\theta,u)$.
	Expanding
	$h_{2,P}(\theta;u,u')-h_{2,P}(\theta;v,v')
	=\operatorname{Cov}_P(f_u-f_v,f_{u'})+\operatorname{Cov}_P(f_v,f_{u'}-f_{v'})$
	and applying Cauchy--Schwarz together with \eqref{eq:env-lip} gives,
	uniformly over $(\theta,P)\in\mathcal F$,
	\begin{align*}
		|h_{2,P}(\theta;u,u')|\le M_0^2,
		\qquad
		|h_{2,P}(\theta;u,u')-h_{2,P}(\theta;v,v')|
		\le M_0^2\{\rho_a(u,v)+\rho_a(u',v')\}.
	\end{align*}
	Hence $\mathcal H_{2,\mathcal P}$ is uniformly bounded and uniformly
	equicontinuous on the compact quotient of $\mathcal U\times\mathcal U$. By
	the Arzel\`a--Ascoli theorem, it is relatively compact under $d$, and its
	closure $\overline{\mathcal H}_{2,\mathcal P}$ is compact. Uniform limits
	preserve symmetry and positive semidefiniteness, so every element of
	$\overline{\mathcal H}_{2,\mathcal P}$ is a covariance kernel. Such a limit
	need not be attained by any $(\theta,P)\in\mathcal F$. \\
	For a kernel $h$ on $\mathcal U\times\mathcal U$, define the semimetric
	$\rho_h(u,v)\equiv\{h(u,u)-2h(u,v)+h(v,v)\}^{1/2}$.
	\begin{itemize}
		\item \emph{Domination.} For $h\in\mathcal H_{2,\mathcal P}$,
		$\rho_h(u,v)$ is the $L_2(P)$ distance between $f_u$ and $f_v$, so
		$\rho_h\le M_0\rho_a$ by \eqref{eq:env-lip}. Since $\rho_h$ is
		$d$-continuous in $h$, the bound extends to
		$\overline{\mathcal H}_{2,\mathcal P}$.
		\item \emph{Entropy.} The proof of Lemma~\ref{L:D-manageability-revised}
		shows $N(\varepsilon,\mathcal U,\rho_a)\le\lambda(2\varepsilon)$ for
		$\varepsilon\in(0,1/2]$, and
		$\int_0^1\sqrt{\log\lambda(\xi)}\,d\xi<\infty$ by
		Lemma~\ref{L:D-entropy-revised}.
	\end{itemize}
	Dudley's entropy bound \citep{dudley1967} therefore yields:
	\begin{enumerate}[label=(G\arabic*),leftmargin=*]
		\item for every $h\in\overline{\mathcal H}_{2,\mathcal P}$ there is a
		mean-zero Gaussian process $\nu_h$ on $\mathcal U$ with covariance $h$
		and bounded, uniformly $\rho_a$-continuous sample paths, and
		$\lim_{\delta\downarrow0}\sup_{h\in\overline{\mathcal H}_{2,\mathcal P}}
		\mathbf E\sup_{\rho_a(u,v)<\delta}|\nu_h(u)-\nu_h(v)|=0$;
		\item if $h_r,h\in\overline{\mathcal H}_{2,\mathcal P}$ and
		$d(h_r,h)\to0$, then $\nu_{h_r}\Rightarrow\nu_h$ in
		$\ell^\infty(\mathcal U)$, because the finite-dimensional laws converge
		and (G1) gives asymptotic uniform equicontinuity.
	\end{enumerate}
	For $h\in\overline{\mathcal H}_{2,\mathcal P}$ and
	$h_1:\mathcal U\to[0,\infty)$, define
	\begin{align*}
		T(h_1,h)\equiv\sup_{u\in\mathcal U}
		\left[\frac{\nu_h(u)+h_1(u)}{\sqrt{h(u,u)+\epsilon}}\right]_-^2.
	\end{align*}
	This is a lower semicontinuous functional of the continuous path $\nu_h$,
	hence a random variable. Let $c_0(h_1,h,1-\alpha)$ denote its
	$(1-\alpha)$-quantile. When $h_1$ is data dependent, such as
	$\overline\varphi_n(\theta,\cdot)$, $c_0$ is evaluated with $h_1$ held fixed
	and $\nu_h$ independent of the data. \\
	Finally, let $\nu,\nu'$ be functions on $\mathcal U$ with sup norms at most
	$B$, let $s,s'\ge\sqrt\epsilon$, and let $\mu:\mathcal U\to[0,\infty)$ be
	arbitrary. Because $[x+\mu]_-\le[x]_-$ and $|[a]_--[b]_-|\le|a-b|$,
	\begin{align}
		\label{eq:stat-lip}
		\left|
		\sup_u\Bigl[\tfrac{\nu(u)+\mu(u)}{s(u)}\Bigr]_-^2
		-\sup_u\Bigl[\tfrac{\nu'(u)+\mu(u)}{s'(u)}\Bigr]_-^2
		\right|
		\le\frac{2B}{\epsilon}
		\Bigl(\|\nu-\nu'\|_\infty+\frac{B}{\sqrt\epsilon}\|s-s'\|_\infty\Bigr).
	\end{align}
	The bound does not depend on $\mu$. This is what permits nonnegative
	slackness functions that need not converge. \\
	Theorem 5.1 of \citet{andrews2017} rests on their Lemma D.1, Theorem
	D.3(a) and Lemma D.4. All of these are \emph{stated} for compact subsets of
	the attainable kernel class, and $\overline{\mathcal H}_{2,\mathcal P}$ need
	not be such a set. In their proofs, and in the proofs of Theorem 1 and
	Lemmas A2--A4 of \citet{andrews2013} to which they refer, compactness is
	used once: to pass from a near-extremal sequence of null pairs to a
	subsequence whose kernels converge uniformly to some covariance kernel. The
	subsequent steps use only PS4, PS5, monotonicity and continuity of $S$,
	GMS1, and $\nu_{h_{2,r}}\Rightarrow\nu_{h_{2,0}}$. The following lemma
	repeats those steps along a sequence that is kernel-convergent by
	hypothesis. The only properties of the limit it uses are (G1)--(G2), which
	hold on all of $\overline{\mathcal H}_{2,\mathcal P}$.
	
	\begin{lemma}
		\label{L:sequential-coverage}
		Let the conditions of Theorem~\ref{T:theorem3} hold, let $(n_r)$ be strictly
		increasing, and let $(\theta_r,P_r)\in\mathcal F$ satisfy
		$d\bigl(h_{2,P_r}(\theta_r;\cdot,\cdot),h_{2,0}\bigr)\to0$ for some
		covariance kernel $h_{2,0}$. Then
		\begin{align*}
			\liminf_{r\to\infty}
			P_r^{\,n_r}\bigl\{\theta_r\in CS_{n_r}^{AS}(1-\alpha)\bigr\}\ge1-\alpha.
		\end{align*}
	\end{lemma}
	
	\begin{proof}
		We use the following abbreviations:
		$h_{1,r}\equiv h_{1,n_r,P_r}(\theta_r,\cdot)\ge0$,
		$h_{2,r}\equiv h_{2,P_r}(\theta_r;\cdot,\cdot)$,
		$\nu_r\equiv\nu_{n_r,P_r}(\theta_r,\cdot)$,
		$\overline\varphi_r\equiv\overline\varphi_{n_r}(\theta_r,\cdot)$,
		$s_r(u)\equiv\{\overline\Sigma_{n_r}(\theta_r,u)\}^{1/2}\ge\sqrt\epsilon$,
		$s_{2,r}(u)\equiv\{h_{2,r}(u,u)+\epsilon\}^{1/2}$,
		$s_0(u)\equiv\{h_{2,0}(u,u)+\epsilon\}^{1/2}$, and
		$c_r\equiv c_0(h_{1,r},h_{2,r},1-\alpha)$. Since
		$h_{2,r}\in\mathcal H_{2,\mathcal P}$, the limit satisfies
		$h_{2,0}\in\overline{\mathcal H}_{2,\mathcal P}$. Hence (G1), (G2) and item
		(vi) apply, while $h_{1,r}$ need not converge. Fix $\delta\in(0,\eta)$.
		
		\emph{(a) Oracle upper-tail bound:
			$\limsup_r P_r^{n_r}\{T^{AS}_{n_r}(\theta_r)>c_r+\delta\}\le\alpha$.}
		By PS4, $(\nu_r,s_r)\Rightarrow(\nu_{h_{2,0}},s_0)$. By (G2),
		$\nu_{h_{2,r}}\Rightarrow\nu_{h_{2,0}}$, and $s_{2,r}\to s_0$ uniformly. By
		the almost sure representation theorem \citep[Theorem
		9.4]{pollard1990}, there exist two pairs of random elements:
		\begin{itemize}
			\item $(\tilde\nu_r,\tilde s_r)$ with the law of $(\nu_r,s_r)$, and
			$\tilde\nu_0$ with the law of $\nu_{h_{2,0}}$, such that
			$\|\tilde\nu_r-\tilde\nu_0\|_\infty+\|\tilde s_r-s_0\|_\infty\to0$
			almost surely;
			\item $\tilde\nu^G_r$ with the law of $\nu_{h_{2,r}}$, and $\tilde\nu'_0$
			with the law of $\nu_{h_{2,0}}$, such that
			$\|\tilde\nu^G_r-\tilde\nu'_0\|_\infty\to0$ almost surely.
		\end{itemize}
		Let $T_{r,0}\equiv\sup_u[(\nu_{h_{2,0}}(u)+h_{1,r}(u))/s_0(u)]_-^2$. The
		limits have bounded paths, so we may apply \eqref{eq:stat-lip} with
		$\mu=h_{1,r}$ and $B$ equal to the sup norm of the limit path plus one. This
		shows that $T^{AS}_{n_r}(\theta_r)$ and $T(h_{1,r},h_{2,r})$ can each be
		coupled with a copy of $T_{r,0}$ such that the difference tends to zero
		almost surely. Therefore
		\begin{align*}
			P_r^{n_r}\{T^{AS}_{n_r}(\theta_r)>c_r+\delta\}
			\le P\{T_{r,0}>c_r+\delta/2\}+o(1)
			\le P\{T(h_{1,r},h_{2,r})>c_r\}+o(1)
			\le\alpha+o(1).
		\end{align*}
		This is the argument of Theorem D.3(a) of \citet{andrews2017} combined
		with Lemma A2 of \citet{andrews2013}.
		
		\emph{(b) GMS adjustment versus true slackness.}
		Let $R_r\equiv\sup_u|\nu_r(u)|$, which is $O_p(1)$ by PS4 and the continuous
		mapping theorem. Consider a direction $u$.
		\begin{itemize}
			\item If $\xi_{n_r}(\theta_r,u)>1$, then
			$\nu_r(u)+h_{1,r}(u)>\kappa_{n_r}s_r(u)$, so
			$h_{1,r}(u)>\kappa_{n_r}s_r(u)-R_r$, while
			$\overline\varphi_r(u)=B_{n_r}s_r(u)$.
			\item If $\xi_{n_r}(\theta_r,u)\le1$, then
			$\overline\varphi_r(u)=0\le h_{1,r}(u)$.
		\end{itemize}
		Once $\kappa_{n_r}>B_{n_r}$, since $s_r\ge\sqrt\epsilon$, the inequality
		$\overline\varphi_r\le h_{1,r}$ therefore holds simultaneously over
		$\mathcal U$ on the event
		$E_r\equiv\{R_r\le(\kappa_{n_r}-B_{n_r})\sqrt\epsilon\}$. By GMS1,
		$(\kappa_{n_r}-B_{n_r})\sqrt\epsilon
		=\kappa_{n_r}(1-B_{n_r}/\kappa_{n_r})\sqrt\epsilon\to\infty$, so
		$P_r^{n_r}(E_r)\to1$. On $E_r$, every path of $\nu_{h_{2,r}}$ satisfies
		$[\nu_{h_{2,r}}+\overline\varphi_r]_-\ge[\nu_{h_{2,r}}+h_{1,r}]_-$
		pointwise. Hence $T(\overline\varphi_r,h_{2,r})\ge T(h_{1,r},h_{2,r})$
		pathwise for the same Gaussian process, and
		\begin{align*}
			c_0(\overline\varphi_r,h_{2,r},1-\alpha)\ge c_r
			\qquad\text{on }E_r.
		\end{align*}
		Replacing the true slackness by the smaller GMS adjustment weakens the
		inequalities in the Gaussian experiment. This enlarges the statistic and
		its quantile, which is the sense in which GMS is conservative.
		
		\emph{(c) Bootstrap critical value:
			$P_r^{n_r}\{c^{AS}_{n_r,1-\alpha}(\theta_r)
			<c_0(\overline\varphi_r,h_{2,r},1-\alpha)+\delta\}\to0$.}
		By PS4, $d(\widehat h_{2,n_r}(\theta_r),h_{2,0})\overset{p}{\to}0$. It
		therefore suffices to argue along a further subsequence of an arbitrary
		subsequence on which this convergence holds almost surely. Fix such a sample
		path. Conditionally on it, $\overline\varphi_r$ is a fixed nonnegative
		function, and by PS5, $(\nu_r^*,s_r^*)\Rightarrow(\nu_{h_{2,0}},s_0)$, where
		$\nu^*_r\equiv\sqrt{n_r}(\overline m^*_{n_r}-\overline m_{n_r})(\theta_r,\cdot)$
		and $s^*_r(u)\equiv\{\overline\Sigma^*_{n_r}(\theta_r,u)\}^{1/2}$. Repeat the
		argument of (a) with $(\nu_r^*,s_r^*,\overline\varphi_r)$ in place of
		$(\nu_r,s_r,h_{1,r})$, and let $P^*$ denote bootstrap probability and
		$T^*_r$ the bootstrap statistic. For every $\xi>0$ and all constants $x_r$,
		\begin{align*}
			\limsup_r\bigl[P^*\{T^*_r\le x_r\}
			-P\{T(\overline\varphi_r,h_{2,r})\le x_r+\xi\}\bigr]\le0.
		\end{align*}
		Now take $x_r\equiv c_0(\overline\varphi_r,h_{2,r},1-\alpha)+\delta-\eta$
		and $\xi\in(0,\eta-\delta)$.
		\begin{itemize}
			\item Since $x_r+\xi$ lies below the $(1-\alpha)$-quantile,
			$P\{T(\overline\varphi_r,h_{2,r})\le x_r+\xi\}<1-\alpha$.
			\item Hence $P^*\{T_r^*\le x_r\}<1-\alpha+\eta$ for all large $r$, so
			$x_r$ lies below the conditional $(1-\alpha+\eta)$-quantile.
			\item Therefore $c^{AS}_{n_r,1-\alpha}(\theta_r)>x_r+\eta
			=c_0(\overline\varphi_r,h_{2,r},1-\alpha)+\delta$ for all large $r$.
		\end{itemize}
		Bounded convergence yields the claim. This is Lemma D.4 of
		\citet{andrews2017}; it uses the fixed $\eta>\delta$.
		
		Combining (a), (b) and (c),
		\begin{align*}
			P_r^{n_r}\{\theta_r\notin CS^{AS}_{n_r}(1-\alpha)\}
			\le P_r^{n_r}\{T^{AS}_{n_r}(\theta_r)>c_r+\delta\}
			+P_r^{n_r}(E_r^c)+o(1)
			\le\alpha+o(1).
		\end{align*}
	\end{proof}
	
	Finally, let us establish uniformity by contradiction. Suppose the coverage claim of Theorem~\ref{T:theorem3} fails. Then there exist $\delta_0>0$, a strictly increasing sequence $(n_r)$, and pairs
	$(\theta_r,P_r)\in\mathcal F$ with
	$P_r^{n_r}\{\theta_r\in CS^{AS}_{n_r}(1-\alpha)\}\le1-\alpha-\delta_0$ for
	all $r$. By relative compactness of $\mathcal H_{2,\mathcal P}$, a
	subsequence satisfies $d(h_{2,P_r}(\theta_r;\cdot,\cdot),h_{2,0})\to0$ for
	some $h_{2,0}\in\overline{\mathcal H}_{2,\mathcal P}$, which
	Lemma~\ref{L:sequential-coverage} contradicts. Hence
	\begin{align*}
		\liminf_{n\to\infty}\inf_{(\theta,P)\in\mathcal F}
		P^{\,n}\{\theta\in CS_n^{AS}(1-\alpha)\}\ge1-\alpha.
	\end{align*}
	The kernel convergence used here is extracted from relative compactness,
	which rests only on \eqref{eq:env-lip}. Neither $\theta_r$ nor $h_{2,0}$
	needs to correspond to a limiting pair, so no continuity of
	$\theta\mapsto h_{2,P}(\theta;\cdot,\cdot)$ is used. \\
	The second claim of Theorem~\ref{T:theorem3} concerns power against fixed alternatives. Fix $P\in\mathcal P$ and $\theta\in\Theta\setminus\Theta_I^*(P)$. Theorem
	6.1 of \citet{andrews2017} is a fixed-alternative result, not restricted
	to compact kernel sets. It is a corollary of their Theorem E.1, whose
	conditions all hold here:
	\begin{itemize}
		\item PS4 and PS5, by item (vi) applied to the constant sequence;
		\item MFA by item (v), and SIG2 by item (iv);
		\item CI and MQ, because the single constant instrument detects the
		violation in (v) and the weight function is a point mass;
		\item S1, S3 and S4, because $S_1$ is nonincreasing, positive if and only
		if its argument is negative, and homogeneous of degree two.
	\end{itemize}
	Therefore
	\begin{align*}
		P^{\,n}\bigl\{T_n^{AS}(\theta)>c_{n,1-\alpha}^{AS}(\theta)\bigr\}
		\longrightarrow1.
	\end{align*}
	This proves Theorem~\ref{T:theorem3}.
	\qed

	\subsection{Finite and truncated direction classes}
	
	\begin{proof}[Proof of Corollary~\ref{C:as-truncations}]
		Throughout, direction sets are nonempty, and the restricted statistics
		and conditional bootstrap quantiles are assumed measurable.
		
		(a) Consider a strictly increasing sequence $(n_r)$ and a sequence
		$(\theta_r,P_r,\mathcal V_r)$ with $\mathcal V_r\subseteq\mathcal U$ and
		$(\theta_r,P_r)\in\mathcal F_{\mathcal V_r}\subseteq\mathcal F^+$.
		The envelope bound, the covariance Lipschitz bound and the domination
		$\rho_h\le M_0\rho_a$ used in the proof of Theorem~\ref{T:theorem3} hold
		uniformly on $\mathcal F^+$. Consequently, the covariance-kernel family
		$\{h_{2,P}(\theta;\cdot,\cdot):(\theta,P)\in\mathcal F^+\}$ is relatively
		compact by the same Arzel\`a--Ascoli argument, and (G1)--(G2) hold on its
		closure. Suppose that the kernels $h_{2,P_r}(\theta_r;\cdot,\cdot)$ converge
		uniformly.
		
		Apply PS4 and PS5, which are stated on $\mathcal F^+$, to the empirical and
		bootstrap processes on the full index set $\mathcal U$. The comparison bound
		\eqref{eq:stat-lip} remains valid when its supremum is restricted to
		$\mathcal V_r$. Its right-hand side is bounded by the same full-space
		process and covariance-estimation errors, independently of $\mathcal V_r$.
		Thus these errors vanish along the sequence even when $\mathcal V_r$ changes
		with $r$. Moreover, $(\theta_r,P_r)\in\mathcal F_{\mathcal V_r}$ gives
		$h_{1,n_r,P_r}(\theta_r,u)\geq0$ for every $u\in\mathcal V_r$. This is the
		nonnegativity used in the oracle-tail and GMS comparisons in Steps (a) and
		(b) of Lemma~\ref{L:sequential-coverage}. The bootstrap-quantile comparison
		in Step (c) uses only $\overline\varphi_{n_r}\ge0$. These comparisons
		therefore apply with every supremum restricted to $\mathcal V_r$, including
		in the Gaussian statistic $T(\cdot,\cdot)$ and its quantile $c_0$. The bound
		$R_r=\sup_{u\in\mathcal U}|\nu_r(u)|=O_p(1)$ in Step (b) is still taken over
		the full index set. The fixed critical-value adjustment $\eta$ supplies the
		same quantile margin as in that lemma.
		
		It follows that the sequential coverage bound holds along every such triple
		sequence with uniformly convergent kernels. If the uniform bound in (a)
		failed, there would exist $\delta_0>0$, a strictly increasing sequence
		$(n_r)$, and triples as above whose coverage probabilities are at most
		$1-\alpha-\delta_0$ for all $r$. Relative compactness supplies a
		kernel-convergent subsequence, contradicting the sequential bound. This
		proves (a).
		
		(b) Fix $P$, $\theta\notin\Theta_O(\mathcal V)(P)$, and a direction
		$u_0\in\mathcal V$ with
		$\mathbf E_P[\widetilde m(W^O,\theta,u_0)]<0$. The law of large
		numbers makes the test statistic over $\mathcal V$ diverge at order
		$n$. Using the same per-direction GMS adjustments and bootstrap
		draws, each bootstrap statistic over $\mathcal V$ is bounded by
		its counterpart over $\mathcal U$. The latter has an $O_p(1)$
		conditional critical value under the fixed law $P$, by the
		bootstrap-process bound used in the proof of Theorem~E.1 of
		\citet{andrews2017}. Hence the critical value over $\mathcal V$
		is $O_p(1)$, and the rejection probability tends to one.
		
		(c) Coverage follows from (a), since
		$B_{K(n)}\subseteq\mathcal U$ and
		$\mathcal F\subseteq\mathcal F_{B_{K(n)}}$.
		For power, fix $P$ and $\theta\notin\Theta_I^*(P)$, and choose
		$u_0\in\mathcal U$ such that
		$\mathbf E_P[\widetilde m(W^O,\theta,u_0)]=-2c<0$.
		Put $u_n=\Pi_{K(n)}u_0\in B_{K(n)}$. By \eqref{eq:env-lip},
		\begin{align*}
			\bigl|\mathbf E_P[\widetilde m(W^O,\theta,u_0)]
			-\mathbf E_P[\widetilde m(W^O,\theta,u_n)]\bigr|
			\le d_mC_1\sum_{k>K(n)}a_k
			\longrightarrow 0 .
		\end{align*}
		Thus $\mathbf E_P[\widetilde m(W^O,\theta,u_n)]\leq-c$ for all
		sufficiently large $n$. PS4, applied to the constant sequence $(\theta,P)$
		on $\mathcal U$, gives
		\[
		\sup_{u\in\mathcal U}
		\bigl|\overline m_n(\theta,u)
		-\mathbf E_P[\widetilde m(W^O,\theta,u)]\bigr|=O_p(n^{-1/2}),
		\]
		so $\overline m_n(\theta,u_n)\le-c/2$ with probability approaching one.
		Since $\overline\Sigma_n\leq M_0^2+\epsilon$, the statistic over
		$B_{K(n)}$ is then at least $nc^2/\{4(M_0^2+\epsilon)\}$.
		
		With the same per-direction GMS adjustments and bootstrap draws,
		each bootstrap statistic over $B_{K(n)}$ is bounded by the
		corresponding statistic over $\mathcal U$. Their conditional
		critical values have the same ordering, and the full-space
		critical value is $O_p(1)$ by the bound used in the proof of
		Theorem~E.1 of \citet{andrews2017}. Therefore $\theta$ is
		excluded with probability approaching one, without a rate
		restriction on $K(n)$.
	\end{proof}

	\section{Derivation of the Exact LP Representation}
	\label{sec:appendix-lp}
	
	We derive equations~\eqref{eq:lp-support-representation} and
	\eqref{eq:lp-epigraph}. Throughout, $d_m=d_y=1$, the covariate domain
	is $[0,1]^{d_x}$, and the feasible set for observation $i$ is the
	nonempty rectangle
	\begin{align*}
		D_i=[y_{iL},y_{iU}]\times
		\prod_{q=1}^{d_x}[x_{iqL},x_{iqU}].
	\end{align*}
	The argument applies to rectangular feasible domains on which the
	moment function has the stated corner-extremum property. In particular,
	additional nonrectangular feasibility restrictions cannot be ignored. \\
	The primitive cube indicators and their positive weights are given in
	Appendix~\ref{sec:weigthedcubes}. In the notation of Assumption~\ref{A:assumption2}, the weighted instrument is $g_k=\omega_k h_k$, where $h_k(x)=\mathbf 1\{x\in C_k\}$. For a complete truncation through resolution $R$, the retained
	instruments are indexed by $(a,r)$ with $r_0\le r\le R$, and $K=N_R=\sum_{r=r_0}^{R}(2r)^{d_x}$; the argument also applies if truncation stops partway through a resolution. \\
	All cube partitions use the same boundary convention: each coordinate
	interval is open on the left and closed on the right, except that the
	first interval contains zero. For each coordinate $q$, collect the
	endpoints of all retained cube sides in that coordinate, together with
	$0$ and $1$, and sort the distinct values as
	\begin{align*}
		0=\xi_{q0}<\xi_{q1}<\cdots<\xi_{qL_q}=1.
	\end{align*}
	Define $I_{q1}^*=[0,\xi_{q1}]$ and
	$I_{qj}^*=(\xi_{q,j-1},\xi_{qj}]$ for $j\ge2$.
	The common refined cells are the rectangles
	\begin{align*}
		C_\ell^*\equiv\prod_{q=1}^{d_x}I_{q,\ell_q}^*,
		\qquad \ell=(\ell_1,\ldots,\ell_{d_x}),\quad
		1\le\ell_q\le L_q.
	\end{align*}
	Their total number is $L=\prod_{q=1}^{d_x}L_q$.
	Every original cube $C_k$ is a union of refined cells, and
	$g_k(x)$ is constant on each $C_\ell^*$. \\
	To make the construction explicit when $d_x=1$, the resolution-$r$
	partition consists of $2r$ bins with endpoints $j/(2r)$,
	$j=0,\ldots,2r$. The common refinement is obtained simply by
	sorting the union of these endpoints over the retained resolutions.
	Thus the original multi-resolution instruments become step functions
	on a single refined grid. \\
	Write $g(x)=(g_1(x),\ldots,g_K(x))^\top$, and define
	\begin{align*}
		c_{\ell k}&=\omega_k\mathbf 1\{C_\ell^*\subseteq C_k\},
		&C&\equiv
		\begin{pmatrix}c_1^\top\\\vdots\\c_L^\top\end{pmatrix}
		\in\mathbb R^{L\times K}.
	\end{align*}
	For any $u\in B_K^\infty=[-1,1]^K$, the induced coefficient on
	refined cell $\ell$ is
	\begin{align}
		\label{eq:lp-refined-coefficients}
		\widetilde u_\ell\equiv c_\ell^\top u,
		\qquad \widetilde u=Cu,
		\qquad u^\top g(x)=\widetilde u_\ell
		\quad(x\in C_\ell^*).
	\end{align}
	Crucially, the set of permissible refined coefficients is
	$C[-1,1]^K$, not an independently constrained box in $\mathbb R^L$.
	One may therefore eliminate $\widetilde u$ from the optimization, as
	in~\eqref{eq:lp-epigraph}, or retain it together with the equality
	$\widetilde u=Cu$. \\
	For each observation, intersect its feasible rectangle with each
	refined cell \emph{before} taking a closure:
	\begin{align}
		\label{eq:lp-cell-intersections}
		E_{i\ell}&\equiv D_i\cap(\mathbb R\times C_\ell^*),
		&S_i&\equiv\{\ell:E_{i\ell}\ne\varnothing\},
		&R_{i\ell}&\equiv\operatorname{cl}(E_{i\ell}),
		\quad \ell\in S_i.
	\end{align}
	Each $R_{i\ell}$ is a compact rectangle. Its coordinate-$q$
	endpoints, for $\ell\in S_i$, are
	\begin{align*}
		x_{iq\ell}^L&=\max\{x_{iqL},\xi_{q,\ell_q-1}\},
		&x_{iq\ell}^U&=\min\{x_{iqU},\xi_{q,\ell_q}\},
	\end{align*}
	with the $y$ endpoints unchanged. These formulas describe the closure
	of the intersection only after nonemptiness of $E_{i\ell}$ has been
	checked using the half-open boundary convention. For example, if
	$d_x=1$, $D_i$ restricts $x$ to the singleton $\{b\}$, and a refined
	cell is $(b,c]$, then that cell does not belong to $S_i$, although
	$[b,c]$ intersects $\{b\}$. Closing first would therefore add an
	incorrect cell and could change the support function. \\
	For fixed $\theta$, assume that $m(\cdot,\cdot,\theta)$ is continuous
	and coordinatewise monotone on every nonempty $R_{i\ell}$. The
	direction of monotonicity in each coordinate can depend on $\theta$
	and on the rectangle but is fixed within that rectangle. Maximizing
	or minimizing one coordinate at a time shows that the extrema occur
	at vertices. Write
	\begin{align*}
		m_{i\ell}^-(\theta)
		&=\min_{z\in\operatorname{vert}(R_{i\ell})}m(z,\theta),
		&m_{i\ell}^+(\theta)
		&=\max_{z\in\operatorname{vert}(R_{i\ell})}m(z,\theta).
	\end{align*}
	There are at most $2^{1+d_x}$ vertices of each rectangle; in
	degenerate rectangles some of them coincide. Since $E_{i\ell}$ is
	dense in $R_{i\ell}$ and $m$ is continuous, for every real scalar
	$\widetilde u_\ell$,
	\begin{align}
		\label{eq:lp-cellwise-support}
		\sup_{(y,x)\in E_{i\ell}}
		\widetilde u_\ell m(y,x,\theta)
		&=\max\bigl\{\widetilde u_\ell m_{i\ell}^-(\theta),
		\widetilde u_\ell m_{i\ell}^+(\theta)\bigr\}.
	\end{align}
	The upper corner is relevant for $\widetilde u_\ell\ge0$ and the
	lower corner for $\widetilde u_\ell<0$. A maximizing vertex may lie
	on an excluded boundary of $C_\ell^*$; its contribution is understood
	as a limit from within $E_{i\ell}$, with the coefficient
	$\widetilde u_\ell$ held fixed. It must not be evaluated with the
	instrument vector assigned to an adjacent cell. \\
	By definition,
	$Q_{\theta,K}(W_i^O)=\operatorname{cl}_{\mathbb R^K}
	\{m(y,x,\theta)g(x):(y,x)\in D_i\}$.
	A continuous linear functional has the same supremum over a set and
	its closure. The refined cells partition the covariate domain, so
	using~\eqref{eq:lp-refined-coefficients} and
	\eqref{eq:lp-cellwise-support} gives
	\begin{align*}
		h(Q_{\theta,K}(W_i^O),u)
		&=\sup_{(y,x)\in D_i}m(y,x,\theta)u^\top g(x)\\
		&=\max_{\ell\in S_i}\;
		\sup_{(y,x)\in E_{i\ell}}
		\widetilde u_\ell m(y,x,\theta)\\
		&=\max_{\ell\in S_i}\max_{s\in\{-,+\}}
		\bigl\{m_{i\ell}^s(\theta)c_\ell^\top u\bigr\}.
	\end{align*}
	This proves~\eqref{eq:lp-support-representation}. In particular,
	there is no numerical approximation to the finite-truncation support
	function: its value is the maximum of finitely many linear functions
	of the original direction $u$. \\
	Recall from Section~\ref{sec:finite_approximation} that
	\begin{align*}
		-\widehat T_{n,K}(\theta)
		=\inf_{u\in[-1,1]^K}
		\frac1n\sum_{i=1}^n h(Q_{\theta,K}(W_i^O),u).
	\end{align*}
	For any feasible pair $(u,t)$ in~\eqref{eq:lp-epigraph},
	its constraints and~\eqref{eq:lp-support-representation} imply
	\begin{align*}
		t_i\ge h(Q_{\theta,K}(W_i^O),u)
		\qquad (i=1,\ldots,n).
	\end{align*}
	Thus its objective is at least the support minimum. Conversely,
	for any $u\in[-1,1]^K$, taking
	$t_i=h(Q_{\theta,K}(W_i^O),u)$ satisfies every epigraph
	constraint and gives exactly the sample support average. Minimizing
	over $u$ proves that the optimal value of~\eqref{eq:lp-epigraph}
	is $-\widehat T_{n,K}(\theta)$.
	
	The sample support average is continuous in $u$, since each
	observation-level support function is a finite maximum of linear
	functions. Compactness of $[-1,1]^K$ therefore ensures that the
	minimum is attained. The direction $u=0$ and $t=0$ is feasible,
	so the optimal value is nonpositive and $\widehat T_{n,K}(\theta)
	\ge0$. The epigraph variables $t_i$ must be unrestricted in sign;
	imposing $t_i\ge0$ would generally change the criterion.
	
	The reduced LP has $K+n$ variables, $2K$ box inequalities and
	$2\sum_{i=1}^n|S_i|$ epigraph inequalities. If the induced
	coefficients $\widetilde u$ are explicitly introduced, there are
	$L$ additional variables and $L$ equalities $\widetilde u=Cu$, with no change in the optimal value. Because the sample objective is continuous and convex in $u$, a unique minimum-Euclidean-norm optimizer can be selected from its compact argmin set if a deterministic tie-breaking rule is needed for sample-dependent direction selection. \\
	For a scalar index residual
	$m(y,x,\theta)=y-F(\theta_0+x^\top\theta_1)$, where $F$ is continuous
	and nondecreasing, the moment is nondecreasing in $y$ and monotone in
	each $x$ coordinate (with direction determined by the corresponding
	coefficient). This includes linear, probit and logit residuals on
	bounded rectangular domains. More generally, the argument only needs
	the cellwise extrema to be attained among finitely many known points.
	For vector-valued moments, coordinatewise monotonicity of the
	individual components is not sufficient by itself: a direction-dependent
	linear combination need not have its extrema at a common finite set
	of corners. Such models require a separate finite-support argument
	before an exact LP can be asserted.

	\section{Relation to Alternative Outer Sets}
	\label{sec:appendixE}
\begin{figure}[H]
		\centering
		\includegraphics[width=0.9\linewidth]{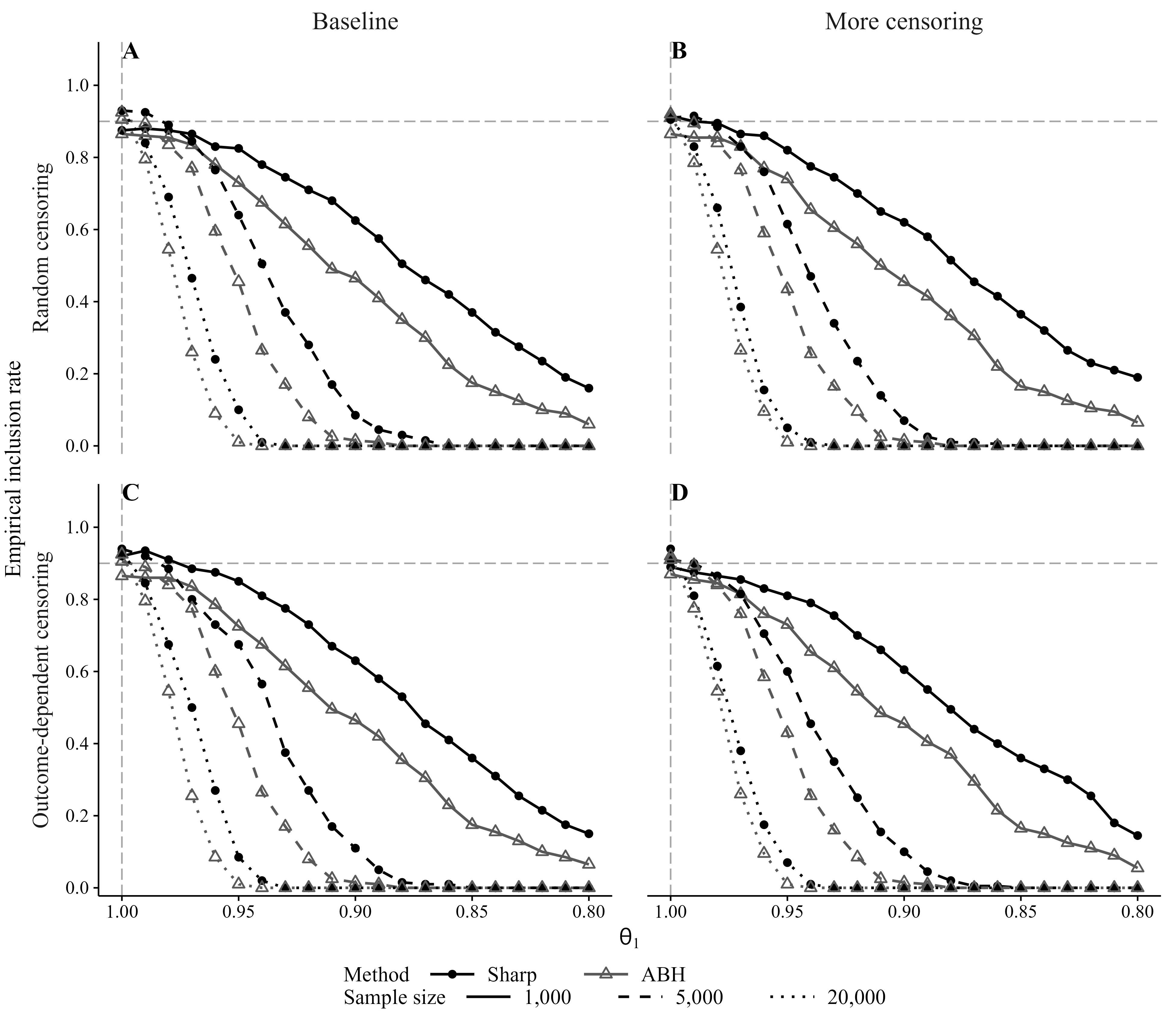}
		\vspace{0.2cm}
		\caption{Empirical Inclusion Frequencies along the Slope Path}
		\label{fig:abhcomparison}
		\vspace{0.2cm}
		\begin{minipage}{\textwidth}
			\scriptsize
			\textit{Note:} Share of 200 Monte Carlo replications in which $(0,\theta_1)$
			lies in the 90\% confidence set, for the adaptive procedure (\emph{Sharp})
			and the ABH confidence set. The true parameter
			is $(0,1)$, so the inclusion rate at $\theta_1=1$ is the coverage and the
			rates at $\theta_1<1$ are one minus the power. The horizontal dashed line
			marks the nominal level of 0.90. Baseline: $p=0.05$; more censoring:
			$p=0.25$.
		\end{minipage}
	\end{figure}

	\citet{aucejo2017} are concerned with the same conditional moment model defined by equation~\eqref{eq:model1} and Assumption~\ref{A:assumption1} above. However, their focus is on the missing data problem, i.e. $[Y_{l},Y_{u}] = \mathcal{Y}$ and $[X_{l},X_{u}] = \mathcal{X}$. Nevertheless, comparing our results with their construction develops further intuition for the identification problem; thus, in this section we discuss both approaches and their relation in detail. For simplicity, let us focus on the one moment case, thus $d_{m}=1$, although both their paper and the present work can deal with the more general case. Similarly, let us only cover the missing covariate case and abstract from fully observed additional covariates. Define $Z=1$ if the covariate is missing and $Z=0$ otherwise. \\
	\indent As mentioned in the main text, \citet{aucejo2017} note that the sharp identified set, $\Theta^{*}_{I}$, is generally too complicated an object to compute and thus they resort to constructing outer sets, i.e. supersets of the sharp identified set. The first step in \citet{aucejo2017} transforms the conditional moment conditions into unconditional moments via instrument functions as we propose for our problem in Lemma~\ref{L:lemma1}. They use the class of boxes $ \mathcal{G}_{\text{box}} \equiv \{ \mathbb{I}[X\in G(x,v)] : \forall (x,v)\in \mathcal{X}\times \mathbb{R}_{++}^{d_x}\}$ with $G(x,v) \equiv \{\tilde{x} \in \mathcal{X} : \{x_j - v_j \leq \tilde{x}_{j} \leq x_j + v_j\}_{j=1}^{d_{x}}\}$ as instrument functions.\footnote{Note that in the definition of $G(x,v)$ we use $\tilde{x} \in \mathcal{X}$ directly, thus $G(x,v) \cap \mathcal{X} \neq \emptyset$ is always true and we can avoid carrying the intersection in the definition of $M_{k}$ below.} Let us denote the collection of boxes by $\{G_{k}\}_{k\geq0} \equiv \{G(x,v) \quad \forall (x,v) \in \mathcal{X}\times \mathbb{R}_{++}^{d_x} \}$. Note that the main point of the following conceptual comparison does not rely on the specific instrument class. They propose two outer sets. \\
	\indent The first outer set in \citet{aucejo2017} uses the following vector-valued auxiliary functions for each member of $\{G_{k}\}_{k\geq0}$:
	\begin{align*}
		M_{k,1}(W^{O}) \equiv 
		\begin{pmatrix}
			-\left( \inf\limits_{y \in \mathcal{Y}, x \in G_{k}} m(y,x,\theta)  \mathbb{I}[Z=1] + m(Y,X,\theta) \mathbb{I}[Z=0, X \in G_{k}] \right) \\
			+ \left( \sup\limits_{y \in \mathcal{Y}, x \in G_{k}} m(y,x,\theta)  \mathbb{I}[Z=1] + m(Y,X,\theta) \mathbb{I}[Z=0, X \in G_{k}] \right)
		\end{pmatrix}
	\end{align*}
	They proceed by defining their outer set as
	$\Theta_{I,1}^{ABH} \equiv \{\theta \in \Theta: \ \mathbf{E}[M_{k,1}(W^{O})] \geq 0 \quad \forall k \geq 0\}$. This solves the moment problem for each box, or in general, each subspace defined by the instrument functions, but not simultaneously for all of them. As \citet{aucejo2017} mention, the difficult part of characterizing the sharp identified set is to ensure that the moment conditions hold \textit{simultaneously} for all boxes and not only individually; thus, $\Theta_{I,1}^{ABH}$ is an outer set. \\
	The second outer set defined in \citet{aucejo2017} is based on moments integrating out the missing covariate. Their auxiliary moment conditions do not depend on $k$ in the case of no additional fully observed covariates and are given by:
	\begin{align*}
		M_{k,2}(W^{O}) \equiv 
		\begin{pmatrix}
			-\left( \inf\limits_{x \in \mathcal{X}} m(Y,x,\theta)  \mathbb{I}[Z=1] + m(Y,X,\theta) \mathbb{I}[Z=0] \right) \\
			+ \left( \sup\limits_{x \in \mathcal{X}} m(Y,x,\theta)  \mathbb{I}[Z=1] + m(Y,X,\theta) \mathbb{I}[Z=0] \right)
		\end{pmatrix}
	\end{align*}
	The second outer identified set is the collection $\Theta_{I,2}^{ABH} \equiv \{\theta \in \Theta: \ \mathbf{E}[M_{k,2}(W^{O})] \geq 0\}$. \\
	\indent Let us now show the relation between $\Theta_{I,1}^{ABH}$, $\Theta_{I,2}^{ABH}$ and our sharp identified set $\Theta_{I}$. First note that the class of boxes is comprehensively revealing (see \citet{andrews2013}), i.e., it fulfills Assumption~\ref{A:assumption2}(i),\footnote{Assumption~\ref{A:assumption2}(ii) is not fulfilled by the class of boxes alone. Instead we would need to augment them with positive and decaying weights. For the identification argument here, the choice of such weights is immaterial so we abstract from them.} so our main identification result applies. Further, we know that any subset $\mathcal{U} \subseteq B_{1}^{\infty}$ defines an outer set. In particular, choose $\mathcal{U}_{point}:=\{\pm e_j : j\geq 1 \}$, where $e_j:=(0,\dots,0,1,0,\dots)$ is the $j$th basis vector. The outer set implied by this choice is
	\begin{align*}
		\Theta_{O}(\mathcal{U}_{point}) = \{\theta \in \Theta:  \mathbf{E}[h(Q_{\theta,0},u)\mathbb{I}[Z=0]] + \mathbf{E}[h(Q_{\theta,1},u)\mathbb{I}[Z=1]] \geq 0 \quad \forall u \in \mathcal{U}_{point}\}
	\end{align*}
	where we use the law of total expectation and divide our theorem into fully observed and missing covariates and denote with a slight abuse of notation the respective random sets as $Q_{\theta,z}$. \\
	Further,  considering the class of boxes, the support function for the missing covariate case only is given by
	$h(Q_{\theta,z},u) = \sup\limits_{x \in [X_{l},X_{u}]} \sum_{k \geq 0} u_{k} m(Y,x,\theta) \mathbb{I}[x \in G_{k}]$. Any choice of $\tilde{u} \in \mathcal{U}_{point}$ simplifies the support function considerably. In particular, choosing $\tilde{u} = e_{k}$ yields $h(Q_{\theta,z},\tilde{u}) = \sup\limits_{x \in [X_{l},X_{u}]} m(Y,x,\theta) \mathbb{I}[x \in G_{k}]$. For the cases for which we fully observe the covariates, $Z=0$, this further collapses to $h(Q_{\theta,0},\tilde{u}) = m(Y,X,\theta) \mathbb{I}[X \in G_{k}]$ with $X=X_{l}=X_{u}$. Thus, the condition summed over missingness patterns for the specific choice $e_{k}$ is
	\begin{align*}
		\mathbf{E}[\sup\limits_{x \in [X_{l},X_{u}]} m(Y,x,\theta) \mathbb{I}[x \in G_{k}] \mathbb{I}[Z=1]] + \mathbf{E}[m(Y,X,\theta) \mathbb{I}[X \in G_{k}] \mathbb{I}[Z=0]] \geq 0 \\
		\iff \mathbf{E}\left[\sup\limits_{x \in \mathcal{X}} m(Y,x,\theta) \mathbb{I}[x \in G_{k}] \mathbb{I}[Z=1] + m(Y,X,\theta) \mathbb{I}[Z=0, X \in G_{k}]\right] \geq 0
	\end{align*}
	where we used $[X_{l},X_{u}] = \mathcal{X}$ for the missing data case. Simultaneously, using $\sup\limits_{a \in A} f(a) = - \inf\limits_{a \in A} \left(- f(a) \right)$, and considering $\tilde{u}=-e_{k}$ we obtain:
	\begin{align*}
		\mathbf{E}\left[ - \left( \inf\limits_{x \in \mathcal{X}} m(Y,x,\theta) \mathbb{I}[x \in G_{k}] \mathbb{I}[Z=1] + m(Y,X,\theta) \mathbb{I}[Z=0, X \in G_{k}] \right)\right] \geq 0
	\end{align*}
	These expressions are almost equivalent to the elements of the vector $M_{k,1}$ defined above; however, whereas we take the supremum over $x \in \mathcal{X}$, \citet{aucejo2017} are considering $y \in \mathcal{Y}, x \in G_{k}$. Additionally, our result includes the indicator $\mathbb{I}[x \in G_{k}]$ in the objective for the missing observations. Notice that taking the supremum over the entire support of $Y$ even for observed values is not using the entire information provided from the data. \\
	The second difference is that instead of maximizing over the full support as in Theorem~\ref{T:theorem1} above, \citet{aucejo2017} maximize over the individual boxes. In combination with also varying $Y$ for the missing $X$ observations this still yields an outer set. However, conceptually it is assuming that for each box, the value of the missing covariate must lie within the specific box. This is stronger than required by the conditional moment restriction. The conditional moment model is only implying that the moment condition has to hold over all subspaces, not that it must hold under all subspaces assuming that all realized values are within that space. Instead, our construction maximizes $m(Y,x,\theta) \mathbb{I}[x \in G_{k}]$ over $x \in \mathcal{X}$, thus it allows for a specific observation not contributing to $G_{k}$.\footnote{Another way to see this point is to consider a simpler setting. Suppose $m(y,x,\theta) = x$ with $\mathcal{X} = [0,1]$, thus the moment condition boils down to $\mathbf{E}[X|X] = 0$. If $X$ is missing for all observations we cannot rule out that $P_{x} = \delta_{0}$, a mass point at zero. The latter is the only distribution of $X$ fulfilling the moment condition. For any box in $[0,1]$ and $u$ vector a feasible solution to our support function approach is zero as we can either choose $x=0$ or another value with associated $u=0$. Instead, considering any box $[a,b]$ with $a \neq 0$, the first moment of the first outer set in \citet{aucejo2017} is negative, thus violated. We thank Federico Bugni for bringing up this example.}  \\
	Taking all of the above into account, the preceding comparison does not by itself establish a general ranking between $\Theta_{I,1}^{ABH}$ and $\Theta_O(\mathcal U_{point})$. The main takeaway is that our identification result also considers combinations of boxes. It is the requirement that the moment conditions hold simultaneously across boxes that makes the sharp identified set difficult to compute. Adding directions to our support-function approach yields tighter outer sets which, under Theorem~\ref{T:theorem1}, converge to the sharp identified set. Since the sharp set is contained in $\Theta_{I,1}^{ABH}$ under the same maintained assumptions, the limiting set is contained in $\Theta_{I,1}^{ABH}$, although the inclusion need not be strict. \\
	For $\Theta_{I,2}^{ABH}$, an equivalence follows from a single box. Because $\mathcal X$ is compact, we can include a box $G_{k_0}=\mathcal X$ in our instrument class. The directions $e_{k_0}$ and $-e_{k_0}$ then yield, after dividing by the positive instrument weight,
	\begin{align*}
		\mathbf E\!\left[
		\sup_{x\in\mathcal X}m(Y,x,\theta)\mathbb I[Z=1]
		+m(Y,X,\theta)\mathbb I[Z=0]
		\right]&\geq0,\\
		\mathbf E\!\left[
		-\inf_{x\in\mathcal X}m(Y,x,\theta)\mathbb I[Z=1]
		-m(Y,X,\theta)\mathbb I[Z=0]
		\right]&\geq0.
	\end{align*}
	These are precisely the two inequalities defining $\Theta_{I,2}^{ABH}$ in the absence of additional fully observed covariates. Thus $\Theta_O(\{e_{k_0},-e_{k_0}\})=\Theta_{I,2}^{ABH}$ and, in particular, $\Theta_O(\mathcal U_{point})\subseteq\Theta_{I,2}^{ABH}$. \\
	We note here that the outer sets proposed by \citet{aucejo2017} can nevertheless be quite informative in certain settings. In particular, in the point-identified design of Section~\ref{sec:simulation}, the power against fixed local alternatives of the confidence sets based on \citet{aucejo2017}, as implemented via the bootstrap of \citet{andrews2017}, compares favorably to our method in finite samples. It seems that in cases in which some parts of the support of the regressor are not affected by censoring, checking the moment condition instrument by instrument or integrating out the unobserved component does not yield a significant loss of information. Figure~\ref{fig:abhcomparison} shows the comparison of our adaptive procedure versus \citet{aucejo2017}, similar to the main text Figure~\ref{fig:pointid_power}.

\end{document}